\documentclass[journal,transmag]{IEEEtran}

\IEEEoverridecommandlockouts                              % This command is only needed if 
\usepackage{hyperref} 

\usepackage{graphicx}
\usepackage{amsmath}
\usepackage{enumitem}
\usepackage{bbm}

\usepackage{cuted}
\usepackage{lipsum}
\usepackage{widetext}
\usepackage{flushend}

\newcommand{\overbar}[1]{\mkern 1.5mu\overline{\mkern-1.5mu#1\mkern-1.5mu}\mkern 1.5mu}
\usepackage{multirow}

\usepackage{mathrsfs}
\usepackage{color,soul}
\usepackage{algorithm}
\usepackage{algpseudocode}
\usepackage{amsmath,amssymb}
\usepackage{amsmath,amsfonts,amssymb}

\usepackage{amsthm}

\usepackage{eurosym}
\usepackage{subfig}
\newtheorem{theorem}{Theorem}[section]

\newtheorem{remk}{Remark}

\newcommand{\cbrac}[1]{\left\{ #1\right\}}
\newcommand{\sbrac}[1]{\left[ #1\right]}
\title{\huge{Storage Optimal Control under Net Metering Policies}}

\author{Md Umar Hashmi$^{1}$, Arpan Mukhopadhyay$^{2}$, Ana Bu\v{s}i\'c$^{1}$, and Jocelyne Elias$^{3}$% <-this % stops a space
	\thanks{$^{1}$M.U.H. and A.B are with INRIA, DI ENS, Ecole Normale Sup\'erieure, CNRS, PSL Research University, Paris, France.}%
	\thanks{$^{2}$A.M is with Department of Computer Science, University of Warwick. }%
	\thanks{$^{3}$J.E. is with DISI, University of Bologna, Italy, E-mail: jocelyne.elias@unibo.it.}%
}
\begin{document}
	%\title{Optimal control of storage under time varying electricity prices}
	%\author{\IEEEauthorblockN{Md Umar Hashmi}
	%\IEEEauthorblockA{INRIA \& ENS Paris}
	%\and
	%\IEEEauthorblockN{Arpan Mukhopadhyay}
	%\IEEEauthorblockA{EPFL Lausanne}
	%\and
	%\IEEEauthorblockN{Ana Bu\v{s}i\'{c}}
	%\IEEEauthorblockA{INRIA \& ENS Paris}
	%\and
	%\IEEEauthorblockN{Jocelyne Elias }
	%\IEEEauthorblockA{Université Paris Descartes }}
	\maketitle
	\begin{abstract}
		Electricity prices and the end user net load vary
		with time. Electricity consumers equipped with energy storage devices can perform energy arbitrage, i.e., buy when energy is cheap or when there is a deficit of energy, and sell it when it is expensive or in excess, taking into account future variations in price and net load.
		Net metering policies indicate that many of the utilities apply a {customer selling} rate lower than or equal to the retail {customer buying rate}
		in order to compensate excess energy generated by end users.
		In this paper, we formulate the optimal control problem for an end user energy storage device 
		in presence of net metering. We propose a computationally efficient algorithm, with worst case run time complexity of quadratic in terms of number of samples in lookahead horizon, that computes the optimal energy ramping rates in a time horizon. The proposed algorithm exploits the problem's piecewise linear structure and convexity properties for the \textit{discretization} of optimal Lagrange multipliers. The solution has a \textit{threshold-based structure} in which optimal control decisions are independent of past or future price as well as of net load values beyond a certain time horizon, defined as a \textit{sub-horizon}.
		Numerical results show the effectiveness of the proposed model and algorithm. Furthermore, we investigate the impact of forecasting errors on the proposed technique. We consider an Auto-Regressive Moving Average (ARMA) based forecasting of net load together with the Model Predictive Control (MPC). We numerically show that adaptive forecasting and MPC significantly mitigate the effects of forecast error on energy arbitrage gains.
	%	\textcolor{cyan}{In this paper, we present parameter dependencies on the energy arbitrage problem with respect to battery variables, consumer load, renewable generation and electricity price. }
		%\par Furthermore, we present an autoregressive moving average (ARMA) based forecasting of net load to be used for model predictive control (MPC). We numerically show using real data that adaptive forecasting and MPC significantly mitigate the effects of forecast error on energy arbitrage gains. 
		%We finally analyze numerically this latter using real data.
	\end{abstract}
	\vspace{-13pt}
	\section{Introduction}
	% \textcolor{blue}{I read the modifications of Arpan (thanks a lot!) in red and they seem good to me. I just did some corrections of typos ...\\}
	
	The share of total energy consumed worldwide by commercial and residential buildings {is} around 20\% and is projected to grow at an average {rate} of 1.4\% per year from 2012 to 2040 \cite{eia2016}. The environmental benefit of connecting more renewable energy sources in meeting the global energy demand is, therefore, irrefutable.
	Nowadays, utilities encourage end users to install distributed generation (DG) sources to satisfy their own energy demands as well as that of others in the {grid~\cite{res_net}}.
	%like solar PV, wind, biogas, biomass, geothermal energy or small hydro
	Any excess power generated by such sources {is} bought back by the grid and compensated at every billing cycle (monthly or yearly). 
	%Nowadays, utilities encourage end users to install distributed generation (DG) sources to satisfy their own energy demands as well as that of others in the grid.  
	%%like solar PV, wind, biogas, biomass, geothermal energy or small hydro
	%\cite{res_net}. Any excess power generated by such sources are bought back by the grid and compensated at every billing cycle (monthly or yearly). 
	This is achieved by \textit{Net Metering}, which facilitates bi-directional power flow to and from an end user.
	Net metering is a policy designed to encourage private investment in renewable energy \cite{wiki_net}. 
	{The fruits of this policy are already visible in California, where more than one million 
		%	over 920,000 
		solar projects with a cumulative capacity exceeding 7 GW \cite{public_power}} are operational.

	{Future adoption of DG is sensitive to the net-metering policy adopted to compensate for a surplus generation. A favorable retail rate would accelerate DG installations \cite{darghouth2016net}. However, incentivizing DG owners pushes utilities to increase charges per kWh for all the customers, creating a disadvantage for non-DG compared to DG owners \cite{eid2014economic}. Therefore, utilities have to find a balance between promoting DG and penalizing non-DG owners with higher electricity rates \cite{public_power}.
		As a result, net metering policies vary significantly by country and by state or province. For instance, in the US each state has its own mechanism 
		for consumer based DGs.
		%to deal with DG by consumers. 
		Net-Energy Metering or NEM can be categorized into NEM 1.0 and NEM 2.0, which are the two
		versions of net-metering policy in California. 
		NEM 1.0 refers to a policy where buying and selling price{s} are equal. 
		When number of users with DG reaches a set cap, additional users are compensated at a rate lower than retail rate. 
		This version of net-metering is referred as NEM 2.0 \cite{cpuc_net,gong2017financial}.
		%and
		%refers to a policy where the selling price is lower than or equal to the buying price \cite{cpuc_net}, \cite{gong2017financial}. 
		%Similar compensation strategies are also applicable for rest of the world.
		Similar compensation mechanisms for surplus consumer generation  are also in use in other regions \cite{wiki_net}.}
	%: (i)
	%\textit{Retail Rate}: The selling price is equal to the buying price of electricity for the end user for each time instant, (ii)
	%\textit{Lower than the Retail rate}: There are various compensation mechanisms where the selling price is lower than the buying price, such as (a.) selling price is a function of buying price, (b.) \textit{Avoided Cost}: is the minimum amount an electric utility is required to pay an end user, (c.) \textit{Generation Rate}: selling price is approximately half the retail rate.
	%\par Note that within a state there could be multiple utilities, the rules for net metering can differ for different utilities, for example, net metering in California is discussed in \cite{cpuc_net}.

	{Combining energy storage with renewables holds several benefits for storage owners.}
	% which would partially meet their consumption and supply the excess energy back to the grid. 
	For example, the Salt Lake Project in Phoenix Arizona (1995) demonstrated that the consumption peak and the solar PhotoVoltaic (PV) generation peak are not aligned.
	Therefore, energy storage provides flexibility in using the surplus generation when the demand is high \cite{palomino1997performance}.
	% the use of storage for load matching is proposed in . 
	Storage also adds {an} economic value to {consumers} by enabling the use of the stored energy based on the electricity prices \cite{ren2016optimal} and the user's own energy demand.  
	This is achieved through energy arbitrage, i.e., 
	buying energy from the grid when the buying price is lower and there is not enough stored energy and selling energy to the grid when the selling price is high and enough stored energy is available.
	%Renewables are intermittent in terms of power generation, and dependent on exogenous parameters like solar irradiance or wind speed. 
	The design of electricity prices indicates that energy arbitrage will be more profitable
	% due to the increased volatility caused by
	with greater integration of renewables leading to more volatile electricity prices \cite{Borenstein_Econ_2005,hashmi2018effect}.
	
	%The variability in electricity price is essential for making energy storage financially viable in coming future. This implies a price based demand response for grid with large percentage of renewables will be much more volatile. 
	%Volatility in price will be brought down by all the responsive loads and generation. There will be a point of minimum price volatility above which the installation of PV-battery will be profitable \cite{mulder2013dimensioning}.
	%
	%\par Net metering facilitates bi-directional power flow to and from an end user. Utilities encourage end users to install distributed generation (DG) sources like solar PV, wind, biogass, biomass, geothermal energy or small hydro \cite{res_net}, any excess power generated by such sources are bought back by the grid and compensated at every billing cycle (monthly or yearly).

	{In this paper, we 
		formulate the energy arbitrage problem as an optimization
		problem under NEM 2.0 net-metering policies (with unequal buying and selling electricity prices).
		This formulation also 
		applies to the NEM 1.0 policy as a special case.
		%(with equal buying and selling electricity prices).
		%generalize the threshold based structure of storage operation 
		%introduced in \cite{hashmi2017} for NEM 1.0 to both NEM 1.0 and NEM 2.0 net-metering  policies.
		%We use a similar search strategy for finding the shadow price levels in a sub-horizon.
		%Sub-horizon denotes a smaller portion of the longer time horizon where storage control actions are coupled.
		%These levels of shadow price required to identify a sub-horizon are equivalent to the derivative of consumption cost paid by the consumer with respect to the optimization variable.
		%In this paper we use the change in storage charge level as the optimization variable, therefore, the shadow price defined takes the value which is governed by the price levels in the sub-horizon and storage charging and discharging efficiency.
		%We formulate the storage arbitrage problem applicable for storage control under net-metering NEM 2.0, where 
		%We consider a consumer with inelastic load and renewable generation. 
		%Based on the selection of optimization variable, we show the convexity of the objective function of operating energy storage in the presence of inelastic load and renewable generation.
		%The cost function under NEM 2.0 has a convex piecewise linear structure, therefore,
		%%The objective function is not strictly convex, so 
		%The cost function for the optimal arbitrage problem for storage with inelastic load and renewable generation is convex and piecewise linear \cite{hashmi2019lp}. 
		Time of operation of the storage is divided into discrete intervals called {\em instants} where the 
		prices remain constant.
		%The cost of energy arbitrage is the sum of the costs in all these instants.
		We show that the cost function of the arbitrage problem (the sum of the costs of all instants)
		is convex and piecewise linear.
		However, as shown later in the paper, standard linear programming (LP) based algorithms 
		and convex optimization solvers are not very efficient for real time
		operations of storage  especially when large volumes of price data need to be handled.
		We, therefore, propose an algorithm based on the Lagrangian dual of the original arbitrage problem.
		Exploiting the fact that the objective function is piecewise linear we derive the 
		optimal action at each instant as a function of the current state of the battery
		and the value of an {\em accumulated Lagrange multiplier} at that instant which captures the information
		about electricity prices at all future instants. We observe that the mapping between
		the optimal action and {the} accumulated Lagrange multiplier at each
		instant has a threshold structure. Based on this threshold structure we propose a fast and efficient way of
		finding the optimal accumulated Lagrange multiplier values and corresponding optimal actions
		at each instant. Numerical comparisons show that the proposed algorithm is orders of magnitude faster
		than standard solvers and is, therefore, suitable for real-time use with large volumes of price data to process
		at each instant. 
		The key contributions of this paper are given below:\\
		$\quad \bullet$ {\em Formulation}: We formulate the optimal arbitrage problem under NEM 2.0 policies as an optimization problem and show that the problem is convex when expressed
		in terms of the energy level differences between consecutive instances.\\
		%\textcolor{red}{In this paper, we apply the Lagrangian theory for performing optimal arbitrage proposed in \cite{cruise2014optimal2,cruise2019control} for a convex piecewise linear cost function for a consumer with inelastic load and renewable generation. Using this structure of the cost function, we develop an efficient algorithm applicable to different net-metering policies.}
		%{for consumer with inelastic load and renewable generation considering storage ramp and capacity constraints}. \\
		$\quad \bullet$ {\em Optimal actions}:
		Exploiting the convexity and the piecewise linear nature of the cost function, we derive a closed form expression
		of the optimal action at each instant as a function of an accumulated Lagrange multiplier that
		captures the information about future electricity prices, battery constrains, and charging and
		discharging efficiencies. This mapping has
		a discrete threshold based structure which allows us to devise an efficient algorithm
		to find the optimal accumulated Lagrange multiplier values and the associated 
		optimal actions. The worst case time complexity of the algorithm is found to be
		$O(N^2)$, where $N$ is the number of discrete time steps. 
		%However,
		%for most practical cases we found the run-time to be growing linearly in $N$, \textcolor{cyan}{refer to numerical case-study presented in Appendix~\ref{appendixcomplexity} in the supplementary material, where we compare the run-time of the proposed algorithm with LP and convex optimization based benchmarks.}\\
		Numerical case-study is presented in Section~\ref{appendixcomplexity}, where we compare the run-time of the proposed algorithm over LP- and convex optimization-based benchmarks.\\
		%Prior works \cite{van2013optimal, qin2012optimal,gast2013impact}, \cite{petrik2015optimal} indicate that the optimal actions for an energy storage device performing energy arbitrage have a threshold-based structure.
		%%The threshold-based structure of the proposed solution provides a set of feasible solutions in a sub-horizon. A backward step is, therefore, used for selecting the optimal trajectory in a sub-horizon. 
		%We provide the conditions for selecting these thresholds \textcolor{red}{with respect to the shadow price\footnote{In optimization, the shadow price is the value of the Lagrange multiplier at the optimal solution.} for the arbitrage problem.
		%Identification of these thresholds is possible due to the convex and piecewise linear structure of the cost function. As the optimal storage operation can be selected from the discrete number of states of the thresholds, thus, the optimization efficiently discretized.}
		%% for storage performing optimal arbitrage problem.
		%{Storage thresholds for optimal arbitrage is a function of (a) price for buying and selling, (b) storage charging and discharging efficiency,
		%	(c) ramping constraint, (d) the relationship between the ratio of selling and buying price and round-trip storage efficiency and
		%	(e) consumption level seen by energy meter.} %Refer to Remark~\ref{rmk:remark2} to \ref{rmk:remark5}.
		%\textcolor{red}{Furthermore, we also provide a framework for searching these shadow price levels.}
		$\quad \bullet$ {\em Sub-horizon selection}: We observe that in order to determine the optimal action 
		in a certain period within the total time horizon
		it is sufficient to consider prices only within a sub-horizon, which is often much smaller than the entire horizon.\\
		%simplifies the problem 
		%We identify
		%\textcolor{red}{We observe that in a sub-horizon, the shadow price remains constant. The shadow prices are selected from a finite set function of electricity price levels and storage efficiencies, unlike \cite{cruise2019control} where a rigorous search is performed.}
		{The length of sub-horizon depends on 
			storage constraints and the electricity price variation.} 
		Using this observation we considerably speed up our algorithm which only needs to look at 
		a small subset of future price data to make optimal decision.\\% at an instant.\\
		%\item {\em Closed form computationally efficient algorithm}: Due to the discrete nature of the 
		%optimization problem, we explicitly characterize the worst case run time complexity
		%of the proposed arbitrage algorithm to be quadratic in the number of time steps. 
		%{The discretization is governed by electricity price levels in the sub-horizon and storage charging and discharging efficiency}.
		%The proposed algorithm is computationally more efficient compared to linear programming \cite{hashmi2019lp} and convex optimization based benchmarks, see Appendix~\ref{appendixcomplexity}. 
		$\quad \bullet$ {\em Numerical evaluation}: 
		We compare our proposed algorithm with other commercial solvers and observe
		orders of magnitude improvement in efficiency. 
		{For real-time implementation of the proposed algorithm}, we apply an
		ARMA-based forecast model with model predictive control (MPC) and 
		numerically analyze the effect of forecast errors on arbitrage gains using 
		real data from Pecan Street \cite{pecan} and ERCOT price data \cite{ENOnline}. \vspace{-12pt}
		\subsection{Related work}
		\vspace{-3pt}
		The problem of optimal energy arbitrage using storage
		has been the subject of many recent works e.g., 
		\cite{van2013optimal,mokrian2006stochastic,anderson2016co,kim2011optimal}.
		In \cite{van2013optimal,ratnam2013optimization, mishra2012smartcharge} a system with 
		rooftop solar PV and energy storage for a residential setting has been considered.
		However, selling of the stored energy to the grid is not allowed in \cite{van2013optimal}, where an MDP-based approach has been taken to tackle the arbitrage problem where the future costs are discounted by a constant factor.
		In \cite{van2013optimal, qin2012optimal}, \cite{petrik2015optimal, gast2013impact} 
		it has been shown that the optimal decisions for arbitrage have a threshold based structure. 
		In this paper, we explore a similar structure for NEM 2.0
		net-metering policies where the buying and selling prices of electricity can be unequal.
		Our work is inspired by the prior work of Cruise {\em et al} ~\cite{cruise2019control,cruise2014optimal}
		where a general convex cost function has been considered for energy arbitrage and the concept
		of accumulated Lagrange multiplier is introduced. The concept of decision horizon is also
		described there and in prior work~\cite{chand2002forecast}.
		However, due to the generality of the problem, finding the optimal accumulated Lagrange multiplier was done through an exhaustive search in a continuous range. In contrast, we deal with a piece-wise linear convex cost function which enables
		us to derive a discrete set of threshold values for the accumulated Lagrange multipliers. These discrete values enable us to design a significantly more efficient algorithm to search for the optimal Lagrange multipliers and their corresponding actions.
		Furthermore, our cost function is non-differentiable due to which we propose
		an algorithm going backward in time to choose the optimal action from an envelope of possible actions. 
		Finally, unlike \cite{cruise2019control} our formulation also considers consumer load and DG.

		We observe that under NEM 2.0 compensation, consumer inelastic load and renewable generation time variation cannot be ignored as the case for NEM 1.0 \cite{xu2017optimal}, \cite{hashmi2017}.
		Modeling  uncertainty in price and net load are essential for real-time operation of energy storage.}
	% with storage control is essential for real-time operations.
	Authors in \cite{wang2018energy} use reinforcement learning for real-time energy arbitrage. In \cite{anderson2016co}, \cite{abdulla2018optimal} a receding horizon dynamic programming method for mitigating uncertainty 
	has been investigated. 
	\cite{hu2010optimal} 
	proposes a deterministic setting for optimal storage control using spot market prices of electricity available one day ahead.
	We use MPC with forecasting for reducing the effect of uncertainty on arbitrage gains.
	%In \cite{anderson2016co,walawalkar2007economics}, the 
	%authors consider the application of energy storage for not only energy arbitrage 
	%but also for providing ancillary services to the grid. 
	
	%
	%================================
	
	{\em Organization}: The paper is organized as follows. Section II introduces the system model. 
	Section III presents a mathematical framework and proposes an algorithm for solving the arbitrage problem. 
	Section IV presents a real-time implementation of the proposed optimal arbitrage algorithm.
	%Section IV presents an online algorithm using the proposed optimal arbitrage algorithm.
	%using ARMA forecasting in the  MPC framework. 
	Section V discuss numerical results and 
	%Section VI provides a discussion on broader aspects of the problem and the proposed solution.
	section VI concludes the paper. \vspace{-14pt}
	\vspace{-5pt}

	\section{System Description}
\begin{figure}[!htbp]
		\center
		\includegraphics[width=2.4in]{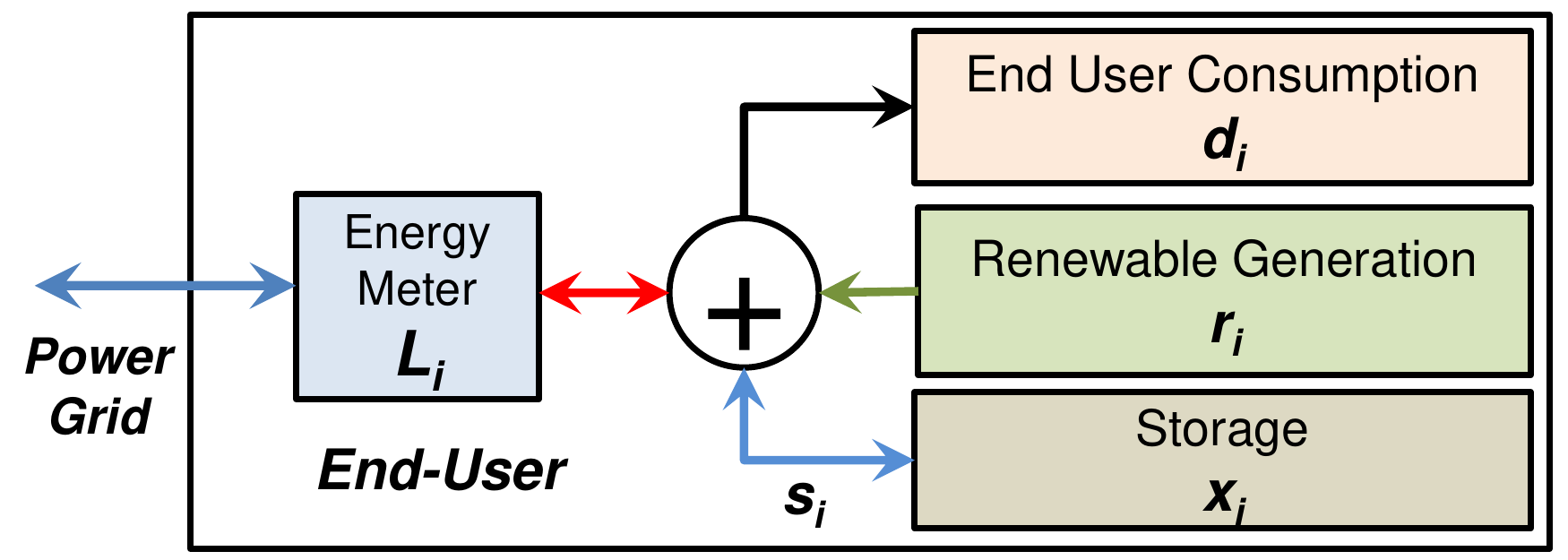} \vspace{-6pt}
		\caption{\small{Behind-the-meter electricity consumer with inelastic consumption, renewable generation and energy storage.}}\label{systemblock}
	\end{figure}

%	\vspace{-10pt}
	\vspace{-4pt}
	We consider the operation of an electricity user over a fixed period of time. 
	The user is assumed to be equipped with renewable generation such as a rooftop solar PV and a battery to store excess generation. It is also connected to the electricity grid
	from where it can buy or to which it can sell energy.
	The objective is to find an efficient algorithm for a user to
	make optimal decisions over a period of varying electricity prices considering variations in the solar generation and end user load. 
	The total duration, $T$, 
	of operation is divided into $N$ steps, indexed by $\{1,...,N\}$, such that in each step the buying and selling price of electricity remains constant. 
	The duration of step $i \in \{1,...,N\}$ is denoted as $h_i$. Hence, $T=\sum_{i=1}^{N} h_i$.
	%The time duration $T$ of the user's operation
	%is typically chosen as one day~\cite{mokrian2006stochastic,hu2010optimal} since the approximate pattern of electricity prices repeats with a period of one day. 
	%In the present work, we consider different buying and selling prices of electricity, denoted as 
	%The price of electricity in the $i^{\text{th}}$ step is defined as
	%\begin{equation}
	%p_{\text{elec}}(i)=
	%\begin{cases}
	%p_b(i) ,& \text{if consumption } \geq 0 ,\\
	%p_s(i) , & \text{otherwise,}
	%\end{cases}
	%\end{equation} 
	%where $p_b$ and $p_s$ denotes buying and selling price respectively.
	The price of electricity, $p_{\text{elec}}(i)$
	%, equals the buying price, $p_b(i)$, if the consumption is positive; otherwise $p_{\text{elec}}(i)$ equals the selling price, $p_s(i)$; 
	is given as 
	\vspace{-5pt}
	\begin{equation}
	p_{\text{elec}}(i)=
	\begin{cases}
	p_b(i) ,& \text{if consumption } \geq 0 ,\\ 
	p_s(i) , & \text{otherwise,} 
	\end{cases}\vspace{-7pt}
	\end{equation} 
	Note $p_{\text{elec}}$ is ex-ante and {the} consumer is a price taker.
	The ratio of selling and buying price is denoted as $\kappa_i = \frac{p_s(i)}{p_b(i)}$.
	%\begin{figure}[!htbp]
	%	\center
	%	\includegraphics[width=2.2in]{sys}
	%	\caption{\small{Power Flow Block Diagram}}\label{a_sys}
	%\end{figure}
	%Fig.~\ref{a_sys} shows the block diagram of energy flow for an end user.
	The end user inelastic consumption is denoted as $d_i$
	%$d_i$ represents inelastic end-user load 
	and 
	renewable generation is given as $r_i$. Fig.~\ref{systemblock} shows the block diagram of the system, i.e., an electricity consumer with renewable generation and battery.
	%generates $r_i$ units of energy through renewable sources in time step $i$. 
	%The quantities $d_i$ and $r_i$ are assumed to be known a priori. 
	Net energy consumption without storage is denoted as $z_i = d_i - r_i ~ \in \mathbb{R} $.

	The efficiency of charging and discharging of the 
	battery are denoted by $\eta_{\text{ch}}, \eta_{\text{dis}} \in (0,1]$, respectively. 
	%This implies that if the amount of energy gain in the battery 
	%is $B$ then the amount of energy bought from the grid or 
	%supplied by the renewables is $B /\eta_{\text{ch}}$. 
	%Similarly, when the battery supplies $B$ amount of energy only  $B  \eta_{\text{dis}}$ reaches out of the battery.  
	We denote the change in the energy level of the battery at $i^{\text{th}}$ instant by $x_i$= $h_i  \delta_i$,
	where $\delta_i$ denotes the ramp rate such that
	% at $i^{\text{th}}$ instant 
	$\delta_i \in [\delta_{\min}, \delta_{max}]$ ~$\forall ~i$ and $\delta_{\min} \leq 0,\delta_{\max} \geq 0$ are the minimum and the maximum ramp rates (kW); 
	$\delta_i > 0$ implies charging and $\delta_i < 0$ implies discharging. 
	Storage energy output 
	%in the $i^{th}$ instant 
	is given as
	\vspace{-8pt}
	\begin{equation}
	s_i =f(x_i)= \frac{1}{\eta_{\text{ch}}}[x_i]^+ - \eta_{\text{dis}}[x_i]^-, \vspace{-6pt}
	\label{finverse}
	\end{equation}
	%\vspace{-5pt}
	%The charge extracted from or supplied to a battery is represented in terms of $s_i$ as
	%We define $s_i$ as a function of $x_i$ represented as $s_i=f(x_i)$.
	where $x_i$ must lie in the range  from $X_{\min}^i=\delta_{\min}h_i$ to $X_{\max}^i={\delta_{\max}h_i}$.
	Note $[x_i]^+=\max(0, x_i)$ and $[x_i]^-= \max(0, -x_i)$.
	Alternatively, we can write $x_i=  \eta_{\text{ch}}[s_i]^+ - \frac{1}{\eta_{\text{dis}}}[s_i]^-$.
	%{Clearly, there is a one-to-one correspondence between $x_i$ and $s_i$ for every $i$.}
	%\end{equation}
	The limits on $s_i$  are given as \vspace{-5pt} 
	\begin{equation}
	s_i \in [S_{\min}^i, S_{\max}^i], \text{where } S_{\min}^i\text{=}\eta_{\text{dis}}X_{\min}^i, S_{\max}^i\text{=} \frac{X_{\max}^i}{\eta_{\text{ch}}}. \vspace{-3pt}
	\end{equation}
	% $ s_i \in [S_{\min}^i, S_{\max}^i]$, where $S_{\min}^i$=$\eta_{\text{dis}}\delta_{\min}h_i$ and $S_{\max}^i$=$ \frac{\delta_{\max}h_i}{\eta_{\text{ch}}}$.
	
	Let $b_i$ denote the energy stored in the battery at the $i^{\textrm{th}}$ step. 
	The battery capacity is defined as \vspace{-5pt}
	\begin{equation}
	b_i = b_{i-1} + x_i, \quad b_i\in [b_{\min},b_{\max}],  \forall i,
	\end{equation}
	%Then, 
	%$b_i = b_{i-1} + x_i$. The capacity of the battery imposes the constraint $b_i\in [b_{\min},b_{\max}],  \forall i$, 
	where $b_{\min}, b_{\max}$ are the minimum and the maximum battery capacity (for avoiding over-charging and over-discharging). 
	% Let $b_i$ denote the energy stored in the battery at the $i^{\textrm{th}}$ step. Then, 
	% $b_i = b_{i-1} + x_i$. The capacity of the battery imposes the constraint $b_i\in [b_{\min},b_{\max}],  \forall i$, 
	% where $b_{\min}, b_{\max}$ are the minimum and the maximum battery capacity. 
	The total energy consumed between time step $i$ and $i+1$ is given as $L_i = z_i+s_i$. 
	% \vspace{-10pt}
	\vspace{-12pt}
	\vspace{-15pt}

	\section{Optimal Arbitrage Problem}
	The optimal energy arbitrage problem is defined as the minimization of the 
	cost of consumption
	%, \textcolor{blue}{XXX $\min \sum_{i=1}^N C_{nm}^{i}(x_i)=\sum_{i=1}^N L_i p_{\text{elec}}(i)$ XXX???},
	subject to the battery constraints.
	It is given as follows:
	\vspace{-4pt}
	\begin{gather}
	\text{($P_{\text{NEM}}$)~~  } 
	\min  \sum_{i=1}^N C_{nm}^{i}(x_i),
	\end{gather}
	\vspace{-15pt}
	\begin{gather*}
	\text{subject to, } 
	b_{\min} - b_0\leq \sum_{j=1}^i x_j \leq b_{\max}- b_0 , ~\forall i \in \{1,..,N\},\\ \vspace{-12pt}
	x_i  \in \left[X_{\min}^i , X_{\max}^i\right]  ~\forall i \in \{1,..,N\},
	%\vspace{-15pt}
	\end{gather*}
	where $C_{\text{nm}}^{i}(x_i)$ denotes the energy consumption cost function at instant $i$ and is given by \vspace{-5pt}
	\begin{equation}
	C_{\text{nm}}^{i}(x_i) = [z_i + f(x_i)]^+ p_b(i) - [z_i + f(x_i)]^- p_s(i).
	\end{equation}

	\begin{theorem}
		\label{thm:convexity}
		If $p_b(i) \geq p_s(i)$ for all $i=\{1,...,N\}$, then problem ($P_{\text{NEM}}$) is convex in $x$. 
	\end{theorem}
	
	{From 
		Theorem~\ref{thm:convexity}, which is proved in  Appendix~\ref{prf:convexity}, we 
		note that $P_{\text{NEM}}$ is convex as long as the
		buying prices are higher than or equal to the selling prices.
		This is generally the case in most practical net metering policies 
		including NEM 2.0 policies \cite{wiki_net}, \cite{nrel_net}.}

	The convexity of ($P_{\text{NEM}}$) established above  helps us exploit the strong duality property with the
	dual problem (D). 
	The Lagrangian of $P_{\text{NEM}}$ is given by $\mathscr{L}({x, \alpha, \beta}) = \sum_{i=1}^N (C_{\text{nm}}^{(i)}(x_i)   +
	\alpha_i ((b_{\min} - b_0) 
	- \sum_{j=1}^i x_j )  + \beta_i (\sum_{j=1}^i x_j - (b_{\max}-b_0) )\vphantom{C_{\text{nm}}^{(i)}}).$
	%\begin{multline}
	%\mathscr{L}({x, \alpha, \beta}) = \sum_{i=1}^N (C_{\text{nm}}^{(i)}(x_i)   +
	%\alpha_i ((b_{\min} - b_0) 
	%- \sum_{j=1}^i x_j )  \\ \vspace{-15pt}
	%+ \beta_i (\sum_{j=1}^i x_j - (b_{\max}-b_0) )\vphantom{C_{\text{nm}}^{(i)}}).
	%\end{multline}
	%	\vspace{-10pt}
	%	\begin{multline}
	%	\mathscr{L}({x, \alpha, \beta}) = \sum_{i=1}^N \Big(C_{\text{nm}}^{(i)}(x_i)   +
	%	\alpha_i \Big((b_{\min} - b_0) \\ \vspace{-20pt}
	%	- \sum_{j=1}^i x_j \Big)  + \beta_i \Big(\sum_{j=1}^i x_j - (b_{\max}-b_0) \Big)\vphantom{C_{\text{nm}}^{(i)}}\Big)
	%	\end{multline}
	The Lagrangian dual of the primal problem is given by
	\vspace{-3pt}
	\begin{gather*}
	\text{(D)} \quad \max  \phi ({ \alpha, \beta}), \quad \text{subject to, }  \alpha_i, \beta_i \geq 0 \quad \forall i,\\ \vspace{-5pt}
	\text{where  }\quad  \phi ({ \alpha, \beta}) = \inf_{f(x_i) \in \left[S_{\min}^i, S_{\max}^i\right]}  \mathscr{L}({x, \alpha, \beta})\vspace{-7pt}
	\end{gather*}
	{To identify the optimal action at each instant as a function of the Lagrange multipliers
		we use the following result from \cite{cruise2019control}. {Note that we have adapted the theorem to our setting in which we require a separate proof (see Appendix~\ref{prf:arbitrage}) because unlike \cite{cruise2019control} we do not require the initial and final level of energy in the storage to be the same.}}
	
	\begin{theorem}
		\label{thm:arbitrage}
		There exists a tuple $(s^*,x^*, \mu^*)$ with $N$-dimensional vectors such that:
		\begin{enumerate}
			\item  $s_i^* = f(x_i^*) \quad \forall  i$.
			
			\item ${x^*} = (x_1^*, ... ,  x_N^*)$ is a feasible solution of the optimal arbitrage problem ($P_{\text{NEM}}$). 
			%	implying $x_i \in \left[X_{\min}^i, X_{\max}^i\right]$ and $b_i^*=b_0+\sum_{j=1}^{i} x_j^*
			%	\in [b_{\min}, b_{\max}]$ for all $i$.\vspace{-1pt}
			
			\item For each $i$, $x_i^*$ minimizes $C_{\text{nm}}^{(i)}(x) -\mu_i^*x$. Here,
			$\mu_i^*$ is the optimal accumulated Lagrange multiplier for time step $i$ and 
			is related to the dual optimal solution $(\alpha^*, \beta^*)$ as follows $\mu_i^* = \sum_{j=i}^N (\alpha_j^* - \beta_j^*) $.
			
			\item The optimal accumulated Lagrange multiplier, $\mu_i^*$, at any time step $i$ satisfies 
			the following recursive conditions: \vspace{-2pt}
			\begin{itemize}
				\item $\mu_{i+1}^* = \mu_{i}^*  , \text{ if   }   b_{\min} <b_i^* < b_{\max}$,\vspace{-1pt}
				\item $\mu_{i+1}^* \leq \mu_{i}^*  , \text{ if   }   b_i^* = b_{\min}$,\vspace{-1pt}
				\item $\mu_{i+1}^* \geq \mu_{i}^*  , \text{ if   }   b_i^* = b_{\max}$.\vspace{-1pt}
			\end{itemize}
			
			\item Additionally $\mu_N^*$ at the last instant $N$ satisfies \vspace{-2pt}
			\begin{itemize}
				\item $\mu_{N}^* = 0  , \text{ if   }   b_{\min} <b_N^* < b_{\max}$,\vspace{-1pt}
				\item $\mu_{N}^* \geq 0  , \text{ if   }   b_N^* = b_{\min}$.\vspace{-1pt}
				%\item $\mu_{N}^* \leq 0  , \text{ if   }   b_N^* = b_{\max}$
			\end{itemize}
		\end{enumerate}		

		For any tuple $(s^*, x^*,\mu^*)$ satisfying the above conditions, ${s^*}$ solves 
		the optimal arbitrage problem ($P_{\text{NEM}}$).\vspace{-2pt}
	\end{theorem}
	
	The proof of Theorem~\ref{thm:arbitrage} is provided in Appendix~\ref{prf:arbitrage}.
	
	{
		\begin{remk}
			{\em From the above theorem we note that the {optimal} accumulated Lagrange multiplier
				$\mu_i^*$ at the $i$th instant is defined as $\mu_i=\sum_{j=i}^{N}(\alpha_j^*-\beta_j^*)$.
				Since $\alpha_j^*$ and $\beta_j^*$ depend on the capacity constraints at the $j$th instant, it is clear
				that $\mu_i^*$ depends on the decisions made in the future instants $i$ through $N$.
				In other words, $\mu_i^*$ is the reduction in the cost due to satisfying the constraints
				in all future instants.
				Thus, given the value of $\mu_i^*$, in order to find the optimal action at instant $i$ it is sufficient to
				look for the minimizer of the cost function $C_{nm}^{(i)}(x)-\mu_i^*x$ in the range $x \in [X_{\min}^i,X_{\max}^i]$
				ignoring the battery capacity constraints for all future instants.}
		\end{remk}}
		
		We note that the optimality conditions stated in Theorem~\ref{thm:arbitrage} do not depend on the particular structure of the cost function in $P_{\text{NEM}}$ and are valid as long as it is a convex function of $x$. 
		In the next subsection we exploit the specific cost function in our setting to further
		characterize the relationship between the optimal actions and the optimal accumulated Lagrange multipliers.
		In particular we identify that for the specific cost function in $P_{\text{NEM}}$ there exists
		a threshold structure in the optimal solution.
		\vspace{-12pt}
		
		\subsection{Threshold Based Structure of the Optimal Solution}
		\vspace{-4pt}
		Statement (3) of Theorem~\ref{thm:arbitrage} shows that the optimal control decision $x_i^*$ in the $i$th instant 
		is the minimizer of $C_{\text{nm}}^{(i)}(x) -\mu_i^*x$ for $f(x_i) \in \sbrac{S_{\min}^i , S_{\max}^i}$.  
		This implies $x_i^*$ or equivalently $s_i^*$ is a function of {the} accumulated Lagrange multiplier $\mu_i$ for time instant $i$.
		%In all these cases we find out that the expression of $s_i^*(\mu)$ can be directly mapped to $x_i^*(\mu)$.
		In the theorem below, we show that the relationship between $s_i^*$ and $\mu$ is 
		based on different threshold values of $\mu$.
		\vspace{-5pt}
		{
			\begin{theorem}
				\label{thoremthresholds}
				%	Using the convex piecewise linear structure of the cost function the thresholds are identified.
				Let $x_i^*(\mu)$ be the minimizer of the function 
				$C_{\text{nm}}^{(i)}(x) -\mu^*x$ in  $f(x_i) \in \sbrac{S_{\min}^i , S_{\max}^i}$. 
				Furthermore, let $s_i^*(\mu)=f(x_i^*(\mu))$. Then $s_i^*(\mu)$ is a set valued map and is given by \vspace{-1pt}
				%	The optimal control decision $s_i^*$ 
				%	minimizes the function 
				%	$C_{\text{nm}}^{(i)}(x) -\mu_i^*x$ for $f(x)=s \in \sbrac{S_{\min}^i, S_{\max}^i}$.
				%	The optimal decision $s_i^*(\mu)$ 
				%	is given by Eq.~\ref{remk4eq}.
				\vspace{-10pt}
				\begin{equation}
				\label{remk4eq}
				\begin{aligned}
				s_i^*(\mu) =
				\begin{cases}
				~\{S_{\min}^i\},
				\quad \quad \text{if }   \mu < \mu_1^i, \text{\small \bf (Region 1)} \\
				\sbrac{S_{\min}^i,0 \wedge (-z_i \vee S_{\min}^i)},
				\text{if }   \mu = \mu_1^i, \text{\small \bf (Region 2)}\\
				\{0 \wedge (-z_i \vee S_{\min}^i)\},
				\text{if }   \mu\in (\mu_1^i,  \mu_2^i \wedge \mu_3^i), \\
				\hspace{3cm}\text{\small \bf (Region 3)}\\
				[0 \wedge (-z_i \vee S_{\min}^i), \mathbb{I}_{(\zeta_i<1)}\big( 0 \wedge (-z_i \vee S_{\min}^i)+\\ 0 \vee (-z_i \wedge S_{\max}^i) \big)],
				\text{if }   \mu= (\mu_2^i \wedge \mu_3^i), \text{\small \bf (Region 4)} \\
				%\begin{cases}
				\{\mathbb{I}_{(\zeta_i<1)}\big( 0 \wedge (-z_i \vee S_{\min}^i)+0 \vee (-z_i \wedge S_{\max}^i) \big)\},
				\\ \text{if }  \mu \in (\mu_2^i \wedge \mu_3^i,  \mu_2^i \vee \mu_3^i),  \text{\small \bf (Region 5)} \vspace{4pt}\\
				
				[ \mathbb{I}_{(\zeta_i<1)}\big( 0 \wedge (-z_i \vee S_{\min}^i)+0 \vee (-z_i \wedge S_{\max}^i) \big), \\ 0 \vee (-z_i \wedge S_{\max}^i)],
				\text{if }   \mu =  (\mu_2^i \vee \mu_3^i),  \text{\small \bf (Region 6)} \vspace{4pt}\\
				
				\{0 \vee (-z_i \wedge S_{\max}^i)\}, 
				\text{if }   \mu\in ((\mu_2^i \vee \mu_3^i), \mu_4^i), \\
				\hspace{2cm}\text{\small \bf (Region 7)} \\
				
				\sbrac{0 \vee (-z_i \wedge S_{\max}^i), ~S_{\max}^i},
				\text{if }   \mu =  \mu_4^i, \text{\small \bf (Region 8)} \vspace{1pt}\\
				
				~\{S_{\max}^i\},
				\quad \quad \text{if }   \mu >  \mu_4^i,  \text{\small\bf(Region 9)}\\ 
				\end{cases} \vspace{-10pt}
				\end{aligned}
				\end{equation} 
				where  the four threshold points (based on which nine regions are defined, see Fig.~\ref{case_fig} and Fig.~\ref{case_fig2} in Appendix~\ref{proofremark1}) 
				are given as
				(1) $\mu_1^i= \eta_{\text{dis}} p_s(i)$,
				(2) $\mu_2^i= \frac{p_s(i)}{\eta_{\text{ch}}}$,
				(3) $\mu_3^i= \eta_{\text{dis}} p_b(i)$,
				(4) $\mu_4^i= \frac{p_b(i)}{\eta_{\text{ch}}}$,
				%$\zeta_i$ is defined as the ratio of $\kappa_i$ and roundtrip battery efficiency, given as
				and $\zeta_i = \frac{\kappa_i }{\eta_{\text{ch}}\eta_{\text{dis}}}$.
				Here $(a\vee b)$, $(a\wedge b)$ denote $\max(a,b)$ and $\min(a,b)$, respectively. \vspace{-4pt}
			\end{theorem}}
			
			The proof of Theorem~\ref{thoremthresholds} is provided in Appendix~\ref{thoremthresholdsproof}. Based on the value of $\zeta_i$ the thresholds are identified.
			\vspace{-6pt}
			{
				\begin{remk}
					{\em Note that $s_i^*(\mu)$ is a set-valued function in $\mu$. 
						Depending on the value of $\mu$ the set $s_i^*(\mu)$
						can take nine different values in regions shown as Region 1 through Region 9
						in Fig.~\ref{case_fig} and Fig.~\ref{case_fig2} in Appendix~\ref{proofremark1} which correspond to the cases
						$\zeta_i < 1$ and $\zeta_i \geq 1$, respectively. For each $i$
						there are four threshold values
						$\mu_1^i,\mu_2^i, \mu_3^i,\mu_4^i$ at which $s_i^*(\mu)$ changes.
						From {their} expressions given above, it is clear that 
						$\mu_1^i, \mu_2^i, \mu_3^i,$ and $\mu_4^i$ correspond to
						(1) the effective discharging cost under selling price,
						(2) the effective charging cost under selling price,
						(3) the effective discharging cost under buying price, and 
						(4) effective charging cost under buying price,
						respectively.}
				\end{remk}}
				
				\vspace{-18pt}

				\begin{figure}[!htbp]
					\center
					\includegraphics[width=2.7in]{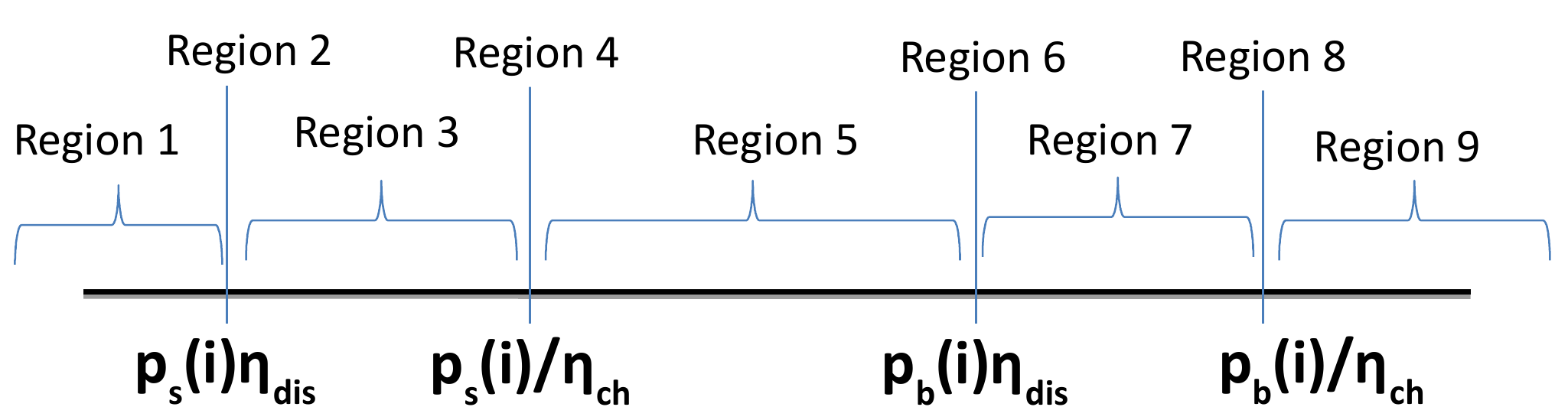} \vspace{-7pt}
					\caption{Regions based on levels of $\mu$ for $\zeta_i< 1$}\label{case_fig}
				\end{figure}
				\vspace{-13pt}
				
				%\textcolor{red}{The four levels in Fig.~\ref{case_fig} and Fig.~\ref{case_fig2} denote 
				%	(a) discharging cost under selling price,
				%	(b) charging cost under selling price,
				%	(c) discharging cost under buying price,
				%	(d) charging cost under buying price. These four levels are functions of storage charging and discharging efficiency and electricity price. 
				%	Based on the ratio of $\kappa_i$ and value of $\mu$, nine regions can be categorized, see Fig.~\ref{case_fig} and Fig.~\ref{case_fig2}.}
				%%	Based on these levels, the storage operation can be divided into 9 regions, shown in Fig.~\ref{case_fig} and Fig.~\ref{case_fig2}.}
				
				\begin{remk}
					{\em We note that $s_i^*(\mu)$ is monotone non-decreasing map in $\mu$
						in the sense that for $\mu_1 \leq \mu_2$ 
						we have $s_i^*(\mu_1) \preceq s_i^*(\mu_2)$,
						%and ii) $s_i^*(\mu_1) \cap s_i^*(\mu_2)\neq \emptyset$
						where for two sets $A$ and $B$
						we say $A \preceq B$ (resp, $A \prec B$) if $a \leq b$ (resp $a < b$)
						for all $a \in A$ and for all $b \in B$.
						As a result, the sets 
						$b_i^*(\mu)$, defined recursively as
						$b_i^*(\mu)=b_{i-1}^*(\mu)+ x_{i}^*(\mu)$ for $i \geq 1$ and $b_{0}^*(\mu)=b_0$
						are also {monotonically} non-decreasing in $\mu$. 
						Here, the addition of two intervals $[a,b]$ and $[c,d]$ denotes the interval $[a+c, b+d]$. }
					%\textcolor{red}{\{a\} denotes [a,a], a singleton set.}
				\end{remk}
				%\vspace{-10pt}
				
				%\vspace{-13pt}

				%The proof of Theorem~\ref{thoremthresholds} is provided in Appendix~\ref{proofremark} and Appendix~\ref{proofremark1}. In Appendix~\ref{proofremark} storage optimal levels are identified for $\kappa_i\in [0,\eta_{\text{ch}}\eta_{\text{dis}})$, see Figure~\ref{case_fig} and 
				%in Appendix~\ref{proofremark1} storage optimal levels are identified for $\kappa_i\in [\eta_{\text{ch}}\eta_{\text{dis}},1)$, see Figure~\ref{case_fig2}. \textcolor{red}{The storage thresholds for $\kappa_i = 1$ is detailed in Section~\ref{kappa1subsec}.}

				%\subsubsection{Optimal thresholds for $\kappa_i =1$}
				%\label{kappa1subsec}
				
				%\textcolor{blue}{In the following Eq (Eg. (\ref{eq:lossy})) we are not using anymore the four threshold of $\mu$ value. Why??\\}
				For NEM 1.0, $p_b(i) = p_s(i) ~ \forall i$, therefore, the thresholds points reduces to $\mu_1^i$ and $\mu_4^i$. Remark~\ref{rmk:remark2} illustrates the threshold based structure for NEM 1.0.
				
				\begin{remk}
					\label{rmk:remark2}
					{\em 
						Note that $\kappa_i=1$ implies buying and selling price for time instant $i$ are the same, which corresponds to the NEM 1.0
						policy. 
						In case of NEM 1.0, the optimal actions as derived in our earlier work \cite{hashmi2017} and can be recovered
						from \eqref{remk4eq} by putting $\kappa_i=1$ and merging the regions together.
						The final expression is given as follows: 
						%	Optimal control decision $x_i^*$ 
						%		minimizes the function 
						%		$C_{\text{storage}}^{(i)}(s) -\mu_i^*x$ for $s \in \sbrac{S_{\min}^i, S_{\max}^i}$.
						%		$s_i^*(\mu)$ 
						%		is given as
						%		%		in Eq.~\ref{eq:lossy}.
						\vspace{-3pt}
						\begin{equation}
						s_i^*(\mu) =
						\begin{cases}
						\{S_{\min}^i\}, & \text{if }  \mu< \mu_1^i =p_s(i) \eta_{\text{dis}}, \\
						\sbrac{S_{\min}^i, 0}  ,& \text{if }  \mu= \mu_1^i,\\
						\{0\} ,& \text{if }   \mu_1^i<\mu< p_b(i) /\mu_4^i ,\\
						\sbrac{0, S_{\max}^i} ,& \text{if }   \mu= \mu_4^i, \\
						\{S_{\max}^i\}  ,& \text{if }   \mu> \mu_4^i = p_b(i) /\eta_{\text{ch}}, \vspace{-5pt}
						\end{cases} \vspace{-1pt}
						\label{eq:lossy}
						\end{equation} 
						Clearly, when $\mu$ lies between $\mu_1^i$ and $\mu_4^i$, 
						the optimal action for the battery is to do nothing, 
						and in this case the cycles of operation that a battery performs can be controlled by introducing a friction coefficient facilitating the elimination of low returning transactions \cite{hashmi2018limiting, hashmi2018long}. }
				\end{remk}
				
				\vspace{-17pt}

				\subsection{Proposed Algorithm}
				\vspace{-3pt}
				In this section, we describe the algorithm we propose to solve the optimal energy arbitrage problem formulated previously.
				The objective of the algorithm is to find a tuple $(s^*,x^*,\mu^*)$ that satisfies conditions (1)-(5) of Theorem~\ref{thm:arbitrage}, and therefore solves ($P_{\text{NEM}}$).
				%The proposed algorithm is developed for Case 1 to 3, described in the previous section, using Remark~\ref{rmk:remark2} and Theorem~\ref{thoremthresholds} 
				%and~\ref{rmk:remark6} 
				%for lossy battery. 
				The presented algorithm can be operated for a variable value of $\kappa_i$, such that $\kappa_i \in [0,1]~ \forall~i$, covering all net-metering compensation scheme. 
				%\textcolor{cyan}{The threshold based structure based on $\kappa_i$ is selected using Theorem~\ref{thoremthresholds} and Remark~\ref{rmk:remark2}.}
				{The threshold-based structure of the optimal solution (versus $\kappa_i$) is selected using Theorem~\ref{thoremthresholds} and Remark~\ref{rmk:remark2}.}
				%In order to identify storage actions we propose a combination of forward algorithm
				In order to identify storage actions, we propose a combination of a forward algorithm (Algorithm \ref{alg:1}) which identifies the sub-horizon and envelope of storage actions and a backward algorithm (Algorithm \ref{alg:2}) which runs one time in the identified sub-horizon to decide the storage control trajectory.
				Algorithm \ref{alg:1} and Algorithm \ref{alg:2} together {perform} optimal arbitrage.\\ \vspace{-7pt}

				%The same structure of the algorithm can be used for other cases. 
				\subsubsection{Description of Algorithm~\ref{alg:1}}
				%Algorithm~\ref{alg:1} illustrates its main steps.  
				{
					According to the value of $\kappa_i$, the lower and upper bound of the envelope are selected, lines 4--5 of the pseudo code of Algorithm~\ref{alg:1}.
				}
				%Based on the value of $\kappa_i$ the lower and upper bound of the envelope is selected, lines 4--5 of the pseudo code of Algorithm~\ref{alg:1}.
				%
				%
				From condition (4) of {Theorem~\ref{thm:arbitrage}}, we can see that $\mu_{i+1}^*$ may differ from the optimal accumulated Lagrange multiplier, $\mu_{i}^*$, only when $b_i^*=b_{\max}$ or $b_i^*=b_{\min}$. 
				%This means that 
				Thus the value of $\mu$ remains constant until the battery charge level at a time lies strictly within the battery capacity limits.
				%
				%Note from Theorem~\ref{thm:arbitrage} condition (3) that $\mu_{i+1}^*$
				%may differ from $\mu_{i}^*$
				%only when $b_i^*=b_{\max}$ or $b_i^*=b_{\min}$. 
				%Hence, if the battery level at the end of a time instant
				%lies strictly within the battery capacity limits, then
				%there is no change in the value of the optimal accumulated Lagrange multiplier.
				%
				%\textcolor{cyan}{Based on this key idea, we define a sub-horizon in Remark~\ref{subhorizonremark}.}
				{Following this key idea, we define the sub-horizon in Remark~\ref{subhorizonremark}.}  \vspace{-3pt}
				\begin{remk}
					\label{subhorizonremark}
					The whole duration~$T$ is divided into $M$~periods, indexed as $\cbrac{1,2,\ldots,M}$. Each period contains a number of consecutive time instants, such that for all instants $i$ belonging to the same period $K \in \cbrac{1,2,\ldots,M}$ the value of the accumulated Lagrange multiplier $\mu_i^*$ remains the same, denoted as $\mu_K$. Each such period is called \textit{sub-horizon}. \vspace{-3pt}
				\end{remk}
				It follows that at the end instant of each sub-horizon, the battery energy level touches either $b_{\max}$ or $b_{\min}$.
				%
				%Using this key idea, in the proposed algorithm, we divide the total duration $T$
				%into groups, indexed as $\cbrac{1,2,\ldots,M}$, of consecutive time instants 
				%such that for all instants $i$ belonging to the same group $K \in \cbrac{1,2,\ldots,M}$ the 
				%value of the accumulated Lagrange multiplier $\mu_i^*$ is the same, denoted as $\mu_K$. 
				%We call each such group as a sub-horizon. 
				%At the end of each sub-horizon, the battery energy level
				%touches either $b_{\max}$ or $b_{\min}$.
				%
				%
				Note that the number of sub-horizons ($M$), the start and end instants of each sub-horizon~$K$, the $\mu_K$ value, and the optimal actions in sub-horizon~$K$ depend on the problem instance and are determined recursively, as described below. 
				
				Assume that we have already identified the first $K-1$ ($K \geq 1$) sub-horizons and determined the values of $s_{i}^*$ and $\mu_{i}^*$ in all instants $i$ belonging to these sub-horizons (for $i \in [1, i_{K-1}]$). The index $i_{K-1}$ denotes the last instant in the $(\!K-1\!)$th sub-horizon. Hence: \vspace{-3pt}
				\begin{itemize}
					\item[] If $i_{K-1}=N$, then we have already covered the whole period $T$ and the algorithm terminates. \vspace{-1pt}
					\item[] If $i_{K-1} < N$, then we proceed to identify the next sub-horizon~$K$, i.e., the values of $i_K$ (the last instant in sub-horizon~$K$), {and} $\mu_K$, and the optimal decisions for the time instants $i \in [i_{K-1}+1, i_{K}]$. \vspace{-3pt}
				\end{itemize}
				%
				%
				%Suppose we have identified the first $K-1$ ($K \geq 1$)
				%sub-horizons and the optimal actions in all instants belonging to these sub-horizons. 
				%Call the last instant identified
				%to be in the $(K-1)$th sub-horizon as $i_{K-1}$.  If $i_{K-1}=N$, then
				%we have already covered whole period $T$. If $i_{K-1} < N$,
				%then we proceed to identify the next sub-horizon $K$, i.e.,
				%the values of $i_K$, $\mu_K$, and the optimal decisions for the time instants
				%$i \in [i_{K-1}+1, i_{K}]$.
				%
				To determine sub-horizon $K$, we start with instant $i_{K-1}+1$ and an initial value of $\mu_K \geq 0$ for that sub-horizon\footnote{For the first sub-horizon $K$=1 (that includes the first time instant) the starting guess value of $\mu_1$ is taken to be $0$ and for every other sub-horizon $K > 1$, the starting guess value of $\mu_K$ is taken to be equal to $\mu_{K-1}$. 
					Note that these choices do not affect the solution given by the algorithm.
				}.
				%
				%
				%
				%To identify the sub-horizon $K$, we start with from instant $i_{K-1}+1$ and 
				%a guess value of $\mu_K \geq 0$ for that sub-horizon.\footnote{For the first sub-horizon $K=1$
				%(that includes the first time instant)
				%the starting guess value of $\mu_1$ is taken to be $0$ and for every
				%other sub-horizon $K > 1$, the starting guess value of $\mu_K$ is taken to be equal to $\mu_{K-1}$.  
				%Note that these choices are arbitrary and the algorithm does not depend on these choices.}
				%
				We compute the values of $s_i^*(\mu_K)$ and $b_i^*(\mu_K)$\footnote{$b_i^*(\mu_K)$ is a set containing the values of lower and upper envelope of optimal battery capacity for the scalar value of $\mu_K$.} according to the method described in Remark~\ref{rmk:remark2} and Theorem~\ref{thoremthresholds}, for all consecutive time instants $i > i_{K-1}$ until we reach a time instant $i=i_{\text{break}}$, for which one of the following conditions is satisfied (we call these as the {\em violation conditions}): \vspace{-4pt}
				%
				%Now, for the chosen value of $\mu_K$, the values of $s_i^*(\mu_K)$
				%and $b_i^*(\mu_K)$ are computed as described in Remark~\ref{rmk:remark1}
				%for all consecutive time instants $i > i_{K-1}$ until we reach a time instant $i=i_{\text{break}}$
				%for which one of the following conditions is satisfied (we call these as the {\em violation conditions}):
				%
				\begin{enumerate}
					\item[] C1: $b_{i_{\text{break}}}^*(\mu_K) \prec \cbrac{b_{\min}}$.
					
					\item[]  C2: $\cbrac{b_{\max}} \prec b_{i_{\text{break}}}^*(\mu_K)$.
					
					\item[] C3: $i_{\text{break}}=N, b_{\min} \notin b_{N}^*(\mu_K), \mu_K > 0$.
				\end{enumerate} 
				If no $i_{\text{break}}$ is found even after reaching $i=N$, then $K$ is the last sub-horizon and we set $i_K=N$
				(and lines 31--36 of the pseudo code are executed).
				From condition~(5) of Theorem~\ref{thm:arbitrage}, if $\mu_K >0$, then $b_N^*=b_{\min}$; else $b_N^*$ can take any value in the set $[b_{\min}, b_{\max}) \cap b_{N}^*(\mu_K)$.
				The optimal decisions of $x_i^*$ and $b_{i}^*$, for $i \in [i_{K-1}+1, N]$, are calculated by using Algorithm~\ref{alg:2}, discussed in more detail later. 
				%
				%
				%If no such $i$ is found
				%even after reaching $i=N$, then $K$ is identified as the last sub-horizon and we set $i_K=N$
				%(and lines 24--31 of the pseudo code are executed).
				%If $\mu_K >0$, then $b_N^*=b_{\min}$; else $b_N^*$ is taken to be
				%some value in the set $[b_{\min}, b_{\max}) \cap b_{N}^*(\mu_K)$ to satisfy
				%condition (4) of Theorem~\ref{thm:arbitrage}. The optimal
				%decisions $x_i^*$ and $b_{i}^*$ for $i \in [i_{K-1}+1, N]$ are then found
				%by using the algorithm \texttt{BackwardStep}, shown as Algorithm~\ref{alg:2} below.
				%The proposed algorithm then terminates.
				%The algorithm \texttt{BackwardStep} will be discussed in more detail later.
				%

				{\em \textbf{Tuning $\mu$ value}}:
				Since the cost function in problem~($P_{\text{NEM}}$) is piecewise linear, the optimal values of the accumulated Lagrange multipliers $\mu_{i}^*$ are chosen from a discrete set of values corresponding to buying and selling prices of electricity. This feature transforms ($P_{\text{NEM}}$) from a continuous optimization problem to a discrete one. Therefore, in the proposed solution, we specify how to tune the Lagrange multipliers to these prices to find their optimal values (lines 14, 15 in the pseudo code).  
				%
				%
				%{\em Tuning $\mu$}: Exploiting the piecewise linear
				%cost structure of the arbitrage problem we find that the optimal accumulated Lagrange multipliers 
				%can only take a discrete set of values corresponding to buying and selling prices of electricity. 
				%This transforms the continuous optimization problem
				%into a discrete optimization problem.
				%In the proposed method, we indicate how to tune the Lagrange multiplier variables
				%to these prices to find their optimal values
				%
				%
				%    
				%\vspace{-10pt}
				%\vspace{-10pt}
				%
				%\textcolor{red}{\st{I HAVE A GENERAL QUESTION ON TUNING $\mu$ VALUE ACCORDING TO THE FOLLOWING: $\min \{p > \mu: \ldots \}$ or $\max \{p < \mu: \ldots\}$ for any $\mu$??? Isn't somehow related to $\mu_K$ value?? The same holds for lines 18-20 of Algorithm 1.} Yes the jumps of $\mu$ should be with respect to $\mu_K$ CHANGED}\\
				%
				%\textcolor{cyan}{The mechanism to update the value of $\mu_k$ is decided by the condition of violation and detailed in Remark~\ref{muupdateremark}.} \vspace{-3pt}
				{The mechanism to update the value of $\mu_K$ is decided by the violation condition and is detailed in Remark~\ref{muupdateremark}.} \vspace{-3pt}
				\begin{remk}
					\label{muupdateremark}
					Now, if condition C1 holds; for the chosen value of $\mu_K$, the battery capacity limit is violated from below at instant $i_{\text{break}}$ since the set $b_{i_{\text{break}}}(\mu)$ lies strictly below $b_{\min}$. The strategy here is to increase $\mu_K$ value to the \vspace{-3pt}
					\begin{equation}
					\mu_K = \min\{p >\mu: p\in \mu_1^i, \mu_2^i, \mu_3^i, \mu_4^i; {i \in( i_{K-1}, i_{\text{break}}]}\}. \vspace{-3pt}
					\label{musearch1}
					\end{equation}
					% $\min \{p > \mu: p\in {p_{\text{s}}(i)\eta_{\text{dis}}, p_{\text{s}}(i)/\eta_{\text{ch}}, p_{\text{b}}(i)\eta_{\text{dis}}, p_{\text{b}}(i)/\eta_{\text{ch}}; i_{K-1} < i \leq i_{\text{break}}}\}$.
					Otherwise, if condition C2 or C3 holds, 
					then decrease $\mu_K$ to
					%	then $\mu_K$ is decreased to 
					\vspace{-6pt}
					\begin{equation}
					\mu_K = \max \{p < \mu: p\in \mu_1^i, \mu_2^i, \mu_3^i, \mu_4^i; { i \in( i_{K-1}, i_{\text{break}}]}\}. \vspace{-3pt}
					\label{musearch2}
					\end{equation}
				\end{remk}
				\vspace{-5pt}
				% $\max \{p < \mu: p\in {p_{\text{s}}(i)\eta_{\text{dis}}, p_{\text{s}}(i)/\eta_{\text{ch}}, p_{\text{b}}(i)\eta_{\text{dis}}, p_{\text{b}}(i)/\eta_{\text{ch}}; i_{K-1} < i \leq i_{\text{break}}}\}$.
				After updating $\mu_K$ value, we repeat the same process as before until we reach a new time instant $i_{\text{break}}$, for which one of the above conditions is satisfied. 
				Note there is a one-to-one mapping of $s_i$ and $x_i$, therefore, $x_i^*(\mu_K)$ and consequently $b_i^*(\mu_K)$ are monotonically non-decreasing
				functions in $\mu_K$, the potential effect of the update of $\mu_K$ is that $i_{\text{break}}$ is pushed to a later instant.
				%According to Remark~\ref{rmk:remark1}, increasing or decreasing $\mu_K$,  
				%has the effect of potentially increasing $i_{\text{break}}$. 
				The update of $\mu_K$ is repeated as long as $i_{\text{break}}$ increases (or remains the same), compared to its previous value, stored in~$i_{\text{mem}}$.
				%
				%
				%If condition C1 above is satisfied, then
				%for the chosen value of $\mu_K$, the battery capacity limit
				%is violated at the instant $i_{\text{break}}$ since the set $b_{i_{\text{break}}}(\mu)$
				%lies strictly below $b_{\min}$. 
				%The value of $\mu_K$ is then
				%increased to $\min \{p > \mu: p\in {p_{\text{s}}(i)\eta_{\text{dis}}, p_{\text{s}}(i)/\eta_{\text{ch}}, p_{\text{b}}(i)\eta_{\text{dis}}, p_{\text{b}}(i)/\eta_{\text{ch}}; i_{K-1} < i \leq i_{\text{break}}}\}$.
				%Otherwise, if C2 or C3 above is satisfied, then $\mu_K$ is decreased to
				%$\max \{p < \mu: p\in {p_{\text{s}}(i)\eta_{\text{dis}}, p_{\text{s}}(i)/\eta_{\text{ch}}, p_{\text{b}}(i)\eta_{\text{dis}}, p_{\text{b}}(i)/\eta_{\text{ch}}; i_{K-1} < i \leq i_{\text{break}}}\}$.
				%With the updated value of $\mu_K$ we again repeat the same process 
				%as discussed above to obtain a new value of $i_{\text{break}}$.
				%Since $x_i^*(\mu_K)$ and $b_i^*(\mu_K)$ are monotonically non-decreasing
				%functions in $\mu_K$, the potential effect of the update of $\mu_K$
				%is that $i_{\text{break}}$ is pushed to a later instant.
				%%According to Remark~\ref{rmk:remark1}, increasing or decreasing $\mu_K$,  
				%%has the effect of potentially increasing $i_{\text{break}}$. 
				%The update $\mu_K$ is  repeated so long as $i_{\text{break}}$
				%increases (or remains the same) as compared to its old value (stored in $i_{\text{memory}}$). 
				%
				This part of the proposed algorithm is mirrored in lines 12--16 of the pseudo-code of Algorithm~\ref{alg:1}. If the value of $i_{\text{break}}$ decreases after updating $\mu_K$, then for the previous value $\mu_{\text{mem}}$ of $\mu_K$ there must have been an instant $i \in [i_{K-1}+1, i_{\text{mem}}]$, where $b_{\max} \in b_i^*(\mu_K)$ (violation occurred due to C1) or $b_{\min} \in b_i^*(\mu_K)$ (violation occurred due to C2 or C3). 
				This is due to the fact that both $\mu_{\text{mem}}$ and $\mu_K$ always lie in the range $\{\mu_1^i, \mu_2^i, \mu_3^i, \mu_4^i\}$, for all $i > i_{K-1}$.
				At this point of the algorithm (lines 17--29 of the pseudo-code), $\mu_K$ and $i_{\text{break}}$ are switched back to their previous values, which are saved, respectively, in $\mu_{\text{mem}}$ and $i_{\text{mem}}$. This value of $\mu_K$ is selected as the final value of the optimal accumulated Lagrange multiplier corresponding to sub-horizon~$K$.
				The end instant of sub-horizon~$K$, $i_K$, is set equal to the latest time instant $i \in [i_{K-1}+1, i_{\text{break}}]$, for which $b_{\min} \in b_i^*(\mu_K)$ or $b_{\max} \in b_i^*(\mu_K)$, and $b_{i_{K}}^*$ takes the value $b_{\min}$ in the former case and $b_{\max}$ in the later case. \\ \vspace{-6pt}
				%
				% 
				%If the value of $i_{\text{break}}$ decreases with an updated value of $\mu_K$,
				%then for the previous value $\mu_{\text{memory}}$ of $\mu_K$
				%there must have been an instant $i \in [i_{K-1}+1, i_{\text{memory}}]$,
				%where $b_{\max} \in b_i^*(\mu_K)$ 
				%(if the violation occurred due to C1) or $b_{\min} \in b_i^*(\mu_K)$
				%(if violation occurred to due to C2 or C3). 
				%This is because  both $\mu_{\text{memory}}$ and $\mu_K$
				%always lie in the range $\{p_{\text{s}}(i)\eta_{\text{dis}}, p_{\text{s}}(i)/\eta_{\text{ch}}, p_{\text{b}}(i)\eta_{\text{dis}}, p_{\text{b}}(i)/\eta_{\text{ch}}\}$
				%for all $i > i_{K-1}$. 
				%
				%%Since $b_i^*(p_{\text{s}}(i))\cap b_i^*(p_{\text{b}}(i))  \neq \emptyset$
				%%an update of $\mu_K$ cannot cause $b_i^*(\mu_K)$ to completely go above $b_{\max}$ 
				%%or below $b_{\min}$ if $\cbrac{b_{\min}} \prec b_i^*(\mu_{\text{memory}}) \prec \cbrac{b_{\max}}$.
				%
				%At this point in the algorithm, $\mu_K$ and $i_{\text{break}}$ are switched back 
				%to their previous values stored in $\mu_{\text{memory}}$ and $i_{\text{memory}}$, respectively.
				%This value of $\mu_K$ is identified to be the final value of the optimal
				%accumulated Lagrange multiplier in the sub-horizon $K$. 
				%We set $i_K$ to be the latest time instant $i \in [i_{K-1}+1, i_{\text{break}}]$ for which $b_{\min} \in b_i^*(\mu_K)$
				%or $b_{\max} \in b_i^*(\mu_K)$.  The value of $b_{i_{K}}^*$ is chosen to be $b_{\min}$
				%in the former case and $b_{\max}$ in the later case. 
				%
				
				\vspace{-5pt}
				\subsubsection{Description of Algorithm~\ref{alg:2}}
				%\vspace{-4pt}
				Using Algorithm~\ref{alg:1} we find the optimal battery capacity in a sub-horizon, $b^*(\mu_K)$. $b^*(\mu_K) \cap [b_{\min}, b_{\max}]$ contains information of the lower and upper feasible envelope of {the} battery capacity; $\cap$ denotes intersection of regions.
				%\textcolor{blue}{XXX Please explain the meaning of the symbol $\cap$ in the previous expression!!!} 
				We propose a novel method based on backward step to find the optimal solution among infinite possibilities, described as \texttt{BackwardStep}, in Algorithm~\ref{alg:2}. Note that the \texttt{BackwardStep} algorithm is implemented only \textit{one time for a sub-horizon}.
				For each $i$ in the range $i_{K-1}+1\leq i < i_{K}$,
				the optimal battery level $b_i^*$ is found from $b_{i+1}^*$ 
				through the function \texttt{BackwardStep} which uses the backward recursion
				$b_i^*=(b_{i+1}^*-x_{i+1}^*(\mu_K))\cap b_i^*(\mu_K) \cap [b_{\min}, b_{\max}]$.
				%\textcolor{blue}{XXX Please explain the meaning of the symbol $\cap$ in the previous expression (see my previous comment)!!!} 
				If the above backward recursion returns a set, then any arbitrary value in the set
				is chosen to be the optimal battery level. 
				%
				%\begin{remk}
				%	The proposed optimal arbitrage algorithm will return a feasible solution for cases in which at least one feasible solution exists, this is ensured by convexity of the problem. For case where no solution exists, the Algorithm~\ref{alg:1} due to \textit{while} loop will not end (infinite loop). 
				%\end{remk}
				A stylized example demonstrating the operation of the proposed optimal arbitrage algorithm is presented in Appendix~\ref{example}.
				Note that the proposed arbitrage algorithm will return a feasible solution for cases in which at least one feasible solution exists, and this is ensured by the convexity of the problem. 
				The optimal solution to $P_{\text{NEM}}$ needs not be unique since its objective function is not strictly convex.  
				For the case where no solution exists, the Algorithm~\ref{alg:1} due to the \textit{while} loop will not end (infinite loop). 
				%Note that the proposed algorithm uses a \textit{while} loop which will end only if the optimization problem has a solution and which can only be ensured if the cost function is convex.
				\vspace{-14pt}
				\subsection{Properties of {the} Optimal Arbitrage Solution}
				\vspace{-4pt}
				{
					%\textcolor{blue}{I commented the following sentence }
					%The threshold-based presented in Theorem~\ref{thoremthresholds} and Remark~\ref{rmk:remark2}.
					We observe that for some values of $\mu$ there is a set of solutions which are possible. These points are the sub-gradient of the cost function. At these points, we observe that \textit{intermediate ramp rates can be optimal}, assuming the storage ramp rates can only be changed at decision epochs and not in between the sampling time. 
					%	This observation is contrary to \textcolor{blue}{a claim} in \cite{petrik2015optimal,wang2018energy, graves1999opportunities}, which assumes \textcolor{blue}{that the} optimal battery action lies in \textcolor{blue}{three} states: maximum, minimum and zero ramp rate.
					We demonstrate this using a numerical case study in Section~\ref{intermediaterampcase}.
				}
				
				% \vspace{-8pt}
				
				%\vspace{-5pt}
				
				% \vspace{-14pt}
				
				{In Theorem~\ref{thoremthresholds} and Remark~\ref{rmk:remark2}
					we provide the conditions for selecting the thresholds with respect to the accumulated Lagrange multiplier or the \textit{shadow price} for the arbitrage problem.
					In optimization, the shadow price is the value of the Lagrange multiplier at the optimal solution.
					%Identification of these thresholds is possible due to the convex and piecewise linear structure of the cost function. As the optimal storage operation can be selected from the discrete number of states of the thresholds, thus, the optimization efficiently discretized.
					{From Eq.~\ref{remk4eq} it is easy to see that storage thresholds for optimal arbitrage are function of the following different parameters: (a) price for buying and selling, (b) storage charging and discharging efficiency, (c) ramping constraint, (d) the relationship between the ratio of selling and buying price and round-trip storage efficiency ($\mu_2^i \leq \mu_3^i$  or $\mu_3^i < \mu_2^i$), and (e) the consumption level as seen by the energy meter.} 
					The shadow prices are selected from a finite set which is a function of electricity price and storage efficiencies, see Eq.~\ref{musearch1} and Eq.~\ref{musearch2}.}
				\vspace{-7pt}
				\begin{algorithm}
					\small{\textbf{Inputs}: {$N$, $T$, $h=(h_1,\ldots,h_N)$, $p_b=(p_b^1,\ldots,p_b^N)$, $p_s=(p_s^1,\ldots,p_s^N)$, $b_0$, $z=(z_1,...,z_N)$}}, 	{$ b_{\max}, b_{\min}, \delta_{\max}, \delta_{\min}, \eta_{\text{ch}}, \eta_{\text{dis}}$}\\
					\textbf{Outputs}: {$s^*$=$(s_1^*,s_2^*,..,s_N^*)$, $b^*$=$(b_1^*,b_2^*,..,b_N^*)$, $\mu^*$=$(\mu_1^*,\mu_2^*,..,\mu_K^*)$}\\
					\textbf{Initialize}: {K=1; $\mu_K$=$\mu_{\text{mem}}$=0; $i_{K-1}$=$i_{K}$=$i_{\text{mem}}$=0; $\mathtt{BreakFlag}$=0}
					\begin{algorithmic}[1]
						\While{$i_K < N$}
						\For{$ i= i_{K-1}+1 \text{ to } N$}
						\State Compute $\kappa_i = p_s(i)/p_b(i)$
						\If{$\kappa_i = 1$}
						Find $s_i^*(\mu_K)$ using Remark~\ref{rmk:remark2}, Eq.~\ref{eq:lossy}
						\Else ~
						Find $s_i^*(\mu_K)$ using  Theorem~\ref{thoremthresholds}, Eq.~\ref{remk4eq}
						%		\ElsIf{$\kappa_i\in [\eta_{\text{ch}}\eta_{\text{dis}}, 1)$}
						%		Find $s_i^*(\mu_K)$ using Remark~\ref{rmk:remark5}, Eq.~\ref{remk5eq}
						%		\ElsIf{$\kappa_i = \eta_{\text{ch}}\eta_{\text{dis}}$}
						%		Find $s_i^*(\mu_K)$ with Rmk.~\ref{rmk:remark6}, Eq.~\ref{remk6eq}
						\EndIf
						\State $x_i^*(\mu_K)$=$f^{-1}(s_i^*(\mu_K))$ and {$b_i^*(\mu_K)$=$b_{i-1}^* (\mu_K)+x_i^*(\mu_K)$}
						\If{C1 or C2 or C3 holds}
						$\mathtt{BreakFlag} \gets 1$; $i_{\text{break}} \gets i$ 
						\State \textbf{Break}
						\EndIf
						\EndFor
						%\State Set $S_p= [p(n+1:cnt_{p})\eta_{\text{dis}}; p(n+1:cnt_{p})/\eta_{\text{ch}}]$ 
						%        \item[]
						\If{$\mathtt{BreakFlag=1}$ and $i_{\text{break}} \geq i_{\text{mem}}$}
						\State $\mathtt{BreakFlag} \gets 0$; $i_{\text{mem}} \gets i_{\text{break}}$; $\mu_{\text{mem}} \gets \mu_K$
						\If{$b_i^*(\mu_K) \prec \cbrac{b_{\min}}$}
						$\mu_{K}$ $\gets$ using Eq.~\ref{musearch1}, % $\min\{p >\mu: p\in \mu_1^i, \mu_2^i, \mu_3^i, \mu_4^i$; ${i \in( i_{K-1}, i_{\text{break}}]}\}$
						\Else  ~~$\mu_{K} \gets$ using Eq.~\ref{musearch2}, %\max \{p < \mu: p\in \mu_1^i, \mu_2^i, \mu_3^i, \mu_4^i$; ${ i \in( i_{K-1}, i_{\text{break}}]}\}$
						\EndIf
						\ElsIf{$\mathtt{BreakFlag=1}$ and $i_{\text{break}} < i_{\text{mem}}$}
						\If{C1 is True}
						\State $i_K \gets \max\{i \in [i_{K-1}+1, i_{\text{mem}}]: b_{\max}\in b_i^*(\mu_{\text{mem}})\}$
						\State Assign $b_{i_K}^*=b_{\max}$
						\ElsIf{C2 or C3 is True}
						\State $i_K \gets \max\{i \in [i_{K-1}+1, i_{\text{mem}}]: b_{\min}\in b_i^*(\mu_{\text{mem}})\}$
						\State Assign $b_{i_K}^*=b_{\min}$
						\EndIf
						\State $\mu_K \gets \mu_{\text{mem}}$; $\mathtt{BreakFlag} \gets 0$; $i_{\text{break}} \gets i_{\text{mem}}$
						\State Update $i_{\text{mem}} \gets i_K$
						\State \texttt{BackwardStep}($\mu_K, i_{K-1}, i_K, b^*, x^*, \mu^*$)
						\State $s^* = f(x^*)$ using Eq~\ref{finverse}
						\State $\mu_{K+1} \gets \mu_{K}$; $K \gets K+1$
						\Else
						\State $i_{K} \gets N$; 
						\If{$\mu_{K} >0$}
						~ $b_{N}^* \gets b_{\min}$
						\Else
						~ $b_{N}^* \gets [b_{\min}, b_{\max}) \cap b_{N}^*(\mu_K)$
						\EndIf
						\State \texttt{BackwardStep}($\mu_K, i_{K-1}, i_K, b^*, x^*, \mu^*$)
						\State $s^* = f(x^*)$ using Eq~\ref{finverse}
						\EndIf
						\EndWhile
					\end{algorithmic}    
					\caption{\texttt{OptimalArbitrage}($p_b,p_s, b_0$)}\label{alg:1}
					%	\vspace{-10pt}
				\end{algorithm}
				\vspace{-18pt}
				\subsection{Complexity Analysis}
				%{\em \textbf{Complexity Analysis}}: 
				Using the discrete nature of the optimization problem, we explicitly characterize the worst case run time of the proposed algorithm. 
				For a sub-horizon starting from instant $i$, there may be at most 
				$N-i+1$ more time instants which may be included in the same sub-horizon.
				Hence, in order to find the optimal accumulated Lagrange
				multiplier for the sub-horizon, we may have to update the value of $\mu_K$
				in the sub-horizon at most $4(N-i+1)$
				times (for NEM 2.0 at each instant $i$ 
				%two possible values $\mu_1^i, \mu_4^i$ may be checked for equal buying and selling price of electricity and 
				four possible values $\mu_1^i, \mu_2^i, \mu_3^i, \mu_4^i$ may be checked). 
				%for selling price strictly lower than buying price).
				For each update, a basic set of operations is performed.
				Furthermore, there may be separate sub-horizons starting from every time instant $i \in [1,N]$.
				Hence, a crude upper bound on the number of times the basic set of operations are 
				needed to be repeated is $\sum_{i=1}^{N} 4(N-i+1)$=$O(N^2)$.
				Therefore, the worst case time complexity of the proposed algorithm is $O(N^2)$.
				In most situations, however, the number of instants included in a sub-horizon
				does not grow with $N$ (see Section~\ref{subhoricase} where lookahead horizon is fragmented into sub-horizons; the length of these sub-horizons is governed by battery parameters, sampling time and electricity price). 
				%Hence, in most cases 
				%the proposed algorithm would be performed
				%linearly in time. 
				Empirically, the run-time complexity of the proposed algorithm grows approximately linearly with number of samples in the decision horizon.
				%Run-time comparison of the convex optimization, the linear programming and the proposed algorithm for performing arbitrage is shown in Appendix~\ref{appendixcomplexity}.
				%\textcolor{cyan}{	In Appendix~\ref{appendixcomplexity} we present a numerical case-study comparing our proposed algorithm with LP and convex optimization based benchmarks, we observe that our proposed algorithm outperforms other benchmarks.}
				{
					In Section~\ref{appendixcomplexity}, we present a numerical case-study comparing our proposed algorithm with LP- and convex optimization-based benchmarks and show that our algorithm outperforms these latter.
				}
				%	, it is observed that run-time complexity grows approximately linearly with number of samples in optimization horizon. 
				%	The proposed algorithm outperforms other benchmarks.

				\vspace{-6pt}
				
				\begin{algorithm}
					\small{\textbf{Inputs}: {$\mu_K, i_{K-1}, i_K, b^*, x^*, \mu^*$}}\\
					\textbf{Function}: {Computes components of the optimal vectors $b^*, x^*$ in the range $[i_{K-1}+1, i_K-1]$}.~
					\textbf{Initialize}: {$i \gets i_K-1$}
					\begin{algorithmic}[1]
						\While{$i \geq i_{K-1}+1$}
						\State $b_i^* \gets (b_{i+1}^*-x_{i+1}^*(\mu_K))\cap b_i^*(\mu_K) \cap [b_{\min}, b_{\max}]$,
						\State $x_{i+1}* \gets b_{i+1}^*-b_i^*$,
						\State $\mu_{i}^* \gets \mu_K$, %(initialize $\mu$ for next sub-horizon), %\textcolor{blue}{XXX It is not clear for me where we are using the updated values of $\mu_{i}^*$ in this Algo.!!!}
						{\State Calculate $i=i-1$},
						\EndWhile
					\end{algorithmic}
					\caption{\texttt{BackwardStep}($\mu_K, i_{K-1}, i_K, b^*, x^*, \mu^*$)}\label{alg:2}
				\end{algorithm}
				\vspace{-17pt}

				\subsection{General applicability}
				\vspace{-4pt}
				{
					The proposed algorithm is generally applicable for a system with finite capacity constraints and ramp rate constraints, and operating under piecewise linear convex costs.
					Since buying and selling decisions in a market are taken in discrete time-intervals with a fixed commodity price in an interval, in most cases, the cost function is piecewise linear. The convexity of the cost function needs an appropriate selection of the decision variable. Once the convexity of the cost function is ensured, the proposed algorithm can be applied.
					For example, in a market of commodities, a participant buys a commodity and stores it in its inventory, to sell it in the future to make a profit. Performing such a transaction has a coupling in buying and selling decisions due to the finite size of the inventory, much like storage performing arbitrage.
					At any time, due to infrastructure and/or capital constraint, the market participant can buy or sell no more than a given limit, much like the ramp rate constraint of a battery.
					%Usage of inventory incurs some cost which decides the efficiency of buying and selling decisions, analogous to battery efficiency.
					Applications such as integrating renewables for self-sufficiency as in \cite{fares2017impacts}, \cite{hashmi2019energy} and controlling excess generation as in \cite{mueller2018evaluating, hashmiThesisPhd} can use the proposed algorithm for identifying storage control decisions.
					More general problems, such as the wheat trading model as presented in \cite{hartl1986forward}, can be solved using the proposed algorithm. 
					%This problem has some aspects analogous to ours in the sense that it has constraints analogous to battery constraints and the objective function is profit maximization, similar to $P_{\text{NEM}}$.
				}
				\vspace{-12pt}

	\section{Real-Time Implementation}
	%The problem of real time energy arbitrage consists of \textcolor{blue}{two coupled problems}: forecasting future parameters like electricity price, end user demand and solar PV generation \textcolor{blue}{required for performing optimal arbitrage PLEASE SAY CLEARLY WHAT IS THE SECOND PROBLEM!! [FORECASTING ... AND PERFORMING ...??]}.
	{The problem of real time energy arbitrage consists of two coupled subproblems: (i) the optimal energy arbitrage and (ii) the forecasting of future parameters (i.e., the electricity price, the end user demand and the solar PV generation), required for performing optimal arbitrage}.
	Due to the mismatch in forecasted and actual parameter values, the arbitrage gains will likely be lower than deterministic optimal arbitrage gains. In this section we present an online algorithm which uses incrementally improving forecasting of parameters along with Model Predictive Control (MPC) to decide optimal control actions.
	MPC is used to optimize the decisions in current time slots, while taking into account future time slots. In {the receding horizon} the forecast is updated and MPC is implemented again, till end time is reached.
	%In our proposed optimal arbitrage algorithm, \texttt{OptimalArbitrage}, we emphasized that the optimal control decisions are independent of past or future values of price, solar generation and consumer inelastic demand beyond the considered sub-horizon.
	The definition of sub-horizon presented earlier
	suggests that we only need to accurately forecast for time instants in proximity to the current time instant, as optimal actions beyond the current sub-horizon are not influenced by parameters in future sub-horizons.
	%compared to distant time instants which will not affect the optimal decisions, as optimal actions beyond the current sub-horizon are not influenced by parameters in future sub-horizons.
	However, quantifying the length of a sub-horizon is challenging since it is governed by storage parameters and variation of prices.
	{A sub-horizon denotes the optimal lookahead in future intervals for selecting the control decisions. In Section~\ref{subhoricase} we present a heuristic-based analysis to quantify the required lookahead based on storage type. We observe that the optimal lookahead for performing arbitrage depends on the ratio of ramp rate over capacity, efficiency and ratio of buying and selling price. 
		%For batteries with high ratio the required lookahead is smaller compared to those with lower ratio. 
		%batteries with lower ratio of ramp rate over capacity.
	}

	%The deterministic arbitrage gains is denoted as $V_a^*$. This denotes the maximum value obtained by arbitrage for actual value of parameters under complete information about future variations of parameters. 
	%It is given as  $V_a^* = \sum_{i=1}^N p_{\text{elec}}(i)(z_i - L_i^{*}(z_i, p_{\text{elec}}(i)))$.
	%%\textcolor{red}{where $L_i^{*}$. PLEASE DEFINE $L_i^{*}$!!!}
	%% Comment: L_i defined in Page 2
	%The real arbitrage gains that the end user would make using optimal actions are calculated using forecasted parameters. 
	%Due to forecast errors, the end user's energy arbitrage gains are affected. 
	%Realistic arbitrage gain,$V_r$, is the energy arbitrage gain made by following the same actions as calculated for forecasted signals and is denoted as
	%$
	%V_r = \sum_{i=1}^N {p}_{\text{elec}} (i) ({z}_i - \hat{L}_i^{*}(\hat{z}_i, \hat{p}_{\text{elec}} (i) ))$.
	%$\hat{L}_i^{*}$ is the optimal end user net consumption for the forecasted price signal, $\hat{p}_{\text{elec}}(i)$, and forecasted load vector, $\hat{z}$. 
	%It is evident that $V_a^*\leq V_r$. In this section we show how adaptive forecasting and MPC can efficiently mitigate the effects of forecast error on user's arbitrage gains.
	
	{
		The deterministic arbitrage gains expressed as $V_a^*= \sum_{i=1}^N p_{\text{elec}}(i)(z_i - L_i^{*}(z_i, p_{\text{elec}}(i)))$, provide upper limits on arbitrage gains under complete information setting.
	}
	%The real arbitrage gains that the end user would make using optimal actions are calculated using forecasted parameters. 
	Due to forecast errors, the end user's energy arbitrage gains are affected. 
	Realistic arbitrage gain,$V_r = \sum_{i=1}^N {p}_{\text{elec}} (i) ({z}_i - \hat{L}_i^{*}(\hat{z}_i, \hat{p}_{\text{elec}} (i) ))$.
	$\hat{L}_i^{*}$ is the optimal end user net consumption for the forecasted price signal, $\hat{p}_{\text{elec}}(i)$, and forecasted load vector, $\hat{z}$. 
	It is evident that $V_a^*\leq V_r$. % In this section we propose online implementation of our arbitrage algorithm in MPC framework with forecasting in effect significantly mitigating the effects of forecast error on user's arbitrage gains.

		\subsection{Forecast Model}
		\label{forecastappen}
		%In this section, we aim at developing a forecast model
		%for net load consumption of the end user without storage where the forecast is updated after each time step. This forecast information is fed to a model predictive control based on the proposed algorithm in Section III. 
		We define the mean behavior of past values of net load without storage at time step $i$ as
		\vspace{-10pt}
		\begin{equation}
		{\overbar{z}}_i=\frac{1}{D} \sum_{p=1}^D z_{(i-pN)} \quad \forall i \in \{k,...,N\}, k\geq 1,
		\end{equation}
		where $N$ is the number of points in a time horizon of 1 day, and $D$ is the number of days in the past whose values are considered in calculating $\overbar{z}$.
		The actual value of net load without storage is given as
		\begin{equation}
		{z}_i= \overbar{z}_i + {X}_i   \quad \forall i \in \{k,...,N\}, k\geq 1,
		\end{equation}
		where ${X}_i$ represents the actual difference from the mean behavior. 
		{The forecasted net load (without storage) is given as}
		\begin{equation}
		\hat{z}_i= \overbar{z}_i + \hat{X}_i   \quad \forall i \in \{k,...,N\}, k\geq 1,
		\label{zhat}
		\end{equation}
		where $\hat{X}_i$ represents the forecasted difference from the mean behavior. We define $\hat{X}_i \quad \forall i \in \{k,...,N\}$ as
		\begin{equation}
		\hat{X}_k=\alpha_1 X_{k-1} + \alpha_2 X_{k-2} + \alpha_3 X_{k-3} + \beta_1 \delta_k^1+\beta_2 \delta_k^2+\beta_3 \delta_k^3,
		\label{xhatk}
		\end{equation}
		where $\delta_k^m=(z_{k-mN} - {\overbar{z}}_{k-mN})$ and $\alpha_i, \beta_i \forall i \in\{1,2,3\}$ are constant. Our forecast model uses the errors in net load without storage for the past three time steps and the error in the same time step for past three days. At time step $i=k-1$ we calculate $\hat{X}_k$ as shown in Eq~\ref{xhatk}. We calculate $\hat{X}_{k+1}$ till $\hat{X}_N$ in order to update the forecast signal to be fed to MPC.
		\vspace{-5pt}
		\begin{gather*}
		\hat{X}_{N}=\alpha_1 \hat{X}_{N-1} + \alpha_2 \hat{X}_{N-2} + \alpha_3 \hat{X}_{N-3} + \sum_{q=1}^3\beta_q \delta_{N}^q.
		\end{gather*}
		%\vspace{-5pt}
		We calculate $\hat{z}$ using Eq.~\ref{zhat}. The vector $\hat{z}$ is fed to MPC for calculating optimal energy storage actions for time step $i=k-1$. At the time step $i=k$, we can calculate $X_k={z}_k- \overbar{z}_k$. Similar steps are done for $i \in\{k+1,...,N\}$, till the end of time horizon is reached.

	\subsection{Model Predictive Control}

	{
		We calculate energy arbitrage gains sequentially with incrementally improving the forecast model. To this aim, we implement the forecast model using the AutoRegressive Moving Average (ARMA) model.
		In this work, we focus on and develop a forecast model for the net load consumption of the end user without storage.
		%\textcolor{blue}{Note that a similar forecasting model can be developed for electricity price.}
		%The details of the forecast model and online algorithm is presented in Appendix~\ref{forecastappen}.
		The details of the forecast model and an online algorithm for real-time implementation of the proposed arbitrage algorithm are presented in Section~\ref{forecastappen}.
	}
	%The forecasted end user consumption without storage is fed to the online algorithm, \texttt{ForecastPlusMPC}, given in Appendix~\ref{algoappendix}. 
%	The online algorithm, \texttt{ForecastPlusMPC} is detailed in Appendix~\ref{algoappendix}.
	%, is executed sequentially and uses MPC to identity the optimal modes of operation of storage.
	The online algorithm for real-time implementation of optimal arbitrage algorithm is given as \texttt{ForecastPlusMPC}.
	\begin{algorithm}
		\small{\textbf{Global Inputs}: {$\eta_{\text{ch}}, \eta_{\text{dis}}, \delta_{\max}, \delta_{\min}, b_{\max}, b_{\min}$}, $b_0$}\\
		\small{\textbf{Inputs}: {$h, N, T,i=0$}}
		\begin{algorithmic}[1]
			\While{$i < N$}
			\State $i=i+1$
			\State Forecast $\hat{p}_{elec}, \hat{z}$ from time step $i$ to $ N$
			\State $s^* = \texttt{Algorithm1}(\hat{p}_{elec}, \hat{z}, h, N, T)$
			\State $b_i^*= b_{i-1}+x_i^* $,
			\State Update $b_0=b_i^*$.
			\EndWhile
		\end{algorithmic}
		\caption{\texttt{ForecastPlusMPC}}\label{alg:3}
	\end{algorithm}

	\vspace{-10pt}

	\section{Numerical Results}
\begin{figure}[!htbp]
		\center
		\includegraphics[width=3.0in]{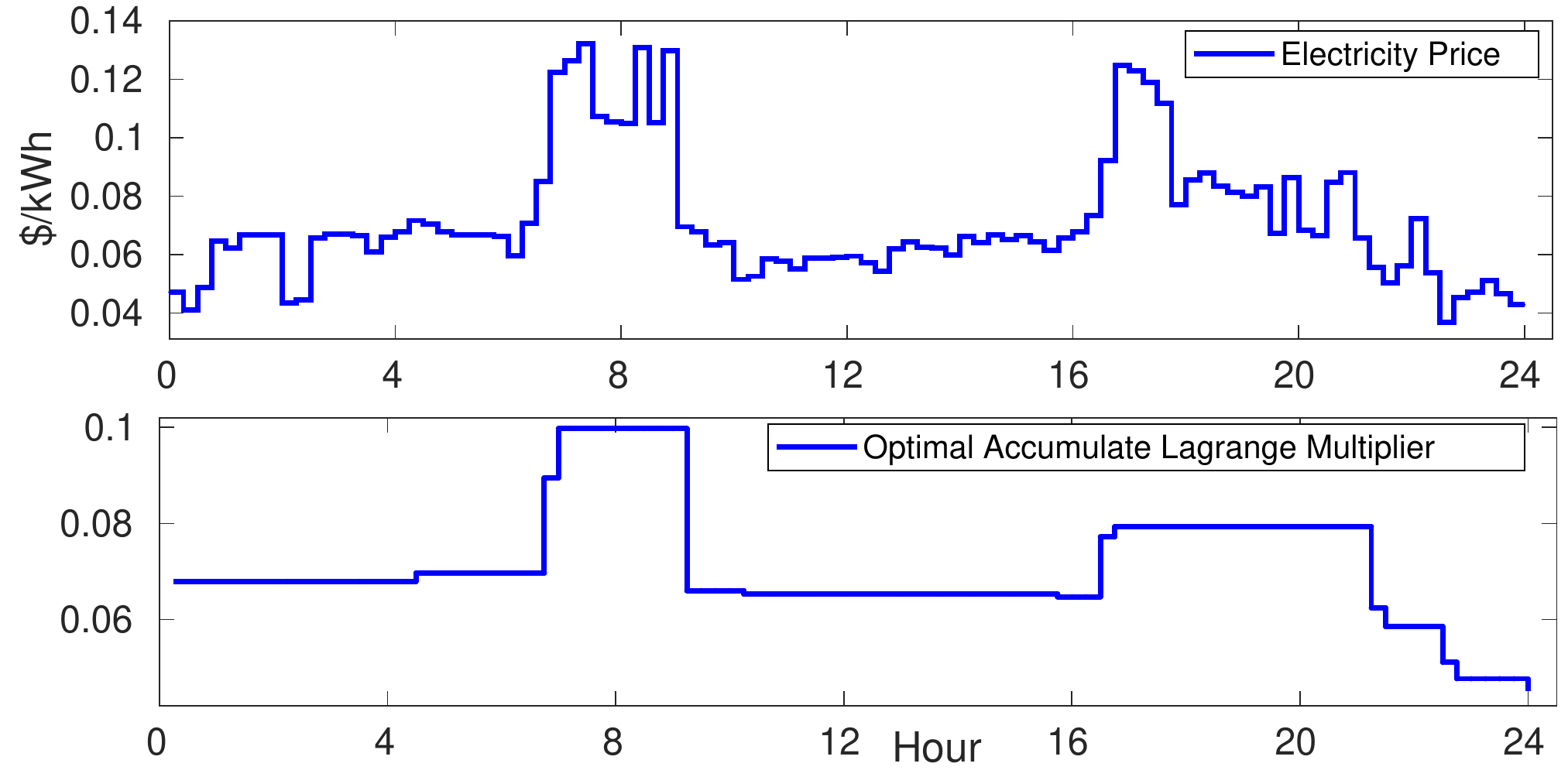}
		\vspace{-8pt}
		\caption{\small{Price Signal and accumulated Lagrange Multiplier}}\label{price}
	\end{figure}

\begin{figure}[!htbp]
		\center
		\includegraphics[width=3.0in]{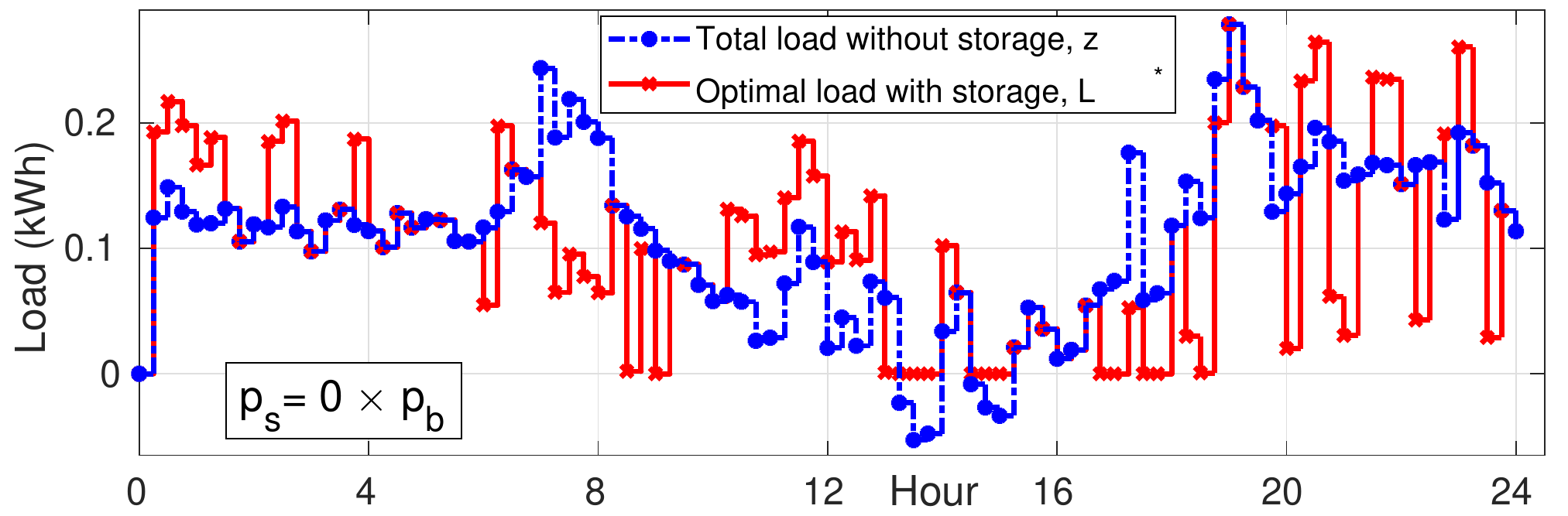} \vspace{-7pt}
		\caption{\small{Demand without storage and with optimal storage control}}\label{load_plot}
	\end{figure}

	For the numerical evaluation, we use a single end user having an inelastic power and energy demand and a rooftop solar generation.
	The end user's consumption data with solar generation data are downloaded from the Pecan Street's online data repository \cite{pecan}. We use the data corresponding to user id 379 for May 2, 2016. This is one of the days when the solar generation exceeds the end user's consumption for some amount of time. 
	%Figure~\ref{pecan} is the plot of resampled data, in order to match the sampling time of electricity price (sampled every 15 minutes). 
	%We solve the optimal arbitrage problem using \texttt{Algorithm 1} described in the previous section.
	The battery parameters are set as follows: $b_{\max}$= 1 kWh,  $b_{\min}$= 0.1 kWh, $\delta_{\max}$= 0.26 kW, $\delta_{\min}$=$- 0.52$ kW.
	Real-time locational marginal pricing data from NYISO \cite{nyiso} is used to calculate the optimal ramping trajectory. The sampling time of the price signal is $h$=15 minutes.
	Simulations are conducted using a laptop PC with Intel Core i7-6600 CPU, 2.6 GHz processor and 16 GB RAM.
	The results obtained by our algorithm for a lossy battery with $\eta_{\text{ch}}$ = $\eta_{\text{dis}}$ = 0.95 and for the case of zero selling price are shown in the figures below.
	%
	%The simulation results below are shown for zero selling price case. 
	Fig.~\ref{price} shows the electricity buying price
	%\footnote{NYISO Real Time Electricity Price Data,  {https://tinyurl.com/2flowo6}} 
	and the shadow price, i.e. $\mu^*$, calculated using the proposed algorithm for initial battery charge level $b_0=0.5$.
	%Figure~\ref{load_plot} shows the load with and without storage. We assumed zero selling price, implying it is not benefitial to supply power back to the grid. 
	%%The net load without storage has no flexibility in shifting the load as 
	%We have assumed non-elastic end user consumption, therefore in the afternoon when the end user generates more than the consumption, it still supplies power back to the grid. 
	Since we assume zero selling price, implying that it is not beneficial for the end user to supply power back to the grid, we can observe from Fig.~\ref{load_plot} that in the afternoon when the end user generates more than the consumption, it still supplies power back to the grid, as the user has no flexibility in form of storage.
	However, with the inclusion of storage the net-load saturates at zero level.
	% we can observe the modification in the total energy consumption curve. 
	% In fact, since it is not beneficial to sell electricity, the total consumption is saturated at zero as the lower limit.
	%\vspace{-10pt}
	Table~\ref{tab:title1} compares the run-time of the proposed algorithm with
	Linear programming \cite{hashmi2019lp},    CVX \cite{cvx}
	% \footnote{CVX Matlab Toolbox, {http://cvxr.com/cvx}}
	, YALMIP and Matlab's Fmincon optimization tool. 
	
	%Additional numerical result plots are shown in Appendix~\ref{numappendix}.

	%It should be noted that the forecast model using ARMA uses mean behavior over the past week as the starting point, however, the mean behavior provides a very crude indicator of future behavior. Even with this model along with MPC, we could significantly mitigate the effects of forecast error on arbitrage gains.
	%\vspace{-7pt}
	%\vspace{-8pt}
	
	%The simulation run time is $0.3524$ seconds, much faster compared to Matlab's Fmincon based optimization which takes $24.427$ seconds.
	%\vspace{-15pt}
	%\vspace{-10pt}
	
%	\vspace{-6pt}
	\begin{table}[!htbp]
		\scriptsize
		\caption{\small{Comparison of runtime}} \label{tab:title1}
		\small
		\vspace{-12pt}
		\begin{center}
			\begin{tabular}{| c |c| } 
				\hline 
				Algorithm Type &Run Time (sec)\\
				\hline 
				\hline 
				Proposed Algo &0.019\\
				\hline 
				Linear Program \cite{hashmi2019lp} &0.045\\
				\hline 
				CVX \cite{cvx}  &0.536\\
				\hline
				YALMIP \cite{Lofberg2004} with Gurobi & 3.700\\
				\hline
				Matlab's Fmincon	&	5.693 \\
				\hline
			\end{tabular}
		\end{center}
	\end{table}
%	\vspace{-10pt}
%	\vspace{-13pt}
%	\vspace{-21pt}
	%\vspace{-6pt}
%	\vspace{-6pt}
%	\vspace{-12pt}
	\begin{table}[!htbp]
		\scriptsize
		\caption {Mean and Standard Deviation of arbitrage gains} \label{tab_gain} \vspace{-12pt}
		\begin{center}
			\small
			\begin{tabular}{| c | c|c |}
				\hline
				Gain Type &  Mean (\$) & STD \\ 
				\hline
				Ideal ($V_a^*$)& 0.05481& 0.04673\\
				\hline
				Actual ($V_r$) & 0.04783& 0.04922\\
				\hline
			\end{tabular}
			\hfill\
		\end{center}
	\end{table}
	
\begin{figure}[!htbp]
			\center
			\includegraphics[width=3.1in]{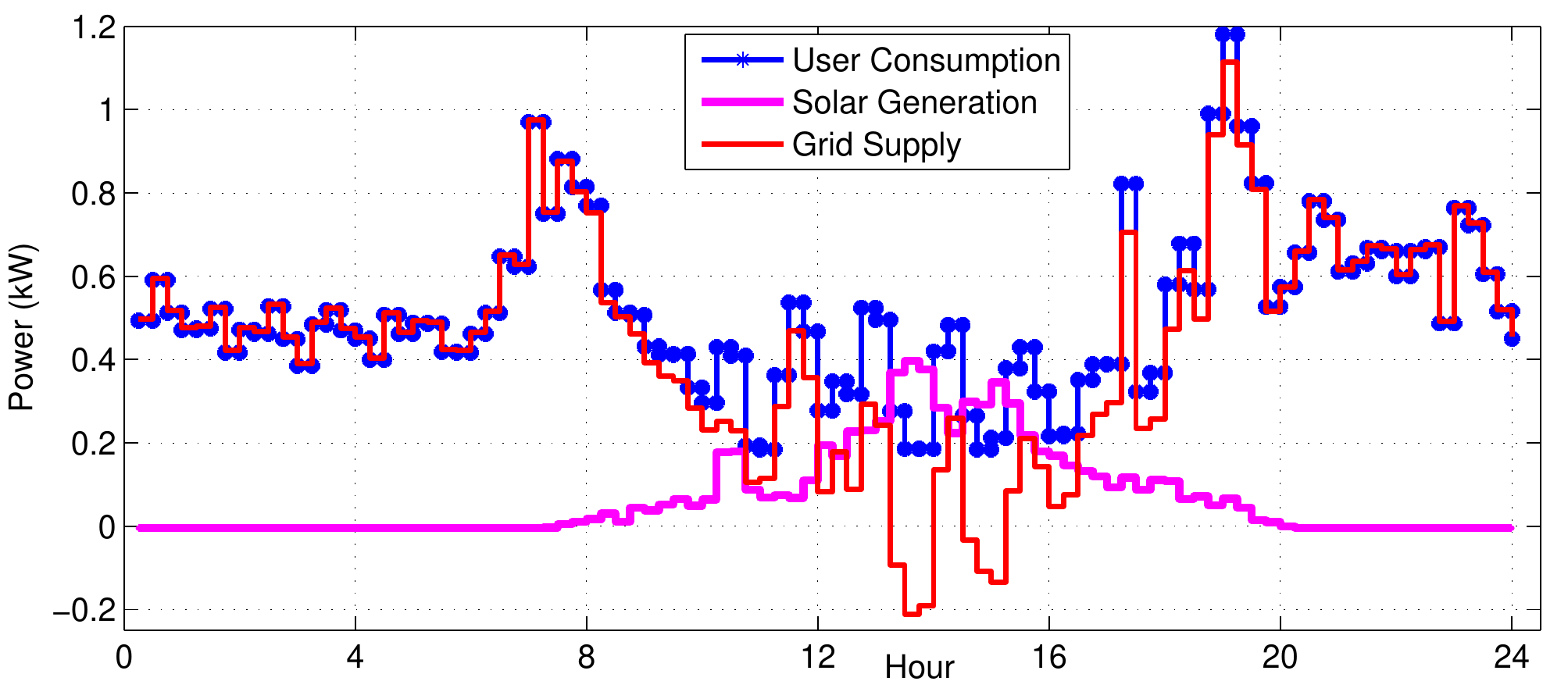}
			\caption{Pecan Street Data of a home with solar generation}\label{pecan}
		\end{figure}

		Fig.~\ref{pecan} shows the solar generation, the end user demand and the energy consumed from the grid.
		Now, consider a hypothetical scenario, when this end user might have had energy storage installed. How much gains could an end user make by installing an energy storage? \\
		
		\begin{figure}[!htbp]
			\center
			\includegraphics[width=3.1in]{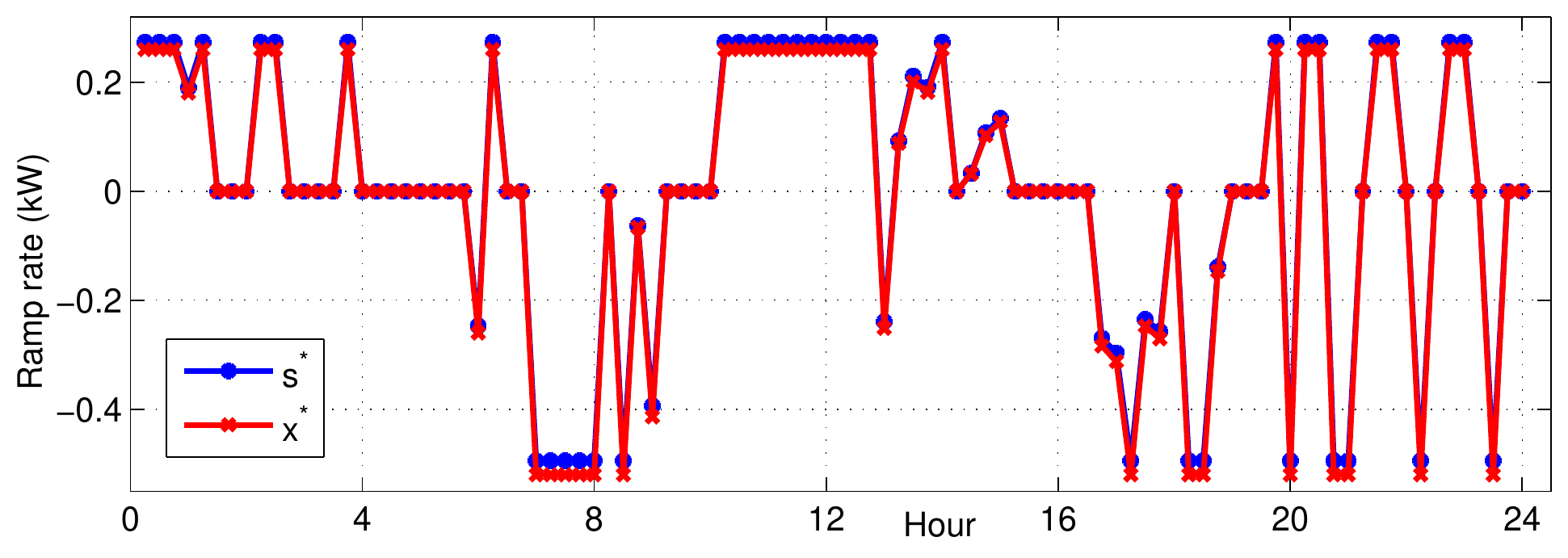}
			\caption{Ramp rate for optimal arbitrage}\label{ramp}
		\end{figure}
		Fig.~\ref{ramp} shows the ramp rate of the battery. It is evident from Fig.~\ref{ramp} that the ramping constraints for the battery are met and intermediate ramp rates could also be optimal. 
		\begin{figure}[!htbp]
			\center
			\includegraphics[width=3.0in]{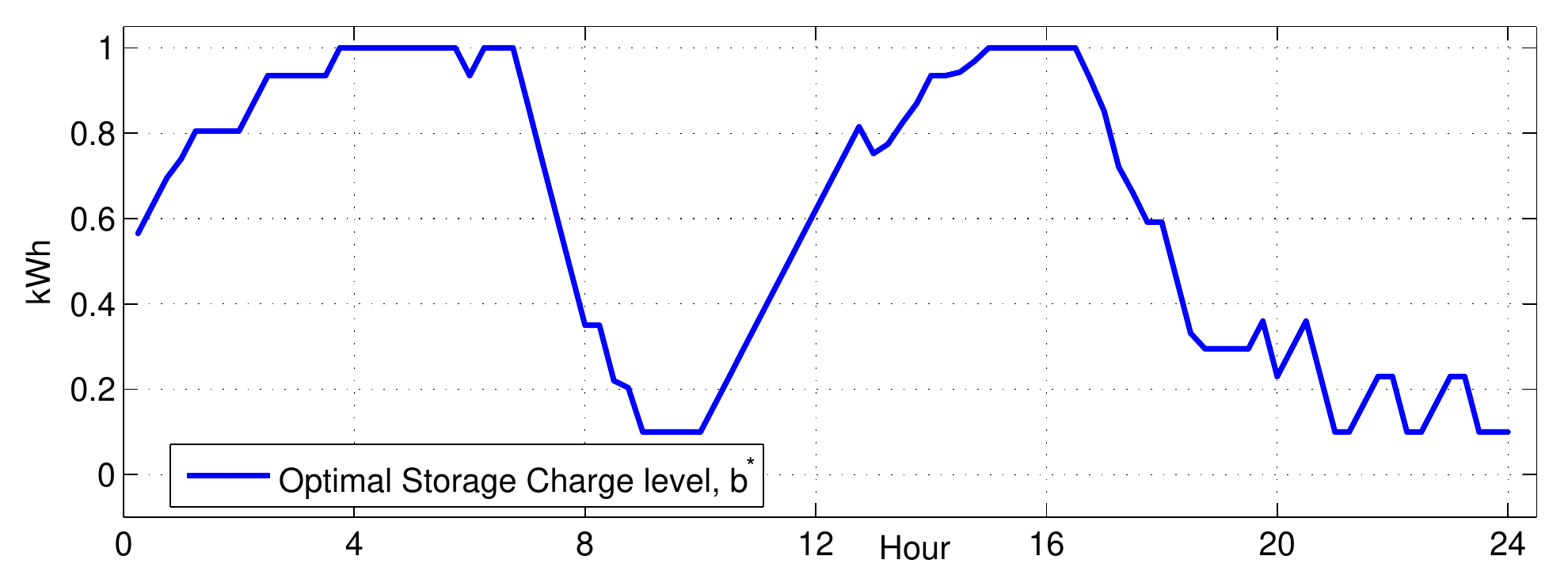}
			\caption{Optimal Energy level}\label{bopt}
		\end{figure}
		%\vspace{-10pt}
		Fig.~\ref{bopt} shows the optimal battery capacity trajectory. Note from Fig.~\ref{price} that the price has two peaks in the whole day, thus the battery does 2 cycles of charge and discharge, as shown in Fig.~\ref{bopt}.
		
		\par The comparison of the change in valuation of only storage and only solar with the variation in the ratio of the selling price and buying price from an end user's perspective is studied. 
		%Here we present numerical results which give an insight into the variation of selling price and valuation of energy storage and renewable energy sources.
		\begin{figure}[!htbp]
			\center
			\includegraphics[width=3.0in]{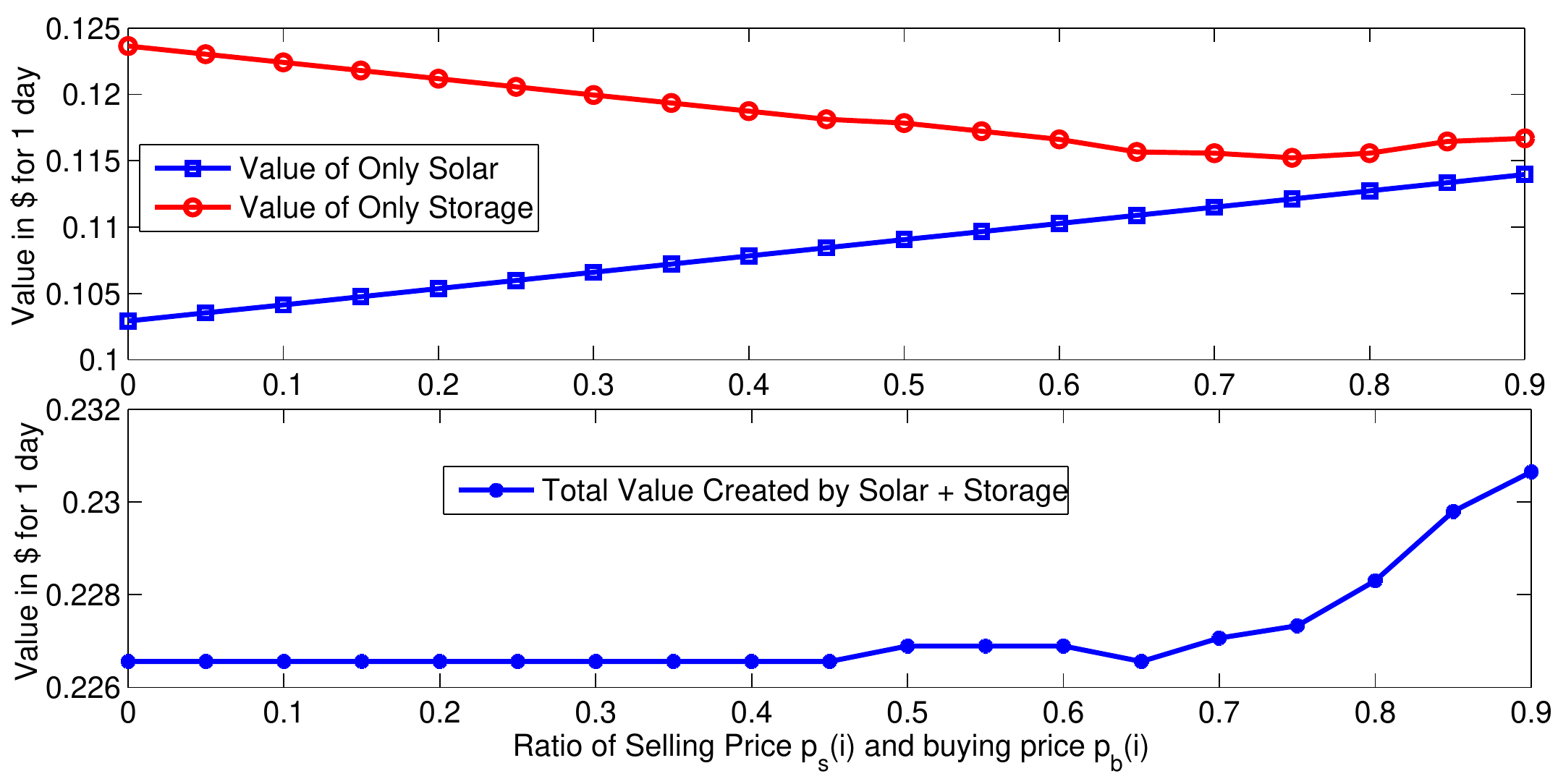}
			\caption{\small{Value of Solar and Storage with the ratio of $p_s$ and $p_b$}}\label{value}
		\end{figure}
		%\vspace{-20pt}
		Fig.~\ref{value} shows the variation of value of only storage and only renewable with the change in the selling price of electricity. We define the value of solar as the difference between cost of consumption with only load and cost of consumption with load and solar. Similarly, the value of storage is defined as the difference between the cost of consumption with load and solar and the cost of consumption with load, solar and battery. 
		The value of storage is defined as 
		$\sum_{i=1}^N\{p_{b}[z_i]^+ - p_{s}[z_i]^- \}-  \sum_{i=1}^N \{ p_{b}[z_i+s_i]^+ - p_{s}[z_i+s_i]^- \}$. 
		We would like to highlight that when selling price is low there is an increase in the storage value which comes at the cost of decrease in the value of renewables connected. 
		%The total value created by solar plus storage remains constant till the selling price increase beyond $0.65 \times p_b(i) \forall i\in [1,...,N]$. However, this threshold will be a function of price curve and may different for a different price signal.
		%Figure~\ref{cost1} shows cost incured to the end user without and with energy storage with variation in selling price of electricity. 
		%\begin{figure}[!htbp]
		%  \center
		%  \includegraphics[width=3.3in]{cost_up.pdf}
		%  \caption{Cost of operation of end user with and without Storage with change in the ratio of $p_s$ and $p_b$}\label{cost1}
		%\end{figure}
		For zero selling price the value of energy storage for the numerical evaluation is \$ 0.1237.

		As the share of renewables connected to power network increases the volatility in electricity prices will increase in order to incentivize users to differ their consumption.  
		%	\vspace{-10pt}
		\begin{figure}[!htbp]
			\center
			\includegraphics[width=3.0in]{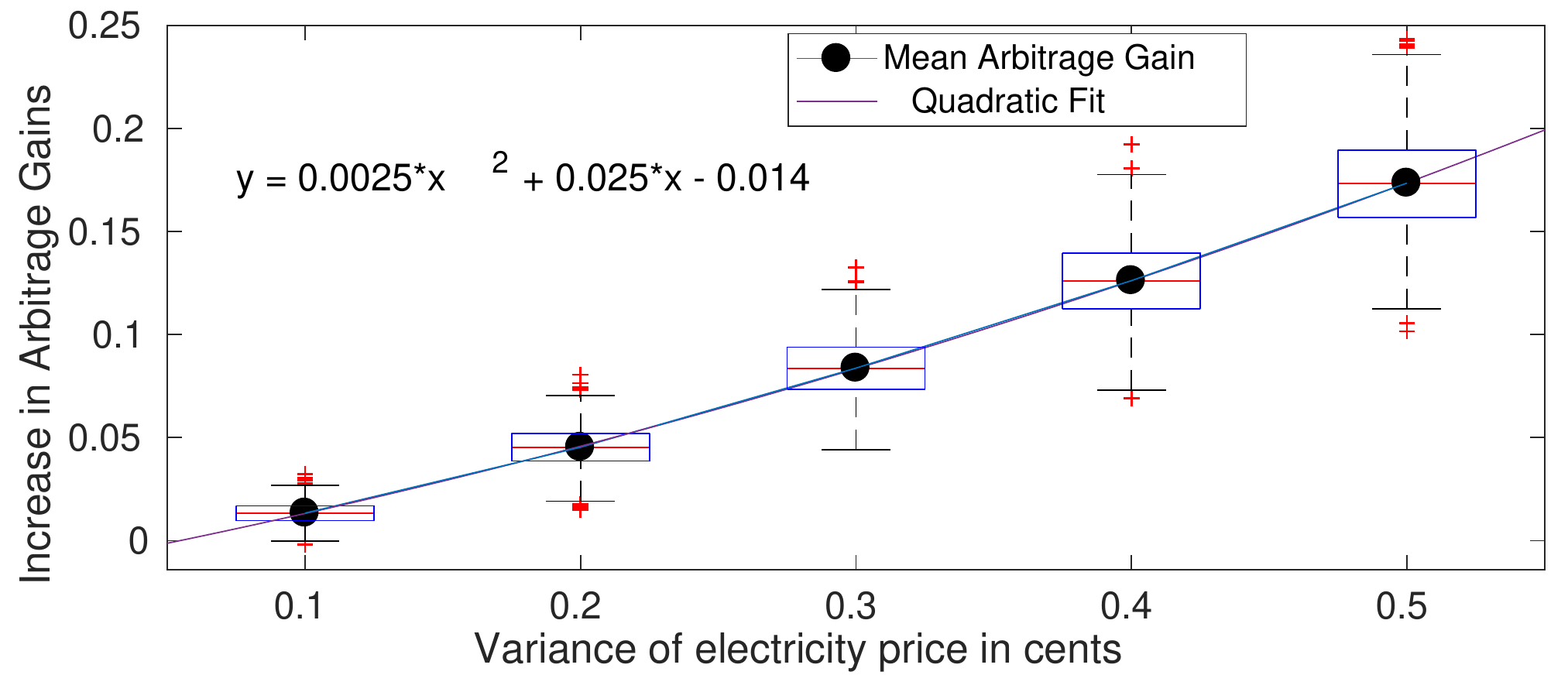}
			\caption{\small{Increase in arbitrage gains with volatility in price}}\label{volatility}
		\end{figure}
		%\vspace{-10pt}
		Fig.~\ref{volatility} shows that as the variance in electricity price increases, i.e. price volatility increases, the amount of arbitrage gains for consumers, with full information about price variation, will increase. The increase in arbitrage gains with respect to variance can be approximated using a quadratic fit. Installing energy storage would provide more financial returns under increased volatility.
		
\begin{figure}[!htbp]
			\center
			\includegraphics[width=2.9in]{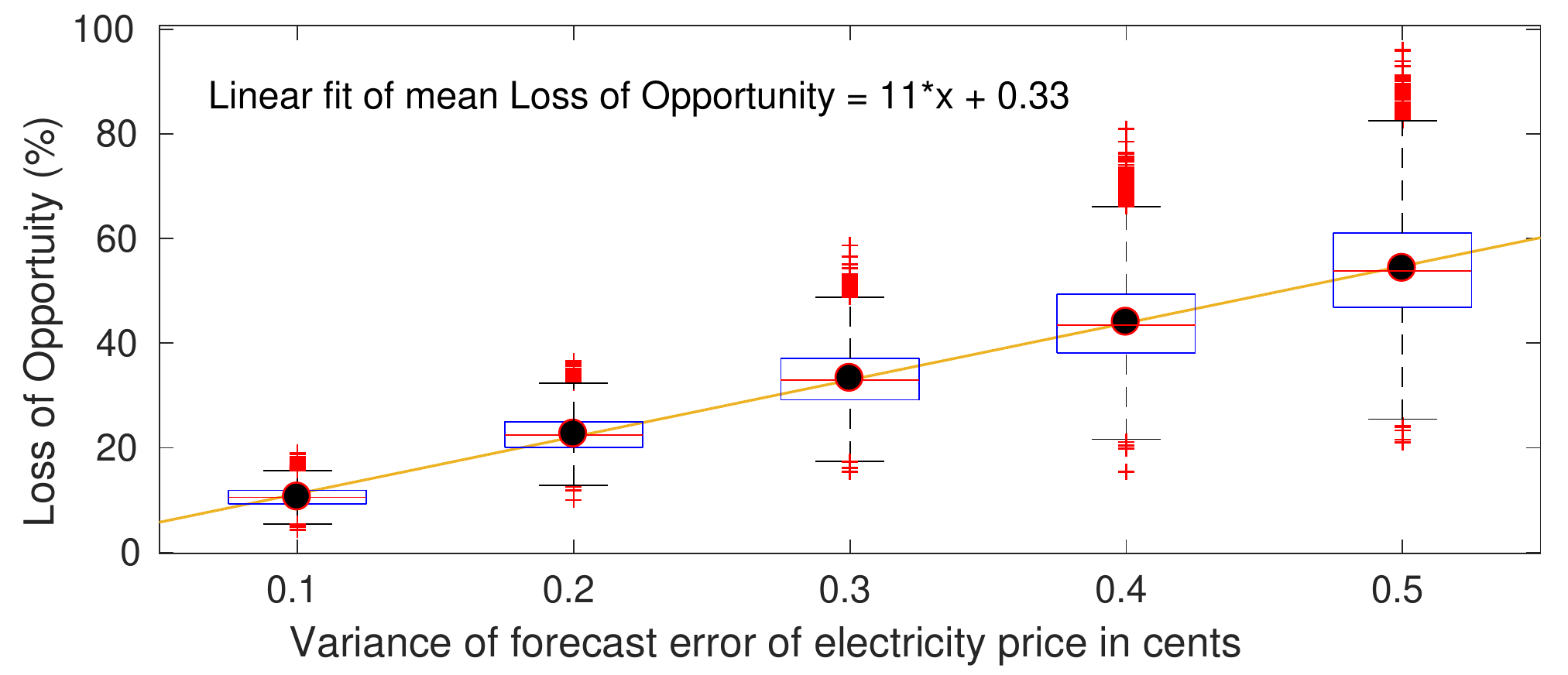}
			\caption{\small{Relationship of Loss of Opportunity and Forecast Error}}\label{perf}
		\end{figure}

		\textit{Forecast Error and Loss of Opportunity:}
		{It is expected that mismatch between forecast and actual values will possibly generate for the user a loss of opportunity. This latter is defined as the per unit variation of ideal versus actual arbitrage gain with respect to ideal arbitrage gains:
			$
			\text{Loss of Opportunity} = \frac{V_a^* - V_r}{V_a^*}.
			$
		}
		In order to understand the effect of forecast error in electricity price on the arbitrage gains we conduct a performance evaluation based on 10,000 simulations for equal buying and selling prices with battery having $95\%$ charging and discharging efficiency for different variance of forecast error (= actual price - forecasted price). Fig.~\ref{perf} shows that with increasing variance of forecast error the loss of opportunity for the user will increase. Black dots in Fig.~\ref{perf} represent the mean value of loss of opportunity corresponding to the variance in forecast error, which could be fitted with a linear function.
		
		%	\subsection{Real-time implementation}

%	\vspace{-23pt}
	\subsection{MPC with incrementally improving forecast}
\begin{figure}[!htbp]
		\center
		\includegraphics[width=2.6in]{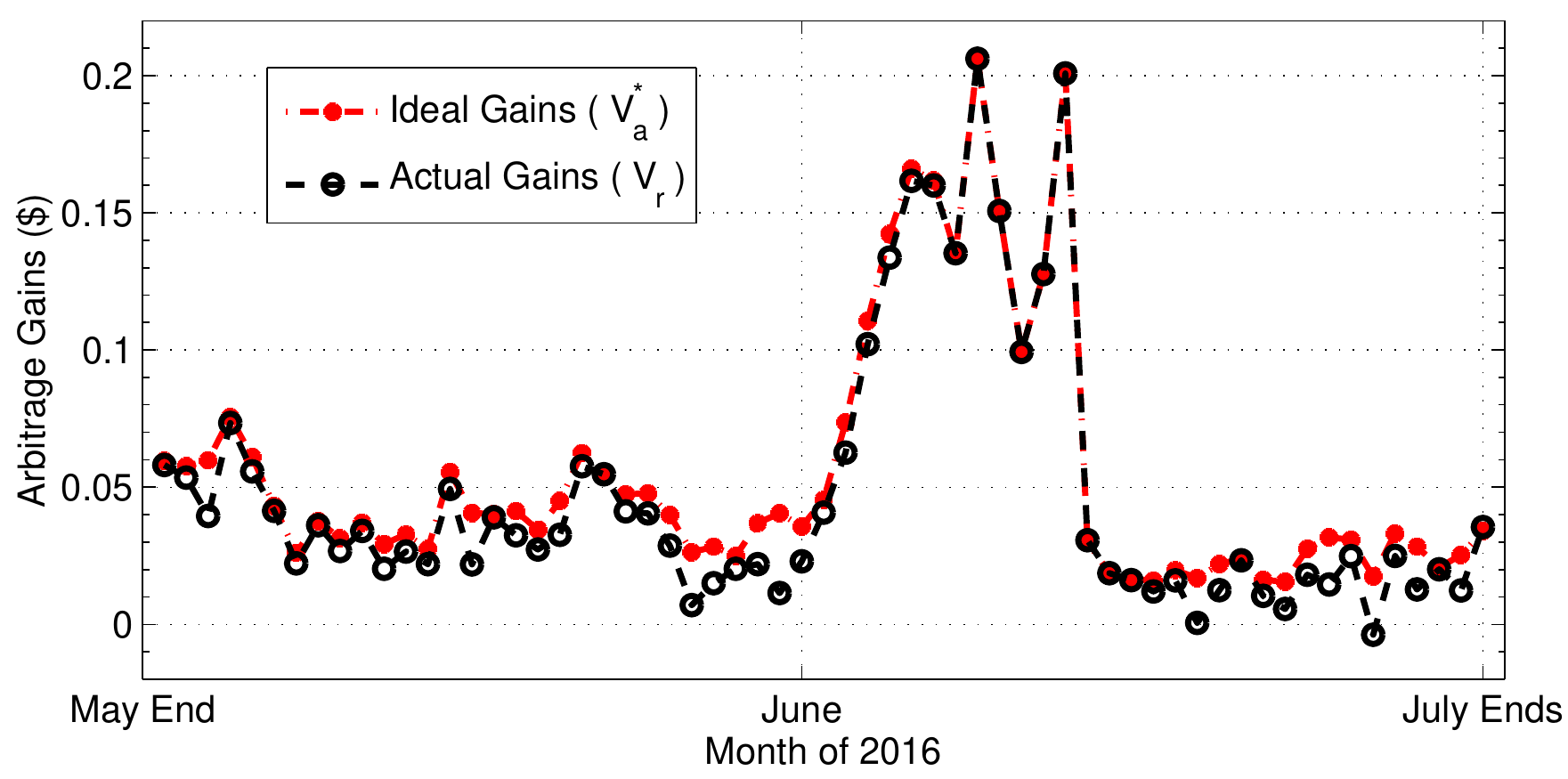} \vspace{-7pt}
		\caption{\small{Comparison of arbitrage gains for ideal and actual case}}\label{mpc1}
	\end{figure}

	\vspace{-3pt}
	%\textcolor{magenta}{@Umar: What do you mean by $x+x^2+x^3$ in the following paragraph? What is the meaning of $x$ here??}
	{
		We use the Pecan Street data \cite{pecan} for hourly consumption for house id 379 for the month of June and July 2016. The day ahead electricity prices in the ERCOT data are used for the same period \cite{ENOnline}. It is assumed that the selling price is half that of the buying price.
		The factors in Eq.~\ref{xhatk} are $\alpha_1=\beta_1=  0.27185$, $\alpha_2=\beta_2= 0.14780$ and $\alpha_3=\beta_3= 0.08036$.
		%note the factors are selected such that $x+x^2+x^3=1$. 
		Parameters $D=3$ days and $N=24$ hours (rolling horizon). We use the same parameters of the battery as described in the previous numerical results. 
		%The run time for forecasting and MPC for the whole period is $9.699$ seconds.
		It can be observed that actual and ideal arbitrage gains are in sync with each other, and that as expected $V_r \leq V_a^*$. Furthermore, Table~\ref{tab_gain} reports the mean and standard deviation of arbitrage gains ($V_a^*$ and $V_r$) over a period of 2 months. Due to inaccuracies in forecasting, end user incurs $\approx 12.7 \%$ of loss in possible opportunity during the period of 2 months. 
	}
	The simulation results for ARMA-based forecasting with MPC are shown in Fig.~\ref{mpc1}.
%	\vspace{-10pt}
	%Since the the forecast model developed, uses only past data and the exogenous parameters like solar insolation are ignored, the forecast error leads to loss of arbitrage gains for end users as shown in Fig.~\ref{block_d}. 
	%A more detailed forecasting model along with MPC will lead to even better performance.
	%A more sophisticated forecasting model along with MPC will lead to even better performance.
	%\vspace{-16pt}

	\section{Case Study I: Quantifying the length of a sub-horizon}
	\label{subhoricase}
	{Identifying optimal lookahead horizon for performing arbitrage would be essential for maximizing the end user gains.} Prior works \cite{mokrian2006stochastic} indicate selecting a time horizon of 1 day is sensible since the electricity pattern repeats with a period of one day approximately, being high during peak consumption hours during the day and low during the night \cite{hu2010optimal}.
	
	In this work we claim that the optimal control actions for energy storage device depends on electricity price and load variations in a smaller part of a larger time horizon and independent of all points in past or beyond the sub-horizon. However, identifying this optimal look-ahead period is challenging as it is governed by variations of electricity price, load and battery parameters. Next we present a case study for understanding the influence of battery parameters on the length of a sub-horizon.
	
	\begin{figure*}[!htbp]
		\center
		\includegraphics[width=6.5in]{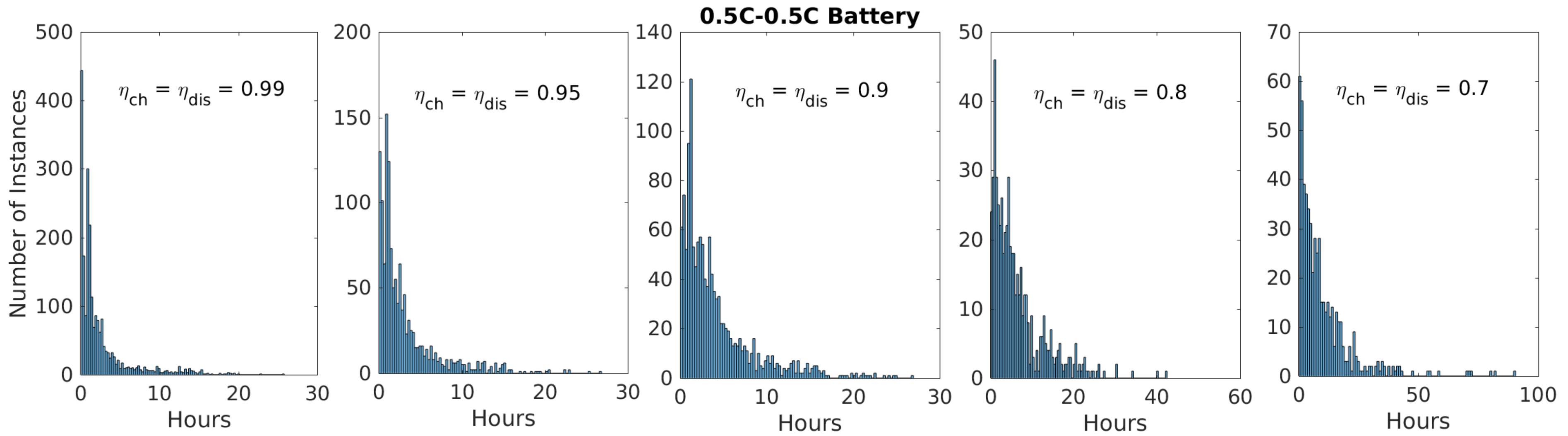}
		\caption{\small{Histogram of sub-horizon spread for 0.5C-0.5C battery for CAISO in 2017}}\label{fig1sc}
	\end{figure*}
	
	\subsection{Case Study: CAISO 2017 for NEM 1.0}
	{
		We consider the electricity price for CAISO of 2017 and identify the variations of the sub-horizon over a year with different energy storage parameters. The parameters setting in this case study is as follows:}
	\begin{itemize}
		\item Electricity price for CAISO in 2017,
		\item Ramp rate of the battery: we consider 1 kWh capacity battery with 3 different ramp rates. xC-yC represents that battery takes 1/x hours to completely charge and 1/y hours to completely discharge. 
		\item Efficiency of the battery: we consider 5 levels of efficiency ($\eta$): 0.99, 0.95, 0.9, 0.8 and 0.7. Here $\eta =\eta_{ch}=\eta_{dis}$. 
	\end{itemize}
	The performance indices used in this case study are:
	\begin{itemize}
		\item $T_{mean}$: denotes the mean length of a sub-horizon over the whole year,
		\item $T_{99\%}$: denotes the 99\% quantile,
		\item $T_{worst}$: denotes the worst case length of a sub-horizon,
		\item \$/cycle: denotes dollars per cycle gain and
		\item Gain: is the total arbitrage gain.
	\end{itemize}

	\begin{table}[!tbph]
		%		\scriptsize
		\caption {Quantifying the length of sub-horizon}
		\label{resulttab1sc}
		\vspace{-10pt}
		\begin{center}
			\begin{tabular}{| c| c| c|c| c|c|}
				\hline
				\multirow{2}{*}{Efficiency} & $T_{mean}$ & $T_{99\%}$ & $T_{worst}$  & \$/cyc & Gains \\ 
				$\eta$& hours	 & hours  & hours	     &        & \$ \\
				\hline
				\multicolumn{6}{|c|}{0.5 C - 0.5C Battery} \\
				\hline
				0.99 &2.45	&15.33	&25.75	&0.037 	&32.80	\\
				0.95 &3.20	&17.50	&26.75	&0.046 	&30.21	\\
				0.9  &4.11	&20.50	&26.92	&0.055 	&27.54	\\
				0.8  &6.68	&26.70	&42.42	&0.068 	&23.21	\\
				0.7  &9.73	&69.83	&90.58	&0.075 	&19.62	\\
				\hline
				\multicolumn{6}{|c|}{1 C - 1 C Battery} \\
				\hline
				0.99 &1.76	&11.83	&18.75	&0.036 	&59.52	\\
				0.95 &2.14	&13.00	&18.75	&0.047 	&54.84	\\
				0.9  &2.79	&14.83	&23.08	&0.059 	&50.04	\\
				0.8  &4.52	&18.67	&33.58	&0.077 	&42.29	\\
				0.7  &6.37	&30.67	&63.50	&0.085 	&35.86	\\
				\hline
				\multicolumn{6}{|c|}{2 C - 2 C Battery}\\
				\hline
				0.99 &1.21	&6.92	&12.75	&0.034 	&103.60	\\
				0.95 &1.61	&9.17	&14.17	&0.048 	&95.28	\\
				0.9  &2.10	&11.00	&16.33	&0.062 	&87.02	\\
				0.8  &3.32	&16.83	&26.42	&0.085 	&73.76	\\
				0.7  &4.60	&23.00	&58.50	&0.093	&62.65	\\
				\hline
			\end{tabular}
			\hfill\
		\end{center}
	\end{table}
	It is evident from Fig.~\ref{fig1sc} and Table~\ref{resulttab1sc} that the length of look ahead required for optimal energy storage arbitrage gains reduces as the ramping rate increases and as the energy storage battery becomes more efficient. As pointed in \cite{hashmi2018limiting} that the efficiency creates a dead band in threshold based structure and increase in efficiency implies battery should not operate during low returning transactions as it would not be profitable. Dollars per cycle calculated using prior work \cite{hashmi2018long} shows increase as the efficiency decreases. For more details refer to \cite{hashmi2018limiting} and \cite{hashmi2018long}. As the ramping of battery increases and battery becomes more efficient the optimal look-ahead period decreases making it more prone to inaccuracies in forecast information. This observation is in sync with \cite{yize2018stochastic}.
	
	Table~\ref{resulttab1sc} compares the look-ahead window required in hours for three batteries. xC-yC battery implies battery takes 1/x hours to charge and 1/y hours to discharge completely. All the three batteries have the same capacity but different ramping rates. The arbitrage gains increases as the battery becomes more efficient and as the ramping rate increases. The \$/cycle calculated here takes into account battery degradation due to operational cycles. Note the look-ahead window for 0.5C-0.C battery with 95\% efficiency (i.e. $ \eta_{ch}=\eta_{dis}=0.95$) is 15.5 hours or below for 99\% of sub-horizons over the whole year (CAISO, 2017), this decrease to 13 hours for 1C-1C battery and further reduces to just 9.17 hours for 2C-2C battery. This case study indicates that the optimal look-ahead window for performing arbitrage is not only governed by price variation but also battery ramping rate and round trip efficiency.
	Fig.~\ref{fig1sc} shows the spread of sub-horizons for 0.5C-0.5 battery with varying efficiency. For a highly efficient battery the spread is much more compact compared to less efficient one.
	
	For instance, faster ramping batteries with same energy capacity require smaller lookahead horizon compared to slower ramping batteries. \cite{cruise2019control} performs numerical studies for a hydro-storage facility in the UK and identifies that the forecast horizon varies between 1 to 15 days. The horizon is so long due to a slow ramping capability of hydro-storage. We observe a similar trend where forecast horizons vary between a few hours to several days for batteries with varying ramping rates, thus making it essential to model price impacts.
	For strictly convex cost function, our proposed algorithm simplifies to the one proposed in \cite{cruise2019control}.
	
	%\section{Comparison}

	\section{Case Study II: Intermediate ramp rate optimality}
	\label{intermediaterampcase}
	Several works on energy storage arbitrage use set thresholds according to which storage operation could be selected from 3 cases, i.e. charge at maximum rate, discharge at minimum rate or stay idle. We believe energy storage performing arbitrage could also have intermediate ramping rates which are neither minimum or maximum nor zero, making the optimal storage ramping selection a continuous set from minimum discharging rate to maximum charging rate.
	
\begin{figure}[!htbp]
		\center
		\includegraphics[width=3.3in]{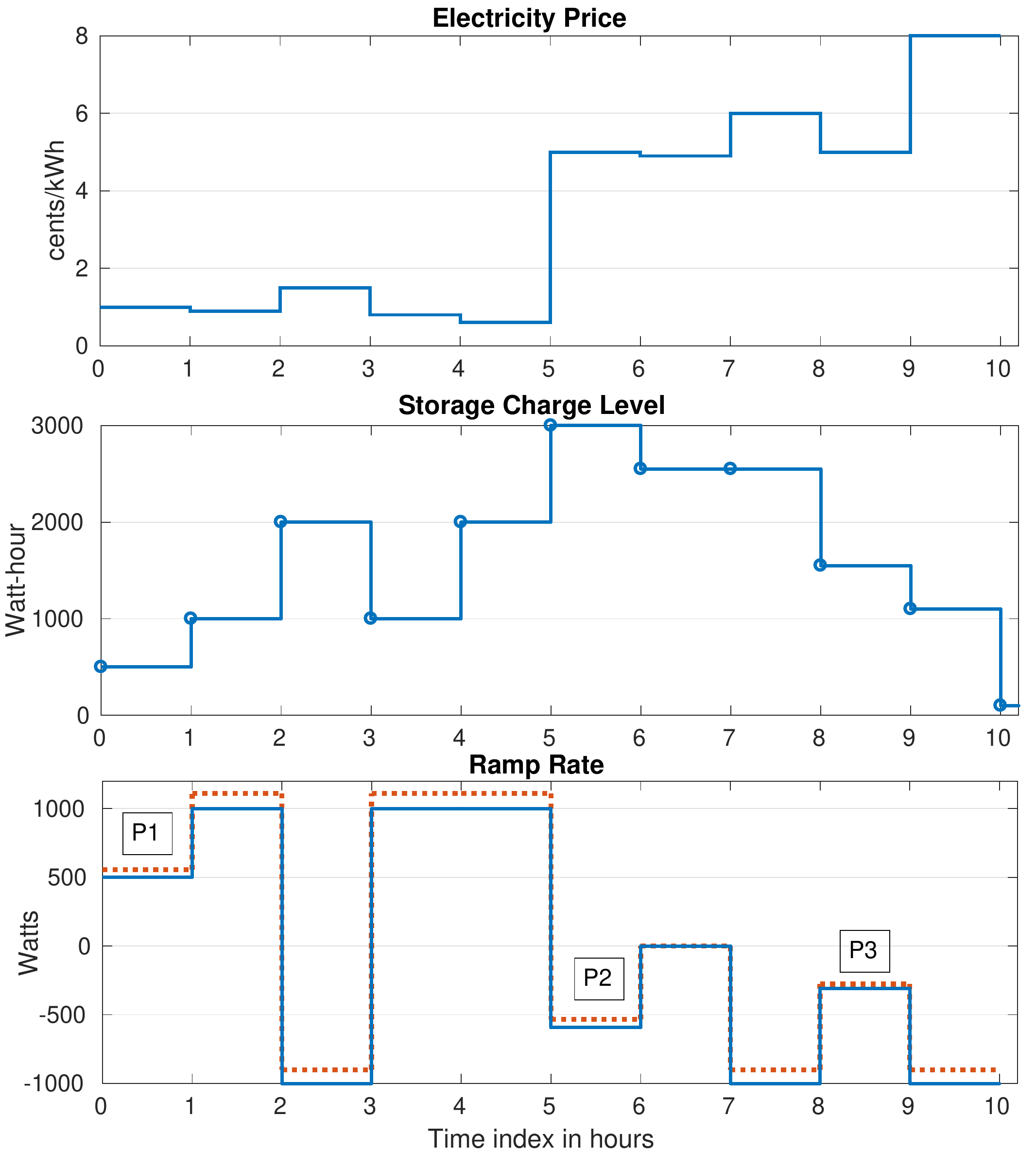}
		\caption{\small{Toy example to show the intermediate ramp rate of energy storage. Plot 1 shows the electricity price. Plot 2 shows the optimal storage charge level. Plot 3 shows the storage ramp rate.}}\label{interramp}
	\end{figure}

	We demonstrate this claim with a stylized example. 
	In this example we consider only storage case with equal buying and selling price of electricity, in order to have analytical tractability.
	Consider the electricity price signal shown in the first plot of Fig.~\ref{interramp}. The battery parameters are as follows: $b_0 = 500$Wh, $b_{\max}=3000$Wh, $b_{\max}=100$Wh, $\eta_{ch}=\eta_{dis}=0.9$, 
	$\delta_{\max}=-\delta_{\min}=1000$W.
	The electricity price values for hour 1 to 10 are provided here: [1, 0.9, 1.5, 0.8, 0.6, 5, 4.9, 6, 5, 8]. We intend to provide all details of the results presented in order to ease the reproducibility of the claims made here.

	The price signal is carefully designed to have values between 1 to 2 cent/kWh for hour 1 to 5 and higher levels of electricity price for hour 6 to 10. This could be analogous to low electricity price during the night and significantly higher price levels during the evening peak. Note that the electricity price for 7th and 9th hour are at the same level, i.e. 5 cents/kWh.
	The second plot of Fig.~\ref{interramp} shows the optimal storage charge level considering electricity price variation and storage parameters.
	The third plot of Fig.~\ref{interramp} shows the ramp rate in $x$ which affects the change in battery charge level in blue and $s$, output power of storage in red which considers the charging and discharging efficiency losses. Points marked P1, P2 and P3 shows ramping of the battery which are neither at maximum, minimum or zero level of ramp rate. 
	
	Clearly, based on the price variation for this example storage needs to be completely charged at the end of 5th hour. In order to discharge during higher price levels for interval 6 to 10 hour. 
	The order of price levels in 0 to 5 hour are in this order:
	$p_{elec}(5) <p_{elec}(4) <p_{elec}(2) <p_{elec}(1) <p_{elec}(3) $. Starting from $b_0 = 500$Wh, storage needs 2500 Wh of energy to be fully charged at the end of 5th hour. The battery charges at a ramp rate of 500 W in hour 1. This level is lower than the max level of ramp rate. The battery reaches a charge level of 1000 Wh. $p_{elec}(2)$ is the third lower price in hour 1 to 5 and the battery charges at maximum rate in order to discharge during hour 3 to capture gains as $p_{elec}(3)$ is the local peak. In subsequent 4th and 5th hour the battery charges at max level to unity state-of-charge at the end of 5th hour.
	
\begin{figure}[!htbp]
		\center
		\includegraphics[width=3.3in]{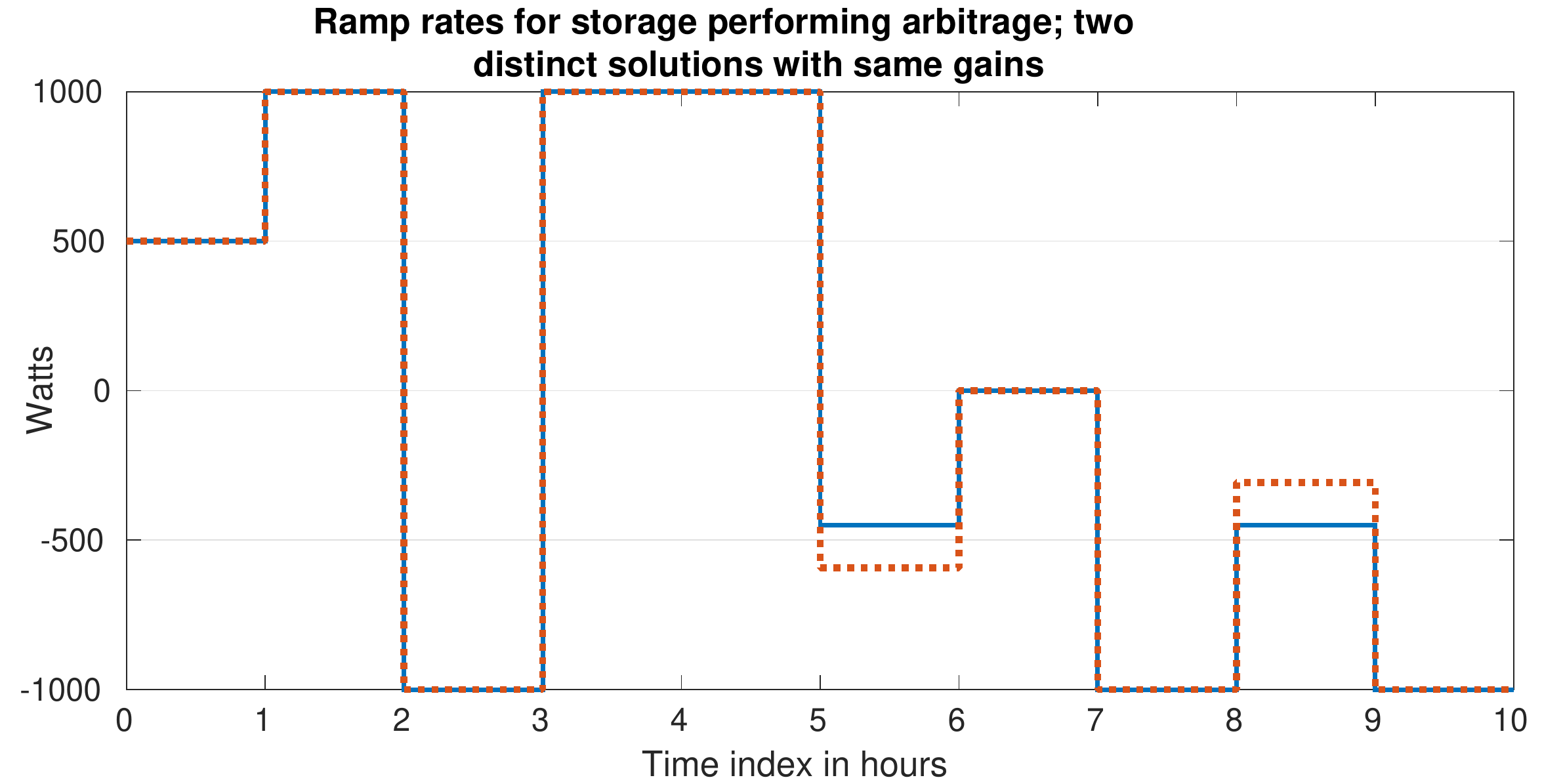}
		\caption{\small{Two distinct optimal ramping solution for performing energy arbitrage for the electricity price signal shown in Fig.~\ref{interramp}.}}\label{distinctsol}
	\end{figure}

	The order of price levels in 6 to 10 hour are in this order:
	$p_{elec}(7) <p_{elec}(6) = p_{elec}(8) <p_{elec}(8) <p_{elec}(10) $.
	Clearly, battery should should discharge maximum possible during 10th hour and then 8th hour. If the battery is still not completely discharged than during 6th and 9th hour. The incentive of discharging during hour 6 and 9 are equal so multiple solutions could be possible if the battery is not discharging at its peak rate. This could be seen in Fig.~\ref{distinctsol} where two distinct solutions are plotted (with infinite other combinations possible). 
	All such combinations provide the same level of arbitrage gains which for this example is 14.89 cents.
	Since hour 7 has the lowest price level in the interval 6 to 10 hour and the battery could be discharged completely in slightly less than 3 hours. Thus storage remains idle during 7th hour.

	%real = [ 1; 0.9; 1.5; 0.8; 0.6; 5; 4.9; 6; 5; 8];
	
	\begin{figure}[!htbp]
		\center
		\includegraphics[width=3.3in]{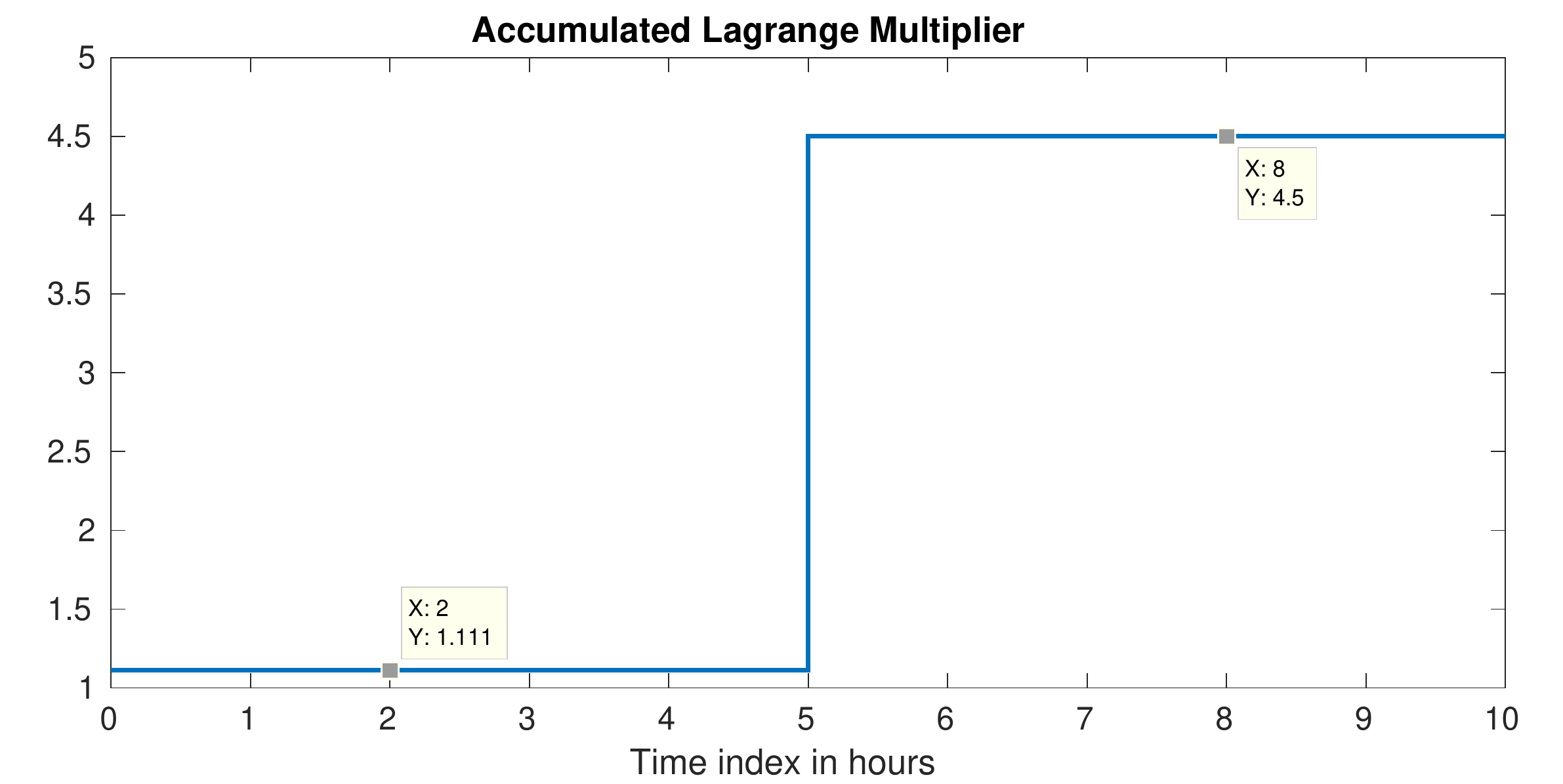}
		\caption{\small{Shadow price (accumulated Lagrange multiplier) for the electricity price signal shown in Fig.~\ref{interramp}.}}\label{muplot}
	\end{figure}
	
	Fig.~\ref{muplot} presents the shadow price (accumulated Lagrange multiplier) for the 2 sub-horizons for this example. First sub-horizon has $\mu_1=1.111$ is applicable for hour 1 to 5 and $\mu_2=4.5$ is applicable for hour 6 to 10.
	Since the intermediate ramp rate is observed at 1st hour with $p_{elec}(1)=1$ cents/kWh and the battery is charging therefore, $\mu_1=p_{elec}(1)/\eta_{ch} = 1.111$.
	Similarly, the intermediate ramp rate in the second sub-horizon is observed at 6th or 9th (as same price level) hour with $p_{elec}(6)=p_{elec}(9)=5$ cents/kWh and the battery is discharging therefore, $\mu_2=p_{elec}(6)\eta_{dis} =p_{elec}(9)\eta_{dis} = 4.5$.
	
	\begin{figure}[!htbp]
		\center
		\includegraphics[width=3.3in]{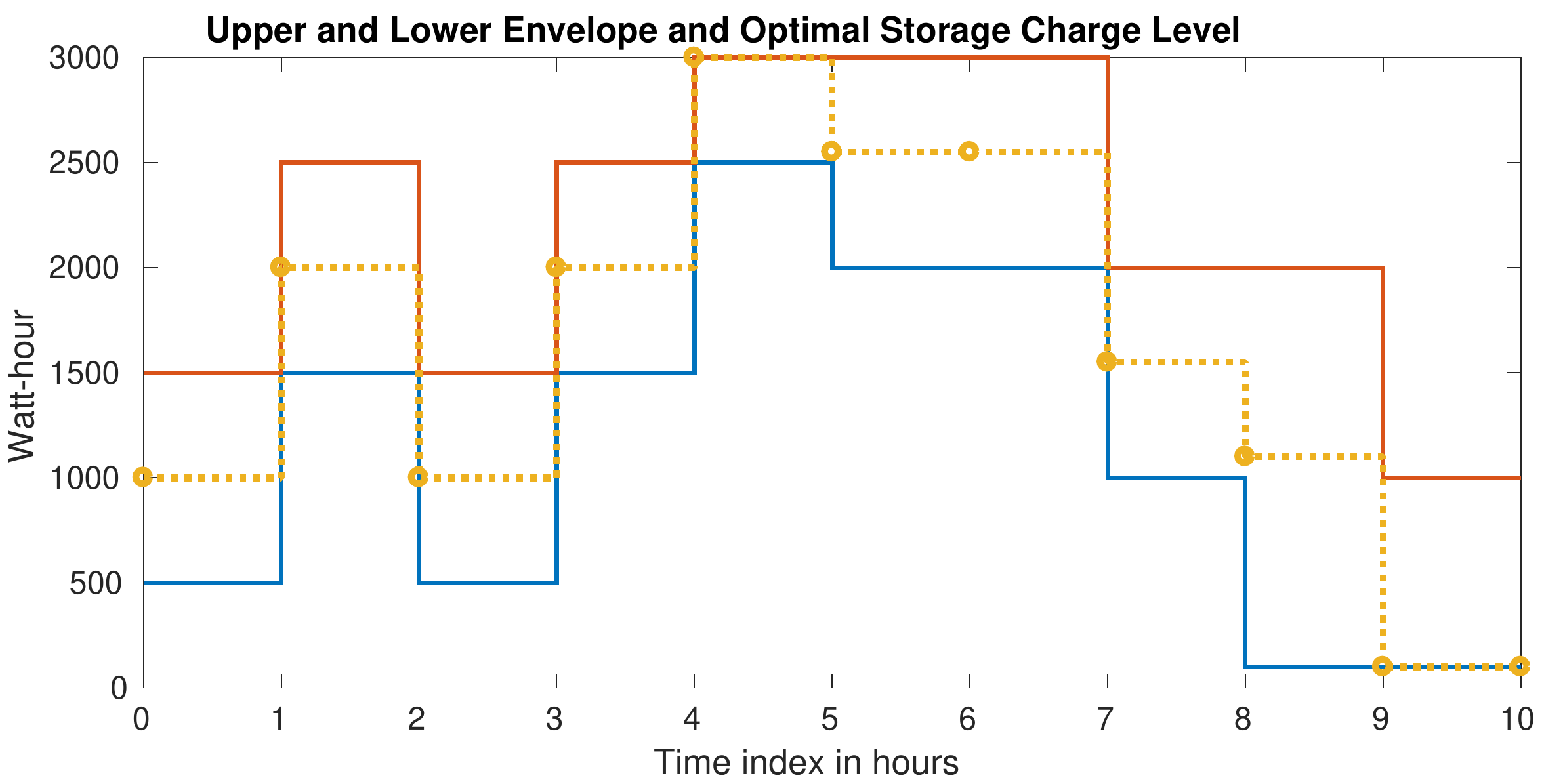}
		\caption{\small{Lower, upper and optimal charge level for the electricity price signal shown in Fig.~\ref{interramp}.}}\label{envelope}
	\end{figure}
	
	Fig.~\ref{envelope} shows the upper and lower envelopes of storage charge level along with the selected optimal storage charge level. The envelope of solutions is due to the piecewise linear cost structure of the cost function which provides sub-gradient like solution at point where the cost function changes its slope.

	\section{Case Study III: Comparing Run-Time of Algorithms}
	\label{appendixcomplexity}
	We compare the run-time of three optimal arbitrage algorithms for a given battery and present the run-times with different number of samples in the time horizon of optimization. The three algorithms compared here are:\\
	(a)\textit{ Proposed algorithm }in this work which shows the structure of optimal arbitrage solution based on price and net-load variation.\\
	(b) \textit{Linear Programming}: We use the LP formulation proposed in \cite{hashmi2019lp}. The LP formulation is possible due to piecewise linear convex cost functions. In this formulation we consider: (i) net-metering compensation (with selling price at best equal to buying price) i.e. $\kappa_i \in [0,1]$, (ii) inelastic load, (iii) consumer renewable generation, (iv) storage charging and discharging losses, (v) storage ramping constraint and (vi) storage capacity constraint. Using numerical results we perform sensitivity analysis of batteries for varying ramp rates and varying ratio of selling and buying price of electricity.\\
	(c) \textit{Convex optimization}: There could be several different ways of formulating optimal arbitrage problem using convex optimization toolbox. We propose one of the many ways of solving optimal arbitrage problem with convex piecewise linear cost function using CVX. Since in the optimization formulation we do not have any binary variable, this optimization problem could be solved using the default solver, SDPT3\footnote{{https://tinyurl.com/yfqclqz}}.
	{
		The decision variable $x_i$ is separated into two variables given as
		$
		x_i = x_i^{ch} - x_i^{ds},
		$
		where $x_i^{ch} \in [0, X_{\max}]$ and $x_i^{ds} \in [0, -X_{\min}]$, denote the charging and discharging values, respectively.\\
	}
	
	For the numerical evaluation we use a battery with initial charge level, $b_0$=500 Wh, $b_{\max}$=3000 Wh, $b_{\min}$=100 Wh, $\eta_{ch}$=$\eta_{dis}$=0.9 and sampling time is equal to 1 hour.
	
	%
	%\begin{figure}
	%	\center
	%	\includegraphics[width=3.2in]{complexitycomparison.pdf}
	%	\caption{\small{Comparison of run-time for CVX based optimization, linear programming and the proposed algorithm with variation of samples in time horizon of optimization for storage performing arbitrage. The run-time without load in blue and with load in red is shown.
	%			}}\label{complexitycomparison}
	%\end{figure}

\begin{figure}[!htbp]
		\center
		\includegraphics[width=3.5in]{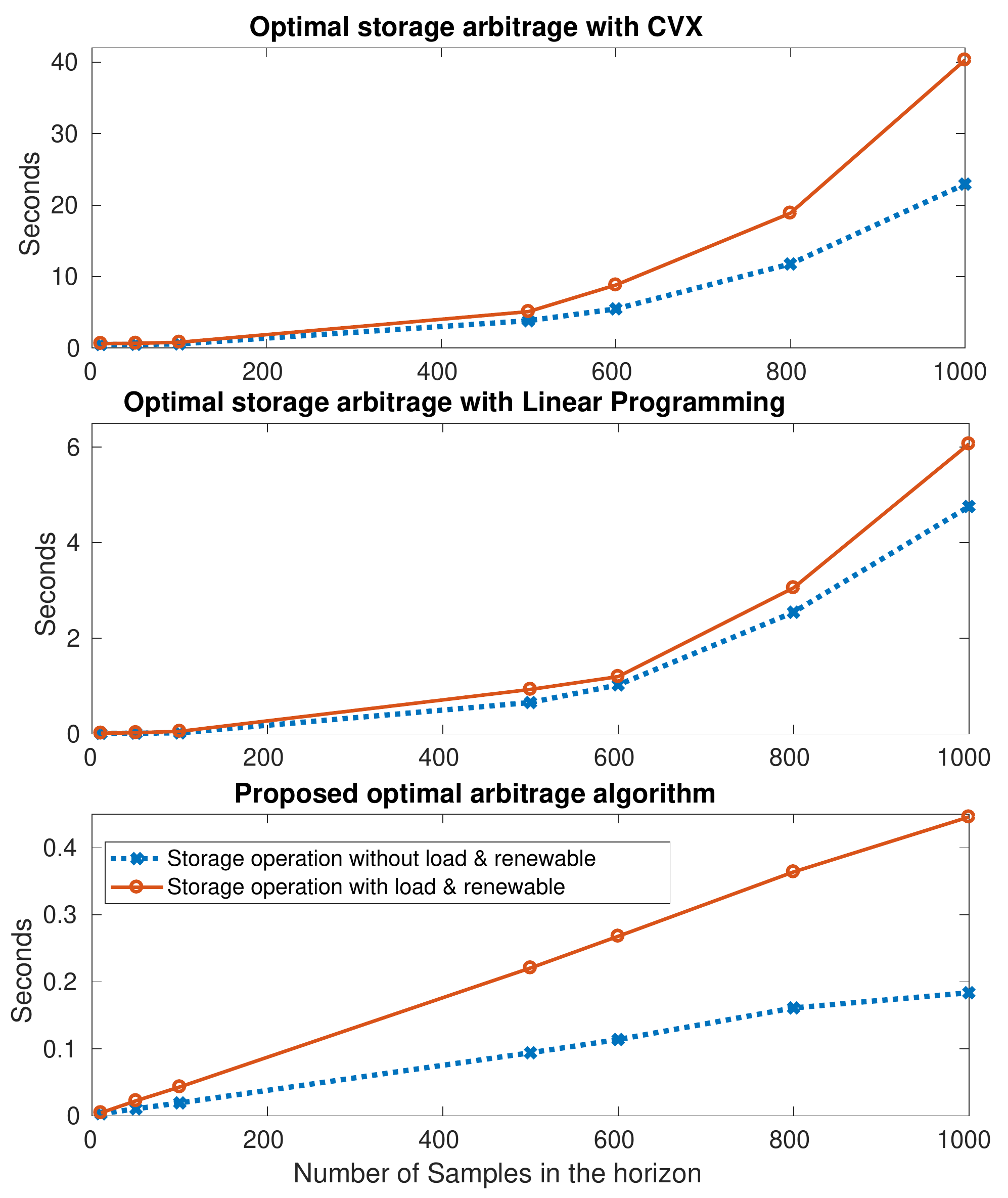}
		\caption{\small{Comparison of run-time for CVX based optimization, linear programming and the proposed algorithm with variation of samples in time horizon of optimization for storage performing arbitrage. The run-time without load in blue and with load in red is shown. 
				%			\textcolor{blue}{UMAR: be careful to the title. Please use "comparison" instead of "comparision"}
			}}\label{complexitycomparison}
		\end{figure}

	{
		Fig.~\ref{complexitycomparison} shows the run-time in seconds for the three described approaches for performing optimal arbitrage decisions. It can be observed that the proposed algorithm greatly outperforms the other two  approaches in terms of computation time.}
	%The complexity of linear programming (LP) based algorithms is polynomial \cite{karmarkar1984new}. 
	With a significantly longer time horizon LP might not be tractable.
	The CVX based convex optimization problem also becomes intractable for optimization horizon greater than $10^{4}$ samples. 
	
\begin{figure}[!htbp]
			\center
			\includegraphics[width=3.5in]{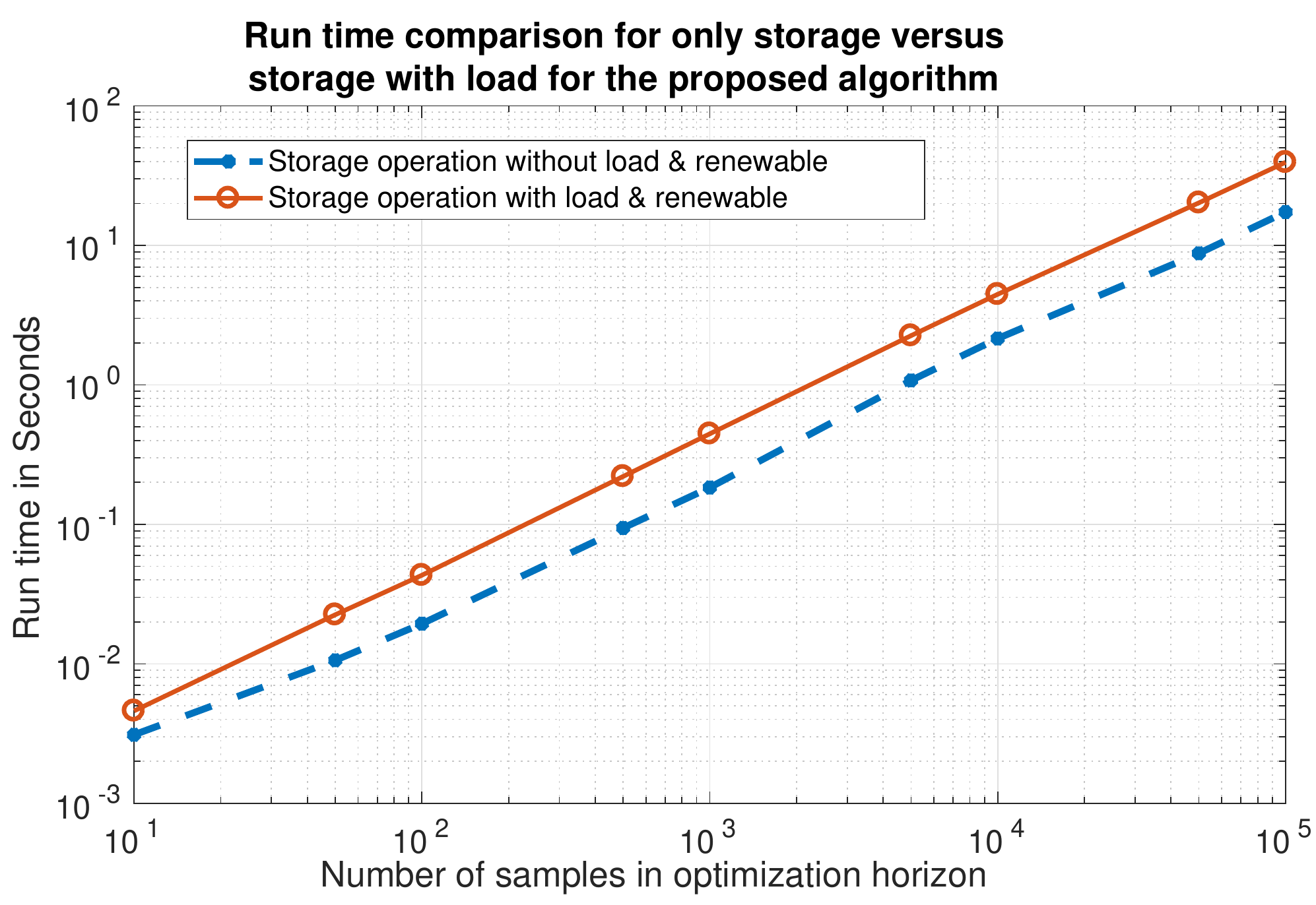}
			\caption{\small{Proposed algorithm run-time comparison for with load (in red) and without load (in blue). 
					%			\textcolor{blue}{UMAR: be careful to the title. Please use "comparison" instead of "comparision"}
				}}\label{complexitylog}
			\end{figure}

	{
		Fig.~\ref{complexitylog} shows the run-time in seconds for our optimal arbitrage algorithm to obtain the optimal solution for the two cases without and with load. As it can be seen in this figure, the complexity of the proposed algorithm grows approximately linearly with the number of samples.}
	The quantification of mean and standard deviation of sub-horizon is shown in Table~\ref{subhorizonproperties}.
	
	\begin{table}[!htbp]
		\caption {\small{Sub-horizon characteristics in number of samples}}
		\vspace{-10pt}
		\label{subhorizonproperties}
		\begin{center}
			\begin{tabular}{| c | c| c|}
				\hline
				Samples in & Mean no. of samples  &  STD of sub- \\ 
				time horizon & in sub-horizon &  horizon length \\ 
				\hline
				10 & 5 & 7.07 \\
				100 & 10 & 6.29 \\
				1000 & 10.99 & 7.19 \\
				10000 & 11.09 & 7.27 \\
				100000 & 11.11 & 7.28 \\
				\hline
			\end{tabular}
			\hfill\
		\end{center}
	\end{table}

	\section{Conclusion}
	We formulated the optimal energy arbitrage problem for storage operation such as batteries for an end user with inelastic load and renewable generation in presence of net-metering policy for compensating excess electricity generation.
	We proposed an efficient algorithm to find an optimal solution, using a method that transforms a continuous, convex optimization problem into a discrete one by exploiting the piecewise linear structure of the cost function.
	We introduced a method for determining the sub-horizons in the whole duration and showed that optimal storage control decisions do not depend on price of electricity and load variations beyond a sub-horizon.
	The proposed algorithm is compared with linear programming and
	convex optimization formulation to demonstrate the computational efficiency.  We show that the worst-case
	run-time complexity is quadratic in the time horizon.  Auto-regressive forecast models are implemented for
	rolling or receding horizon model predictive control to take into account real-world uncertainties
	in electricity price, consumer load and renewable generation.
	
	Theoretical and numerical case studies help us identify the governing parameters of following:
	(i) \textit{Sub-horizon}: is governed by (a) electricity price, (b) charging and discharging efficiency, (c) sampling
	time, (d) ratio of ramp rate and the rated storage capacity and (e) initial storage capacity,
	(ii) \textit{Shadow price}:  is governed by (a) electricity price and (b) charging and discharging efficiency,
	(iii) \textit{Thresholds of storage operation}:  are governed by (a) consumer inelastic load, (b) renewable generation, (c) electricity price, (d) storage ramp rate, (e) charging and discharging efficiency and (f) ratio
	of ramp rate and the rated capacity,
	(iv) \textit{Effect  of  uncertainty on arbitrage gains}:  governed by (a) relationship between sampling time and the ratio of ramp rate and the rated storage capacity, (b) charging and discharging efficiency.
	The dependencies require further exploration to quantify their relationship with each other.  
	
%	Furthermore, to find an optimal solution among infinite possibilities, we described a novel methodology using backward step algorithm, which is done only once in a sub-horizon. The worst case run-time complexity of the proposed algorithm is found to be quadratic in the number of time instants for which price signals are available.
%	\textcolor{blue}{We also presented an online implementation of the proposed algorithm for MPC based control for mitigating the effects of the forecast error on energy arbitrage gains XXX significantly. XXX.}%PLEASE IMPROVE THE WRITING OF THIS FINAL SENTENCE
	
	\vspace{-10pt}
	
	\bibliographystyle{IEEEtran}
	\bibliography{bibfile}
	
%	\pagebreak
	
	\appendices
	\section{Structure of cost function}
%	\section{Proof of Theorem~\ref{thm:convexity}}
	\label{prf:convexity}
	
	\begin{proof}
		Let $\psi(t) =a[t]^+ - b[t]^-$ with $a\geq b \geq 0$. Using $t=[t]^+- [t]^-$ we have $\psi(t) = (a-b)[t]^+ + bt$. 
		Since both $[t]^+$ and $t$ are convex in $t$ and $a-b, b\geq 0$ we have that $\psi$ is convex since it is the positive sum of two convex functions. 
		
		Now let $f(x)= \frac{1}{\eta_{ch}} [x]^+ - \eta_{dis}[x]^-$ (as defined in \eqref{finverse}) and $h_i(s) =[z_i+s]^+p_b(i) -[z_i+s]^-p_s(i)$.
		Then by the above reasoning we have that for $p_b(i) \geq p_s(i) \geq 0$ and $\eta_{ch}, \eta_{dis} \in (0,1]$,
		$h_i$ is convex in $s$ and $f$ is convex in $x$. Also, note that $h_i$ is non-decreasing in $s$. Hence, for $\lambda \in [0,1]$ we have

		\begin{align}
		h_i\big(f(\lambda x +(1-\lambda)y)\big) &\leq h_i\big(\lambda f(x) + (1-\lambda)f(y)\big)\\
									&\leq \lambda h_i(f(x)) + (1-\lambda)h_i(f(y))
		\end{align}
		In the above, the first inequality follows from the convexity of $f$ and non-decreasing nature of $h_i$
		and the second inequality follows from convexity of $h_i$. Therefore, we have that $h_i\cdot f=h_i(f())$ is 
		a convex function in $x$. This shows that the objective function of (P) is convex in $x$ since $C_{nm}^i=h_i\cdot f$.
		Finally, since the constraints are linear in $x$, we have that problem (P) is convex.
	\end{proof}
	
%	\textcolor{blue}{I do modifications from here to the end of this appendix.\\ \\}
	The cost function of the optimization problem (P) is plotted for the sake of visual inspection of its convexity.
	The cost function is denoted as $C_{nm}(i)$ which equals $[z_i + s_i]^+ p_b^i - [z_i + s_i]^- p_s^i$. Reiterating our convention: consumed electricity is considered to be positive, thus for $x_i>0$ the battery is consuming or in other words charging. The net load without storage is positive means that load seen from the grid is charged for consumption. For plotting the cost function with respect to the optimization variable $x_i$ we consider the following two cases:
	
	\subsection{The net load is positive ($z_i > 0$)}
\begin{figure}[!htbp]
		\center
		\includegraphics[width=2.3in]{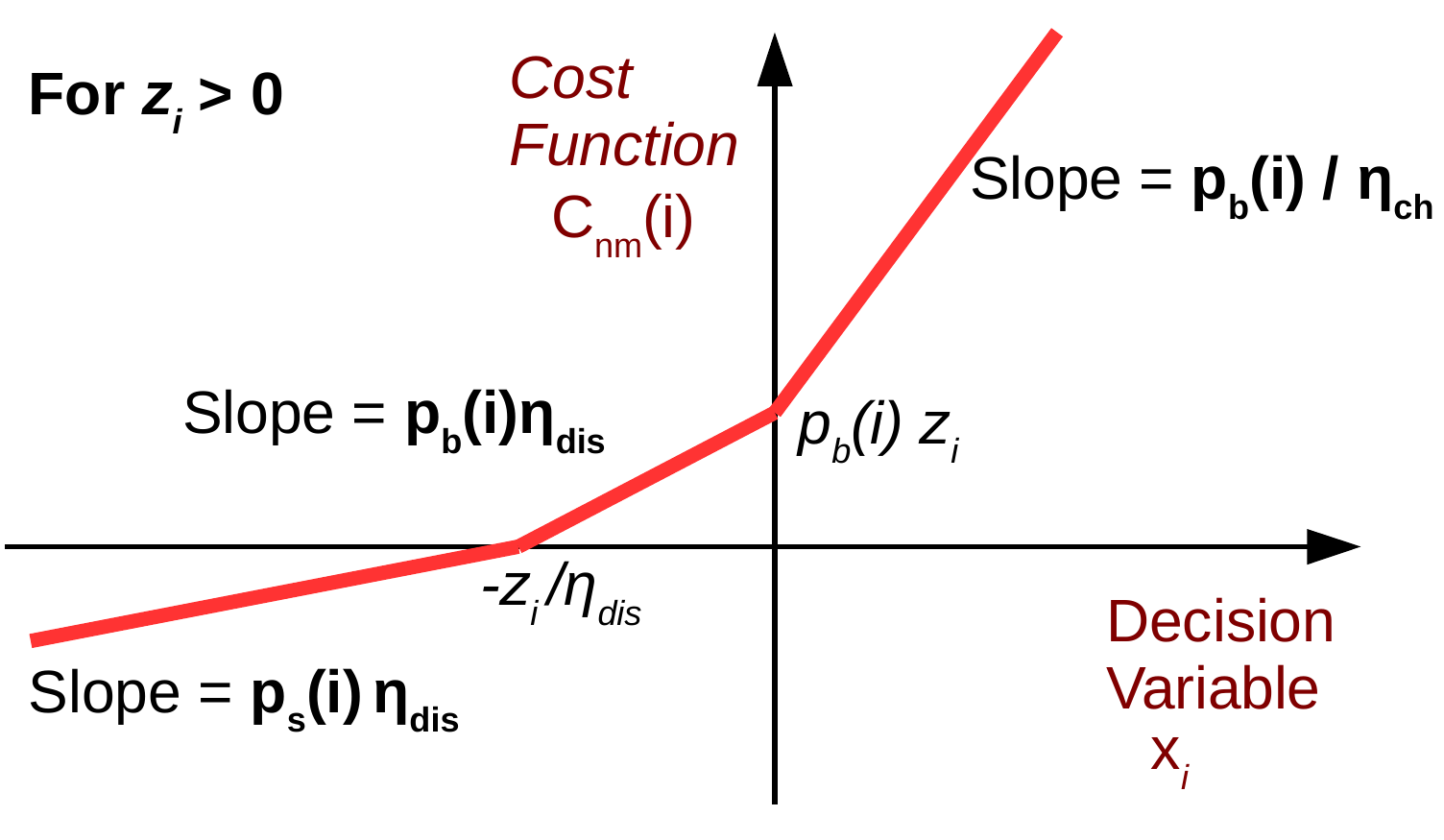}
		\caption{Cost function for $z_i > 0$}\label{cond1cvx}
	\end{figure}

	In this case we have the following cost function versus the storage operation. It is also shown in Fig.~\ref{cond1cvx}.
	\begin{enumerate}
		\item For charging $C_{nm}(i) = [z_i + x_i/\eta_{\text{ch}}]p_b^i$,
		\item For discharging:
		\begin{enumerate}
			\item If $-z_i < x_i \eta_{\text{dis}}$ then $C_{nm}(i) = [z_i + x_i\eta_{\text{dis}}]p_b^i$,
			\item Else $C_{nm}(i) = [z_i + x_i\eta_{\text{dis}}]p_s^i$.
		\end{enumerate}
	\end{enumerate}
	
%%%
%
%

	\subsection{The net load is negative ($z_i < 0$)}
\begin{figure}[!htbp]
		\center
		\includegraphics[width=2.3in]{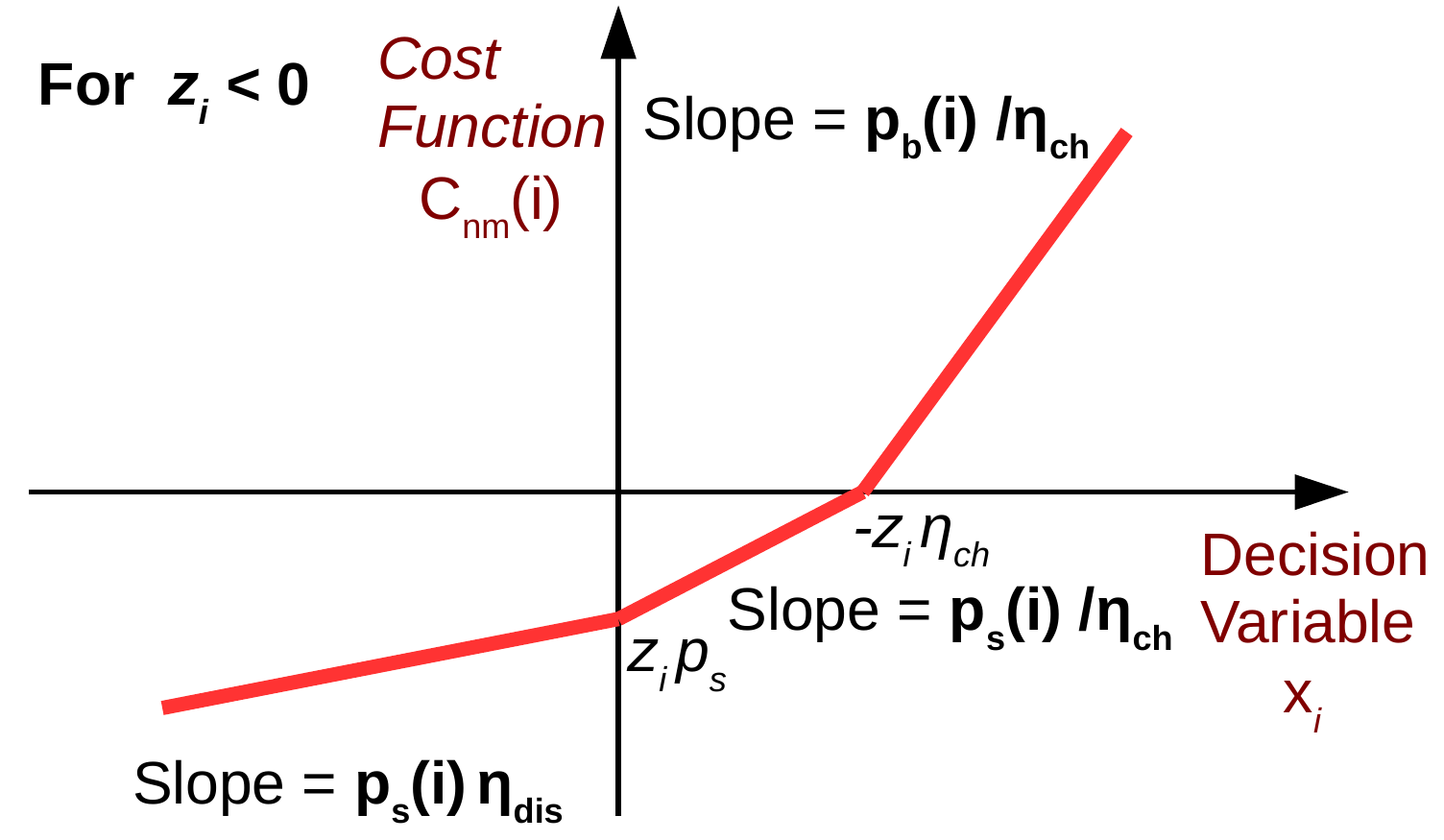}
		\caption{Cost function for $z_i< 0$}\label{cond2cvx}
	\end{figure}

	Here the cost function (versus the storage operation) is expressed as follows. It is illustrated in Fig.~\ref{cond2cvx}.
	
	\begin{enumerate}
		\item For charging we have the following conditions:
		\begin{enumerate}
			\item If $|z_i| < x_i/ \eta_{\text{ch}}$ then $C_{nm}(i) = [z_i + x_i/\eta_{\text{ch}}]p_b^i$,
			\item Else $C_{nm}(i) = [z_i + x_i/\eta_{\text{ch}}]p_s^i$.
		\end{enumerate}
		\item For discharging we have $C_{nm}(i) = [z_i + x_i\eta_{\text{dis}}]p_s^i$.
	\end{enumerate}

	It can be observed from Fig.~\ref{cond1cvx} and Fig.~\ref{cond2cvx} that when $\kappa_i$ exceeds 1, the plots will become concave.
	Therefore, in this work we assume that $\kappa_i \in [0,1]$ and hence this assumption ensures the convexity of the cost function.

	\section{Proof of Theorem~\ref{thm:arbitrage}}
	\label{prf:arbitrage}
	\begin{proof}
		{The optimization problem is convex with respect to the optimization variable $x^*$, using Theorem~\ref{thm:convexity}.} {From Eq.~\ref{finverse}, it is easy to see that there is a one-to-one mapping of $s_i$ and $x_i$.
		We first prove the existence of $(s^*, \alpha^*,\beta^*)$ such that:}
		\begin{enumerate}
			\item $s^*$ is the primal optimal solution,  
			\item $(\alpha^*,\beta^*)$ is the dual optimal solution, and 
			\item the optimality gap is zero (strong duality).
		\end{enumerate}
		Since the constraints of the primal problem are all linear, weak Slater's  constraint qualification conditions (which imply strong duality) 
		follow simply from the feasibility of the primal problem. Clearly, under the assumptions $b_{\min} \leq b_{\max}, \delta_{\min} \leq \delta_{\max}$, 
		$b_0 \in [b_{\min}, b_{\max}]$, $0 \in [\delta_{\min}, \delta_{\max}]$
		a feasible solution exists ($s_i=0$ for all $i=1,2,\ldots,N$ is feasible). Furthermore,
		since the primal objective function is continuous and the constraints define a 
		convex compact set, its minimum must be finite and achieved at the some $s^*$ in the feasibility region.
		According to the strong duality theorem, the above facts imply that the dual problem must be maximized
		at some $(\alpha^*,\beta^*)$ and the duality gap must be zero. 
		\par From the above reasoning it also follows that $(s^*,\alpha^*,\beta^*)$ must be
		the saddle point satisfying the KKT conditions. 
		Hence, using RHS inequality of the Saddle Point conditions, 
		\begin{gather*}
		\mathscr{L}({x^*, \alpha^*, \beta^*})  \leq \mathscr{L}({x, \alpha^*, \beta^*}) \\ \implies
		\sum_{i=1}^N \Big \{ C_{\text{nm}}^{(i)}(x_i^*) + \alpha_i^*(b_{\min} - b_i^*) + \beta_i^*(b_i^* - b_{max}) \Big \}  \\ \leq \sum_{i=1}^N \Big \{ C_{\text{nm}}^{(i)}(x_i)  + \alpha_i^*(b_{min} - b_i) + \beta_i^*(b_i - b_{max}) \Big \}
		%\\ \implies
		%  \sum_{i=1}^N \Big \{ p_{bat}(i) x_i^*  - b_i^*( \alpha_i^* - \beta_i^*) \Big \}   \\ \leq \sum_{i=1}^N \Big \{ p_{bat}(i) x_i  - b_i( \alpha_i^* - \beta_i^*) \Big \}\\ \implies
		%\sum_{i=1}^N \Big \{ p_{bat}(i) x_i^*  - (b_0 +\sum_{j=1}^i x_j ^*)( \alpha_i^* - \beta_i^*) \Big \}  \\ \leq \sum_{i=1}^N \Big \{ p_{elec}(i) x_i  - (b_0 + \sum_{j=1}^i x_j) ( \alpha_i^* - \beta_i^*) \Big \}\\ \implies
		% \sum_{i=1}^N \Big \{ p_{bat}(i) x_i^*  - \sum_{j=1}^i x_j ^*( \alpha_i^* - \beta_i^*) \Big \} \\ \leq \sum_{i=1}^N \Big \{ p_{bat}(i) x_i  - \sum_{j=1}^i x_j ( \alpha_i^* - \beta_i^*) \Big \}
		\end{gather*}
		Substituting $b_i=b_0+\sum_{j=1}^i x_j$ we get,
		\begin{equation}
		\sum_{i=1}^N \Big \{ C_{\text{nm}}^{(i)}(x_i^*)  - \mu_i^* x_i ^*\Big \}  \leq \sum_{i=1}^N \Big \{ C_{\text{nm}}(x_i)  - \mu_i^* x_i  \Big \}
		\end{equation}
		where
		%\begin{equation}
		$ \mu_i^* = \sum_{j=i}^N (\alpha_j^* - \beta_j^*)$.
		%\end{equation}
		$\mu_i^*$ is the accumulated Lagrange multiplier for time instant $i$ to $N$. 
		Hence, 
		%the effect of level of battery in $\mu$ selection
		%\begin{gather*}
		% \mu_k^* = (\alpha_k^* - \beta_k^*) + (\alpha_{k+1}^* - \beta_{k+1}^*) + ... + (\alpha_N^* - \beta_N^*) \\
		%\mu_{k+1}^* = \quad \quad \quad  \quad  (\alpha_{k+1}^* - \beta_{k+1}^*) + ... + (\alpha_N^* - \beta_N^*)
		%\end{gather*}
		\begin{equation}
		\mu_k^* - \mu_{k+1}^* = (\alpha_k^* - \beta_k^*) 
		\end{equation}
		The complementary slackness conditions for the Lagrangian are defined as
		\begin{gather*}
		\alpha_i(b_{\min} - b_i) = 0, \quad  \beta_i(b_i - b_{\max}) = 0, \forall i \text{ s.t. } \alpha_i, \beta_i \geq 0
		\end{gather*}
		{
		Equation (8) derived above and complementary slackness conditions imply the following relation between $\mu_k^*$ and $\mu_{k+1}^*$,
		}
		\begin{gather*}
		\mu_{k+1}^*
		\begin{cases}
		= \mu_{k}^*  ,& \text{if }   b_{min} <b_k^* < b_{max} \quad \text{ as } \quad \alpha_k^*=\beta_k^*=0\\
		\leq \mu_{k}^* ,& \text{if }  b_k^*  =  b_{min}  \quad \text{ as }  \quad \alpha_k^*\geq 0 \text{ and } \beta_k^*=0\\
		\geq \mu_{k}^* ,& \text{if }  b_k^* =  b_{max}   \quad \text{ as } \quad \alpha_k^*=0 \text{ and }  \beta_k^*\geq 0
		\end{cases}
		\end{gather*}
		The accumulated Lagrangian i.e. $\mu$ for the $N^{th}$ (last) instant is $\mu_N^*=\alpha_N^* - \beta_N^*$, therefore
		\begin{gather*}
		\mu_{N}^*=
		\begin{cases}
		= 0^*  ,& \text{if }   b_{min} <b_N^*< b_{max}\\
		\geq 0 ,& \text{if }  b_N^* =  b_{min} \\
		\leq 0 ,& \text{if }  b_N^* =  b_{max} 
		\end{cases}
		\end{gather*}
		
		%where, $b_{end}$ is the end battery level the user specifies, the default value of $b_{end}= b_{min}$ if not specified
		
		Such a ${x^*}$  solves the optimal arbitrage problem (P)  and ${\alpha^*, \beta^*}$ solves the dual problem. 
	\end{proof}

	\section{Proof of Theorem~\ref{thoremthresholds}}
	\label{thoremthresholdsproof}
		For a given $\mu_i^*=\mu$ the optimal decision $s_i^*(\mu)$ 
	is given by minimizing Eq.~\ref{minim}.
	\vspace{-8pt}
	\begin{equation}
	[z_i +s_i]^+p_b(i) - [z_i +s_i]^-p_s(i)  - \mu \Big(\eta_{\text{ch}}[s_i]^+ - \frac{1}{\eta_{\text{dis}}}[s_i]^-\Big)
	\label{minim}
	\end{equation}
	%We already know the following:
	%\begin{itemize}
	%	\item $p_b(i) > p_s(i) \geq 0 \quad \forall i \in \{1,...,N\}$ 
	%	\item $s_i \in [S_{\min}, S_{\max}]$ such that $S_{\min} \leq 0$ and $S_{\max} \geq 0$
	%\end{itemize}
	Hence, in order to minimize Eq.~\ref{minim} we consider the sign of $(z_i+s_i)$ and $s_i$. This will provide the following cases
	\begin{itemize}
		\item[] J1: $s_i (p_b(i) - \mu \eta_{\text{ch}})$ s.t. $z_i + s_i \geq 0$ and $s_i \in [0,S_{\max}]$,
		\item[] J2: $s_i (p_s(i) - \mu \eta_{\text{ch}})$ s.t. $z_i + s_i \leq 0$ and $s_i \in [0,S_{\max}]$,
		\item[] J3: $s_i \Big(p_b(i) - \frac{\mu }{\eta_{\text{dis}}}\Big)$ s.t. $z_i + s_i \geq 0$ and $s_i \in [S_{\min},0]$,
		\item[] J4: $s_i \Big(p_s(i) - \frac{\mu }{\eta_{\text{dis}}}\Big)$ s.t. $z_i + s_i \leq 0$ and $s_i \in [S_{\min},0]$.
	\end{itemize}
	The accumulated Lagrange multiplier, $\mu$, can be viewed as the shadow price of decision making. Based on conditions J1 to J4, the value of $\mu$ will divide the price levels into nine cases.
	Table~\ref{tab:title5} lists the constraints and minimizing conditions, we will use this table to find optimal value of $s_i^*$.
	\begin{table}[!htbp]
		\caption {Conditions to check} \label{tab:title5} \vspace{-7pt}
		\begin{center}
			\small
			\begin{tabular}{| c|c | c |c| c |}
				\hline
				Tag & $[z_i +s_i]$ &  $[s_i]$  & $\min$ Condition & Desired scenario\\ 
				\hline
				\hline
				J1& +ve &  +ve  & $s_i(p_b(i) - \mu \eta_{\text{ch}}) $  & $(p_b(i) - \mu \eta_{\text{ch}})\leq 0$  \\ 
				\hline
				J2&-ve & +ve & $s_i(p_s(i) - \mu \eta_{\text{ch}}) $ & $(p_s(i) - \mu \eta_{\text{ch}}) \leq 0$\\
				\hline
				J3&+ve & -ve &  $s_i(p_b(i) - \frac{\mu}{\eta_{\text{dis}}}) $ & $(p_b(i) - \frac{\mu}{\eta_{\text{dis}}}) \geq 0$\\
				\hline
				J4&-ve & -ve &  $s_i(p_s(i) - \frac{\mu}{\eta_{\text{dis}}}) $ &  $(p_s(i) - \frac{\mu}{\eta_{\text{dis}}}) \geq 0 $ \\
				\hline
			\end{tabular}
			\hfill\
		\end{center}
	\end{table}
	From Table~\ref{tab:title5} we can see the conditions of desired scenarios. Based on conditions J1 to J4, we can observe there will be four distinct levels in price signal which will subsequently divide the real line into nine possible levels for the accumulated Lagrange multiplier ($\mu$) as shown in Fig~\ref{case_fig} and Fig.~\ref{case_fig2}.
	
	\subsection{For $\kappa_i \in [0, \eta_{\text{ch}}\eta_{\text{dis}})$}
	\label{proofremark}

	\textbf{Region 1:} $\mu < \eta_{\text{dis}} p_s(i)$:  The minimizing conditions will be achieved by J3 and J4 as shown below:
	\begin{center}
		\small
		\begin{tabular}{| c|c | c |c | c |c|}
			\hline
			Tag & $[z_i +s_i]$ &  $[s_i]$  & $\min$ Condition & Sign & Comment\\ 
			\hline
			\hline
			J1& +ve &  +ve  & $s_i(p_b(i) - \mu \eta_{\text{ch}}) $  & + (+) & Undesired\\ 
			\hline
			J2&-ve & +ve & $s_i(p_s(i) - \mu \eta_{\text{ch}}) $ & + (+) & Undesired\\
			\hline
			J3&+ve & -ve &  $s_i(p_b(i) - \frac{\mu}{\eta_{\text{dis}}}) $ &  - (+) & Desired\\
			\hline
			J4&-ve & -ve &  $s_i(p_s(i) - \frac{\mu}{\eta_{\text{dis}}}) $ &   - (+) & Desired\\
			\hline
		\end{tabular}
		\hfill\
	\end{center}
	From J4 if $(z_i+s_i) <0$ then $s_i^* = [S_{\min}, S_{\min}]$ and from J3 if $(z_i+s_i) \geq0$ then $s_i^* = [S_{\min}, S_{\min}]$. Therefore, irrespective the sign of $z_i$ the optimal value is $[S_{\min}, S_{\min}]$.
	
	\textbf{Region 2:} $\mu = \eta_{\text{dis}} p_s(i)$:  The minimizing conditions will be achieved by J3 and J4 is a don't care condition with only constraint on $s_i$ being negative or zero.
	\begin{center}
		\small
		\begin{tabular}{| c|c | c |c | c |c|}
			\hline
			Tag & $[z_i +s_i]$ &  $[s_i]$  & $\min$ Condition & Sign & Comment\\ 
			\hline
			\hline
			J1& +ve &  +ve  & $s_i(p_b(i) - \mu \eta_{\text{ch}}) $  & + (+) & Undesired\\ 
			\hline
			J2&-ve & +ve & $s_i(p_s(i) - \mu \eta_{\text{ch}}) $ & + (+) & Undesired\\
			\hline
			J3&+ve & -ve &  $s_i(p_b(i) - \frac{\mu}{\eta_{\text{dis}}}) $ &  - (+) & Desired\\
			\hline
			J4&-ve & -ve &  $s_i(p_s(i) - \frac{\mu}{\eta_{\text{dis}}}) $ &   - (0) & Don't Care\\
			\hline
		\end{tabular}
		\hfill\
	\end{center}
	\textit{Sub-Case 1: } from J3 and J4 if $z_i\geq 0$ then $s_i^* = [S_{\min} , \max(-z_i, S_{\min})]$.
	
	\textit{Sub-Case 2: } from J4 if $z_i< 0$ then $s_i^* = [S_{\min}, 0]$.\\

	\textbf{Region 3:} $\mu \in (\eta_{\text{dis}} p_s(i), \frac{p_s(i)}{\eta_{\text{ch}}})$:  The minimizing conditions will be achieved by minimizing J3. All other conditions, i.e., J1, J2 and J4 are undesired.
	\begin{center}
		\small
		\begin{tabular}{| c|c | c |c | c |c|}
			\hline
			Tag & $[z_i +s_i]$ &  $[s_i]$  & $\min$ Condition & Sign & Comment\\ 
			\hline
			\hline
			J1& +ve &  +ve  & $s_i(p_b(i) - \mu \eta_{\text{ch}}) $  & + (+) & Undesired\\ 
			\hline
			J2&-ve & +ve & $s_i(p_s(i) - \mu \eta_{\text{ch}}) $ & + (+) & Undesired\\
			\hline
			J3&+ve & -ve &  $s_i(p_b(i) - \frac{\mu}{\eta_{\text{dis}}}) $ &  - (+) & Desired\\
			\hline
			J4&-ve & -ve &  $s_i(p_s(i) - \frac{\mu}{\eta_{\text{dis}}}) $ &   - (-) & Undesired\\
			\hline
		\end{tabular}
		\hfill\
	\end{center}
	
	\textit{Sub-Case 1: } from J3 if $z_i\geq 0 $ then $s_i^* = [\max\{-z_i,S_{\min}\}, \max\{-z_i,S_{\min}\}]$.
	
	\textit{Sub-Case 2: } from J2 and J4 $s_i^* = [0, 0]$.\\

	\textbf{Region 4:} $\mu = \frac{p_s(i)}{\eta_{\text{ch}}}$:  The minimizing conditions will be achieved by minimizing J3. J2 is a don't care condition.
	
	\begin{center}
		\small
		\begin{tabular}{| c|c | c |c | c |c|}
			\hline
			Tag & $[z_i +s_i]$ &  $[s_i]$  & $\min$ Condition & Sign & Comment\\ 
			\hline
			\hline
			J1& +ve &  +ve  & $s_i(p_b(i) - \mu \eta_{\text{ch}}) $  & + (+) & Undesired\\ 
			\hline
			J2&-ve & +ve & $s_i(p_s(i) - \mu \eta_{\text{ch}}) $ & + (0) & Don't Care\\
			\hline
			J3&+ve & -ve &  $s_i(p_b(i) - \frac{\mu}{\eta_{\text{dis}}}) $ &  - (+) & Desired\\
			\hline
			J4&-ve & -ve &  $s_i(p_s(i) - \frac{\mu}{\eta_{\text{dis}}}) $ &   - (-) & Undesired\\
			\hline
		\end{tabular}
		\hfill\
	\end{center}
	
	\textit{Sub-Case 1: } from J3 if $z_i\geq 0 $ then $s_i^* = [\max\{-z_i,S_{\min}\}, \max\{-z_i,S_{\min}\}]$.
	
	\textit{Sub-Case 2: } from J2, if $z_i < 0 $ then $s_i^* = [0, \min\{-z_i,S_{\max}\}]$.\\

	\textbf{Region 5:} $\mu \in (\frac{p_s(i)}{\eta_{\text{ch}}}, \eta_{\text{dis}} p_b(i))$:  The minimizing conditions will be achieved by minimizing J2 and J3.
	\begin{center}
		\small
		\begin{tabular}{| c|c | c |c | c |c|}
			\hline
			Tag & $[z_i +s_i]$ &  $[s_i]$  & $\min$ Condition & Sign & Comment\\ 
			\hline
			\hline
			J1& +ve &  +ve  & $s_i(p_b(i) - \mu \eta_{\text{ch}}) $  & + (+) & Undesired\\ 
			\hline
			J2&-ve & +ve & $s_i(p_s(i) - \mu \eta_{\text{ch}}) $ & + (-) & Desired\\
			\hline
			J3&+ve & -ve &  $s_i(p_b(i) - \frac{\mu}{\eta_{\text{dis}}}) $ &  - (+) & Desired\\
			\hline
			J4&-ve & -ve &  $s_i(p_s(i) - \frac{\mu}{\eta_{\text{dis}}}) $ &   - (-) & Undesired\\
			\hline
		\end{tabular}
		\hfill\
	\end{center}
	
	\textit{Sub-Case 1: } from J3 if $z_i\geq 0 $ then $s_i^* = [\max\{-z_i,S_{\min}\}, \max\{-z_i,S_{\min}\}]$.
	
	\textit{Sub-Case 2: } from J2, if $z_i< 0$ then $s_i^* = [\min\{-z_i, S_{\max}\}, \min\{-z_i, S_{\max}\}]$.\\

	\textbf{Region 6:} $\mu = \eta_{\text{dis}} p_b(i)$:  The minimizing conditions will be achieved by J2 and J3 is a don't care condition with only constraint on $s_i$ being negative or zero.
	\begin{center}
		\small
		\begin{tabular}{| c|c | c |c | c |c|}
			\hline
			Tag & $[z_i +s_i]$ &  $[s_i]$  & $\min$ Condition & Sign & Comment\\ 
			\hline
			\hline
			J1& +ve &  +ve  & $s_i(p_b(i) - \mu \eta_{\text{ch}}) $  & + (+) & Undesired\\ 
			\hline
			J2&-ve & +ve & $s_i(p_s(i) - \mu \eta_{\text{ch}}) $ & + (-) & Desired\\
			\hline
			J3&+ve & -ve &  $s_i(p_b(i) - \frac{\mu}{\eta_{\text{dis}}}) $ &  - (0) & Don't Care\\
			\hline
			J4&-ve & -ve &  $s_i(p_s(i) - \frac{\mu}{\eta_{\text{dis}}}) $ &   - (-) & Undesired\\
			\hline
		\end{tabular}
		\hfill\
	\end{center}
	
	\textit{Sub-Case 1: } from J3 if $z_i\geq 0$ then $s_i^* = [\max\{-z_i,S_{\min}\}, 0]$.
	
	\textit{Sub-Case 2: } from J2 if $z_i< 0$ then $s_i^* =  [\min\{-z_i, S_{\max}\}, \min\{-z_i, S_{\max}\}]$.\\

	\textbf{Region 7:} $\mu \in (\eta_{\text{dis}} p_b(i), \frac{p_b(i)}{\eta_{\text{ch}}})$:  The minimizing conditions will be achieved by J2. All other cases will be undesirable.
	
	\begin{center}
		\small
		\begin{tabular}{| c|c | c |c | c |c|}
			\hline
			Tag & $[z_i +s_i]$ &  $[s_i]$  & $\min$ Condition & Sign & Comment\\ 
			\hline
			\hline
			J1& +ve &  +ve  & $s_i(p_b(i) - \mu \eta_{\text{ch}}) $  & + (+) & Undesired\\ 
			\hline
			J2&-ve & +ve & $s_i(p_s(i) - \mu \eta_{\text{ch}}) $ & + (-) & Desired\\
			\hline
			J3&+ve & -ve &  $s_i(p_b(i) - \frac{\mu}{\eta_{\text{dis}}}) $ &  - (-) & Undesired\\
			\hline
			J4&-ve & -ve &  $s_i(p_s(i) - \frac{\mu}{\eta_{\text{dis}}}) $ &   - (-) & Undesired\\
			\hline
		\end{tabular}
		\hfill\
	\end{center}
	
	\textit{Sub-Case 1: } from J2 if $z_i< 0$ then $s_i^* =  [\min\{-z_i, S_{\max}\}, \min\{-z_i, S_{\max}\}]$.
	
	\textit{Sub-Case 2: } if $z_i \geq 0$ then do nothing, $s_i^* =  [0,0]$. This is because J1 and J3 covers two direction of movement i.e. charging and discharging, both of which will increase the objective function\\

	\textbf{Region 8:} $\mu = \frac{p_b(i)}{\eta_{\text{ch}}}$:  The minimizing conditions will be achieved by J2 and J1 is a don't care condition.
	\begin{center}
		\small
		\begin{tabular}{| c|c | c |c | c |c|}
			\hline
			Tag & $[z_i +s_i]$ &  $[s_i]$  & $\min$ Condition & Sign & Comment\\ 
			\hline
			\hline
			J1& +ve &  +ve  & $s_i(p_b(i) - \mu \eta_{\text{ch}}) $  & + (0) & Don't Care\\ 
			\hline
			J2&-ve & +ve & $s_i(p_s(i) - \mu \eta_{\text{ch}}) $ & + (-) & Desired\\
			\hline
			J3&+ve & -ve &  $s_i(p_b(i) - \frac{\mu}{\eta_{\text{dis}}}) $ &  - (-) & Undesired\\
			\hline
			J4&-ve & -ve &  $s_i(p_s(i) - \frac{\mu}{\eta_{\text{dis}}}) $ &   - (-) & Undesired\\
			\hline
		\end{tabular}
		\hfill\
	\end{center}
	
	\textit{Sub-Case 1: } from J2 and J1 if $z_i< 0$ then $s_i^* =  [\min\{-z_i, S_{\max}\},  S_{\max}]$.
	
	\textit{Sub-Case 2: } from J1 is $z_i \geq 0$ then $s_i^* =  [0, S_{\max}]$.\\

	\textbf{Region 9:} $\mu > \frac{p_b(i)}{\eta_{\text{ch}}}$:  The minimizing conditions will be achieved by J2 and J1.
	
	\begin{center}
		\small
		\begin{tabular}{| c|c | c |c | c |c|}
			\hline
			Tag & $[z_i +s_i]$ &  $[s_i]$  & $\min$ Condition  & Sign & Comment\\ 
			\hline
			\hline
			J1& +ve &  +ve  & $s_i(p_b(i) - \mu \eta_{\text{ch}}) $  & + (-) & Desired\\ 
			\hline
			J2&-ve & +ve & $s_i(p_s(i) - \mu \eta_{\text{ch}}) $ & + (-) & Desired\\
			\hline
			J3&+ve & -ve &  $s_i(p_b(i) - \frac{\mu}{\eta_{\text{dis}}}) $ &  - (-) & Undesired\\
			\hline
			J4&-ve & -ve &  $s_i(p_s(i) - \frac{\mu}{\eta_{\text{dis}}}) $ &   - (-) & Undesired\\
			\hline
		\end{tabular}
		\hfill\
	\end{center}
	
	Irrespective of sign of $z_i$, $s_i^* =  [S_{\max}, S_{\max}]$.

	\subsection{Proof of Theorem~\ref{thoremthresholds} for $\kappa_i \in [\eta_{\text{ch}}\eta_{\text{dis}}, 1)$}
	\label{proofremark1}
	\vspace{-10pt}
	\begin{figure}[!htbp]
		\center
		\includegraphics[width=3in]{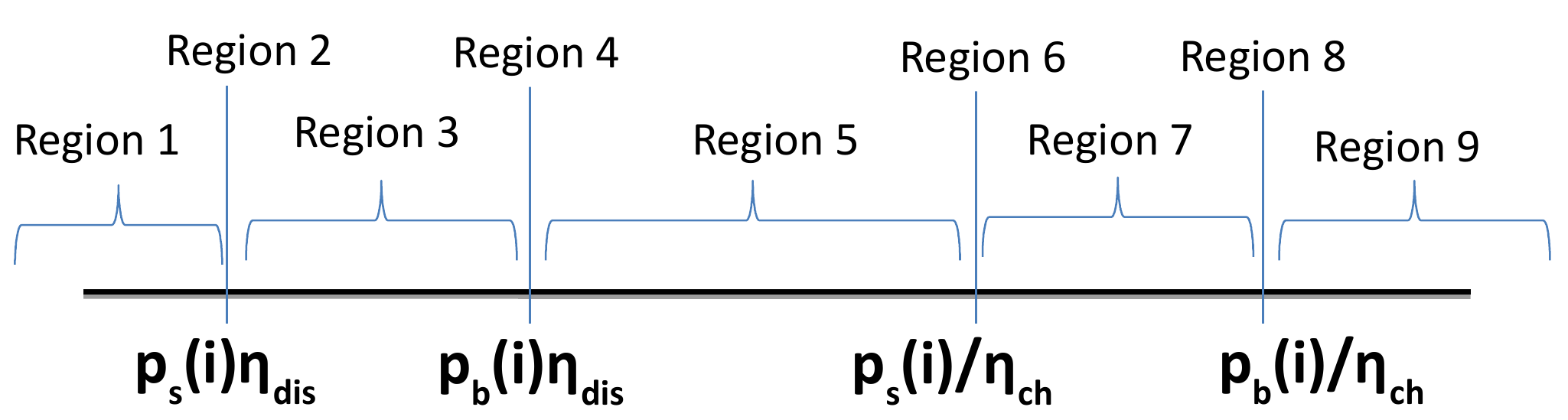} \vspace{-5pt}
		\caption{Regions based on levels of $\mu$ for $\zeta_i \geq 1$}\label{case_fig2}
	\end{figure}
	Based on four levels of prices shown in Fig~\ref{case_fig2}, the range is divided into into nine possible bands for the accumulated Lagrange multiplier.

	\textbf{Region 1:} $\mu < \eta_{\text{dis}} p_s(i)$:  The minimizing conditions will be achieved by J3 and J4 as shown below
	
	\begin{center}
		\small
		\begin{tabular}{| c|c | c |c | c |c|}
			\hline
			Tag & $[z_i +s_i]$ &  $[s_i]$  & min Condition & Sign & Comment\\ 
			\hline
			\hline
			J1& +ve &  +ve  & $s_i(p_b(i) - \mu \eta_{\text{ch}}) $  & + (+) & Undesired\\ 
			\hline
			J2&-ve & +ve & $s_i(p_s(i) - \mu \eta_{\text{ch}}) $ & + (+) & Undesired\\
			\hline
			J3&+ve & -ve &  $s_i(p_b(i) - \frac{\mu}{\eta_{\text{dis}}}) $ &  - (+) & Desired\\
			\hline
			J4&-ve & -ve &  $s_i(p_s(i) - \frac{\mu}{\eta_{\text{dis}}}) $ &   - (+) & Desired\\
			\hline
		\end{tabular}
		\hfill\
	\end{center}
	From J4 if $(z_i+s_i) <0$ then $s_i^* = [S_{\min}, S_{\min}]$ and from J3 if $(z_i+s_i) \geq0$ then $s_i^* = [S_{\min}, S_{\min}]$. Therefore, irrespective the sign of $z_i$ the optimal value is $[S_{\min}, S_{\min}]$.\\

	\textbf{Region 2:} $\mu = \eta_{\text{dis}} p_s(i)$:  The minimizing conditions will be achieved by J3 and J4 is a don't care condition with only constraint on $s_i$ being negative or zero.
	
	\begin{center}
		\small
		\begin{tabular}{| c|c | c |c | c |c|}
			\hline
			Tag & $[z_i +s_i]$ &  $[s_i]$  & min Condition & Sign & Comment\\ 
			\hline
			\hline
			J1& +ve &  +ve  & $s_i(p_b(i) - \mu \eta_{\text{ch}}) $  & + (+) & Undesired\\ 
			\hline
			J2&-ve & +ve & $s_i(p_s(i) - \mu \eta_{\text{ch}}) $ & + (+) & Undesired\\
			\hline
			J3&+ve & -ve &  $s_i(p_b(i) - \frac{\mu}{\eta_{\text{dis}}}) $ &  - (+) & Desired\\
			\hline
			J4&-ve & -ve &  $s_i(p_s(i) - \frac{\mu}{\eta_{\text{dis}}}) $ &   - (0) & Don't Care\\
			\hline
		\end{tabular}
		\hfill\
	\end{center}
	
	\textit{Sub-Case 1: } from J3 and J4 if $z_i\geq 0$ then $s_i^* = [S_{\min}, \max\{-z_i,S_{\min}\}]$.
	
	\textit{Sub-Case 2: } from J4 if $z_i< 0$ then $s_i^* = [S_{\min}, 0]$.\\

	\textbf{Region 3:} $\mu \in (\eta_{\text{dis}} p_s(i), \eta_{\text{dis}} p_b(i) )$:  The minimizing conditions will be achieved by minimizing J3. All other conditions, i.e., J1, J2 and J4 are undesired.
	
	\begin{center}
		\small
		\begin{tabular}{| c|c | c |c | c |c|}
			\hline
			Tag & $[z_i +s_i]$ &  $[s_i]$  & min Condition & Sign & Comment\\ 
			\hline
			\hline
			J1& +ve &  +ve  & $s_i(p_b(i) - \mu \eta_{\text{ch}}) $  & + (+) & Undesired\\ 
			\hline
			J2&-ve & +ve & $s_i(p_s(i) - \mu \eta_{\text{ch}}) $ & + (+) & Undesired\\
			\hline
			J3&+ve & -ve &  $s_i(p_b(i) - \frac{\mu}{\eta_{\text{dis}}}) $ &  - (+) & Desired\\
			\hline
			J4&-ve & -ve &  $s_i(p_s(i) - \frac{\mu}{\eta_{\text{dis}}}) $ &   - (-) & Unesired\\
			\hline
		\end{tabular}
		\hfill\
	\end{center}
	
	\textit{Sub-Case 1: } from J3 if $z_i\geq 0 $ then $s_i^* = [\max\{-z_i,S_{\min}\}, \max\{-z_i,S_{\min}\}]$.
	
	\textit{Sub-Case 2: } from J2 and J4 $s_i^* = [0, 0]$.\\

	\textbf{Region 4:} $\mu = \eta_{\text{dis}} p_b(i)$:  
	\begin{center}
		\small
		\begin{tabular}{| c|c | c |c | c |c|}
			\hline
			Tag & $[z_i +s_i]$ &  $[s_i]$  & min Condition & Sign & Comment\\ 
			\hline
			\hline
			J1& +ve &  +ve  & $s_i(p_b(i) - \mu \eta_{\text{ch}}) $  & + (+) & Undesired\\ 
			\hline
			J2&-ve & +ve & $s_i(p_s(i) - \mu \eta_{\text{ch}}) $ & + (+) & Undesired\\
			\hline
			J3&+ve & -ve &  $s_i(p_b(i) - \frac{\mu}{\eta_{\text{dis}}}) $ &  - (0) & Don't Care\\
			\hline
			J4&-ve & -ve &  $s_i(p_s(i) - \frac{\mu}{\eta_{\text{dis}}}) $ &   - (-) & Undesired\\
			\hline
		\end{tabular}
		\hfill\
	\end{center}
	
	\textit{Sub-Case 1: } from J3 if $z_i\geq 0 $ then $s_i^* = [\max\{-z_i,S_{\min}\}, 0]$.
	
	\textit{Sub-Case 2: } from J2 and J4, if $z_i < 0 $ then $s_i^* = [0, 0]$.\\

	\textbf{Region 5:} $\mu \in (\eta_{\text{dis}} p_b(i), \frac{p_s(i)}{\eta_{\text{ch}}})$:  
	
	\begin{center}
		\small
		\begin{tabular}{| c|c | c |c | c |c|}
			\hline
			Tag & $[z_i +s_i]$ &  $[s_i]$  & min Condition & Sign & Comment\\ 
			\hline
			\hline
			J1& +ve &  +ve  & $s_i(p_b(i) - \mu \eta_{\text{ch}}) $  & + (+) & Undesired\\ 
			\hline
			J2&-ve & +ve & $s_i(p_s(i) - \mu \eta_{\text{ch}}) $ & + (+) & Undesired\\
			\hline
			J3&+ve & -ve &  $s_i(p_b(i) - \frac{\mu}{\eta_{\text{dis}}}) $ &  - (-) & Undesired\\
			\hline
			J4&-ve & -ve &  $s_i(p_s(i) - \frac{\mu}{\eta_{\text{dis}}}) $ &   - (-) & Undesired\\
			\hline
		\end{tabular}
		\hfill\
	\end{center}
	
	\textit{Sub-Case 1: } from J1 and J3 if $z_i\geq 0 $ then $s_i^* = [0,0]$.
	
	\textit{Sub-Case 2: } from J2 and J4, if $z_i< 0$ then $s_i^* = [0,0]$.\\

	\textbf{Region 6:} $\mu =  \frac{p_s(i)}{\eta_{\text{ch}}}$:  
	
	\begin{center}
		\small
		\begin{tabular}{| c|c | c |c | c |c|}
			\hline
			Tag & $[z_i +s_i]$ &  $[s_i]$  & min Condition & Sign & Comment\\ 
			\hline
			\hline
			J1& +ve &  +ve  & $s_i(p_b(i) - \mu \eta_{\text{ch}}) $  & + (+) & Undesired\\ 
			\hline
			J2&-ve & +ve & $s_i(p_s(i) - \mu \eta_{\text{ch}}) $ & + (0) & Don't Care\\
			\hline
			J3&+ve & -ve &  $s_i(p_b(i) - \frac{\mu}{\eta_{\text{dis}}}) $ &  - (-) & Undesired\\
			\hline
			J4&-ve & -ve &  $s_i(p_s(i) - \frac{\mu}{\eta_{\text{dis}}}) $ &   - (-) & Undesired\\
			\hline
		\end{tabular}
		\hfill\
	\end{center}
	
	\textit{Sub-Case 1: } from J1 and J3 if $z_i\geq 0$ then $s_i^* = [0, 0]$.
	
	\textit{Sub-Case 2: } from J2 if $z_i< 0$ then $s_i^* =  [0, \min\{-z_i, S_{\max}\}]$.\\

	\textbf{Region 7:} $\mu \in (\frac{p_s(i)}{\eta_{\text{ch}}}, \frac{p_b(i)}{\eta_{\text{ch}}})$:  The minimizing conditions will be achieved by J2. All other cases will be undesirable.
	
	\begin{center}
		\small
		\begin{tabular}{| c|c | c |c | c |c|}
			\hline
			Tag & $[z_i +s_i]$ &  $[s_i]$  & min Condition & Sign & Comment\\ 
			\hline
			\hline
			J1& +ve &  +ve  & $s_i(p_b(i) - \mu \eta_{\text{ch}}) $  & + (+) & Undesired\\ 
			\hline
			J2&-ve & +ve & $s_i(p_s(i) - \mu \eta_{\text{ch}}) $ & + (-) & Desired\\
			\hline
			J3&+ve & -ve &  $s_i(p_b(i) - \frac{\mu}{\eta_{\text{dis}}}) $ &  - (-) & Undesired\\
			\hline
			J4&-ve & -ve &  $s_i(p_s(i) - \frac{\mu}{\eta_{\text{dis}}}) $ &   - (-) & Undesired\\
			\hline
		\end{tabular}
		\hfill\
	\end{center}
	
	\textit{Sub-Case 1: } from J2 if $z_i< 0$ then $s_i^* =  [\min\{-z_i, S_{\max}\}, \min\{-z_i, S_{\max}\}]$.
	
	\textit{Sub-Case 2: } if $z_i \geq 0$ then do nothing, $s_i^* =  [0,0]$. This is because J1 and J3 covers two direction of movement i.e. charging and discharging, both of which will increase the objective function\\

	\textbf{Region 8:} $\mu = \frac{p_b(i)}{\eta_{\text{ch}}}$:  The minimizing conditions will be achieved by J2 and J1 is a don't care condition.
	
	\begin{center}
		\small
		\begin{tabular}{| c|c | c |c | c |c|}
			\hline
			Tag & $[z_i +s_i]$ &  $[s_i]$  & min Condition & Sign & Comment\\ 
			\hline
			\hline
			J1& +ve &  +ve  & $s_i(p_b(i) - \mu \eta_{\text{ch}}) $  & + (0) & Don't Care\\ 
			\hline
			J2&-ve & +ve & $s_i(p_s(i) - \mu \eta_{\text{ch}}) $ & + (-) & Desired\\
			\hline
			J3&+ve & -ve &  $s_i(p_b(i) - \frac{\mu}{\eta_{\text{dis}}}) $ &  - (-) & Undesired\\
			\hline
			J4&-ve & -ve &  $s_i(p_s(i) - \frac{\mu}{\eta_{\text{dis}}}) $ &   - (-) & Undesired\\
			\hline
		\end{tabular}
		\hfill\
	\end{center}
	
	\textit{Sub-Case 1: } from J2 and J1 if $z_i< 0$ then $s_i^* =  [\min\{-z_i, S_{\max}\}, S_{\max}]$.
	
	\textit{Sub-Case 2: } from J1 is $z_i \geq 0$ then $s_i^* =  [0, S_{\max}]$.\\

	\textbf{Region 9:} $\mu > \frac{p_b(i)}{\eta_{\text{ch}}}$:  The minimizing conditions will be achieved by J2 and J1.
	
	\begin{center}
		\small
		\begin{tabular}{| c|c | c |c | c |c|}
			\hline
			Tag & $[z_i +s_i]$ &  $[s_i]$  & min Condition & Sign & Comment\\ 
			\hline
			\hline
			J1& +ve &  +ve  & $s_i(p_b(i) - \mu \eta_{\text{ch}}) $  & + (-) & Desired\\ 
			\hline
			J2&-ve & +ve & $s_i(p_s(i) - \mu \eta_{\text{ch}}) $ & + (-) & Desired\\
			\hline
			J3&+ve & -ve &  $s_i(p_b(i) - \frac{\mu}{\eta_{\text{dis}}}) $ &  - (-) & Undesired\\
			\hline
			J4&-ve & -ve &  $s_i(p_s(i) - \frac{\mu}{\eta_{\text{dis}}}) $ &   - (-) & Undesired\\
			\hline
		\end{tabular}
		\hfill\
	\end{center}
	
	Irrespective of sign of $z_i$,  $s_i^* =  [S_{\max}, S_{\max}]$.

	\section{Stylized Example}
	\label{example}
	
	In this appendix we present a stylized example to demonstrate the operation of the proposed optimal arbitrage algorithm which is composed of Alg. 1 and Alg. 2. Alg. 1 is used to identify a sub-horizon and returns the lower and the upper envelope of battery charge level in the sub-horizon. Alg. 2 is implemented once for a sub-horizon to identify the optimal battery charge level.
	
	We consider the case with $\kappa_i=1$ implies buying and selling price for time instant $i$ are the same and charging and discharging efficiency equal to 1.
	For equal buying and selling price, minimizing the total cost of consumption, i.e., $\sum(z_i+s_i)p_{\text{elec}}(i)$, is equivalent to minimizing the cost of operation of storage, $\sum s_ip_{\text{elec}}(i)$. Here the price of electricity, $p_{\text{elec}}(i) = p_s(i) = p_b(i) ~ \forall ~i$ \cite{hashmi2017}.
% 	\vspace{-5pt}
% 	\begin{remk}
% 		\label{rmk:remark1}

		{ The optimal control decision $x_i^*$
			in the $i$th instant minimizes the function 
			$C_{\text{storage}}^{(i)}(x) -\mu_i^*x$ for $x \in \sbrac{X_{\min}^i, X_{\max}^i}$.
			$x_i^*(\mu)$ is given by:
			%		\vspace{-5pt}
			\begin{equation}
			x_i^*(\mu) =
			\begin{cases}
			[X_{\min}^i,X_{\min}^i], & \text{if }  \mu< p_{\text{elec}}(i), \\
			\sbrac{X_{\min}^i, X_{\max}^i} ,& \text{if }   \mu= p_{\text{elec}}(i), \\
			[X_{\max}^i,X_{\max}^i]  ,& \text{if }   \mu> p_{\text{elec}}(i), 
			\end{cases}
			\label{eq:case1}
			\end{equation}
			%		\vspace{-5pt}
			where $C_{\text{storage}}^{(i)}(x_i)= s_i p_{\text{elec}}(i)$.		
			For $\mu=p_{\text{elec}}(i)$, $x_i^*(\mu)$ takes an {\em envelope} of values. This threshold based structure has a sub-gradient. 
			For any other value of $\mu$ it is a singleton set. 
		}
% 	\end{remk}
	
\begin{figure}[!htbp]
		\center
		\includegraphics[width=3.0in]{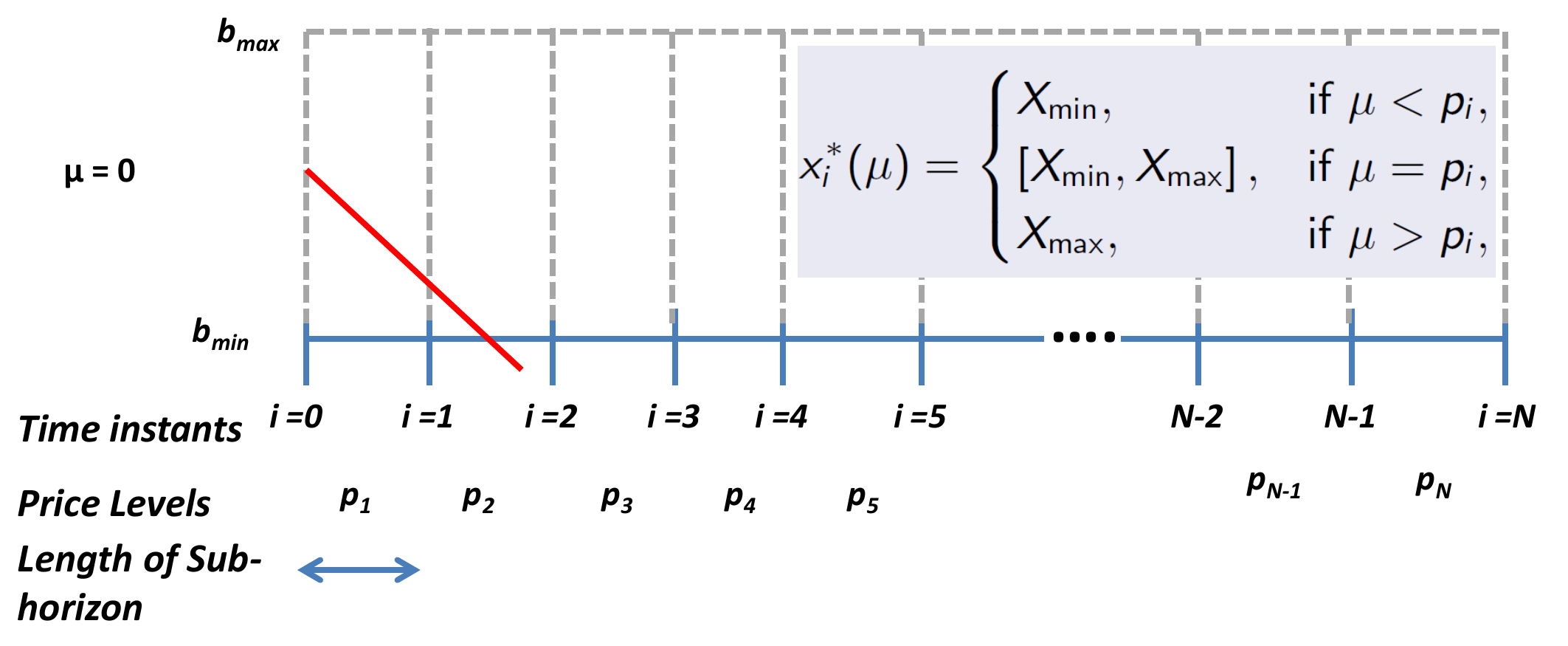}
		\caption{Battery charge level trajectory for $\mu=0$}\label{mu1}
	\end{figure}

	This example considers a lossless battery under equal buying and selling price of electricity. For this example the price of electricity is assumed to be in ascending order (worst case), i.e., $0< p_1< p_2<p_3 ...$ and so on.
	The accumulated Lagrange multiplier ($\mu$) is initiated from zero.
	Fig.~\ref{mu1} shows the battery charge level trajectory for $\mu=0$. The battery charge level has a feasible trajectory till $i=1$. The temporary sub-horizon has sample between $i=0$ and $i=1$. Since $p_1>\mu$ therefore, battery should discharge at maximum rate, based on the threshold based structure.
	
	In the next iteration of the algorithm the accumulated Lagrange multiplier should be increased to the price level in the identified sub-horizon, i.e. $p_1$. The value of $\mu$ is increased because of the lower capacity violation in Fig.~\ref{mu1}, {see Condition (4) of Theorem~\ref{thm:arbitrage}}. With this alteration of $\mu$, the temporary sub-horizon has increased from $i=0$ to $i=2$, as no feasible charge level exists at $i=3$.
	
	\begin{figure}[!htbp]
		\center
		\includegraphics[width=3.0in]{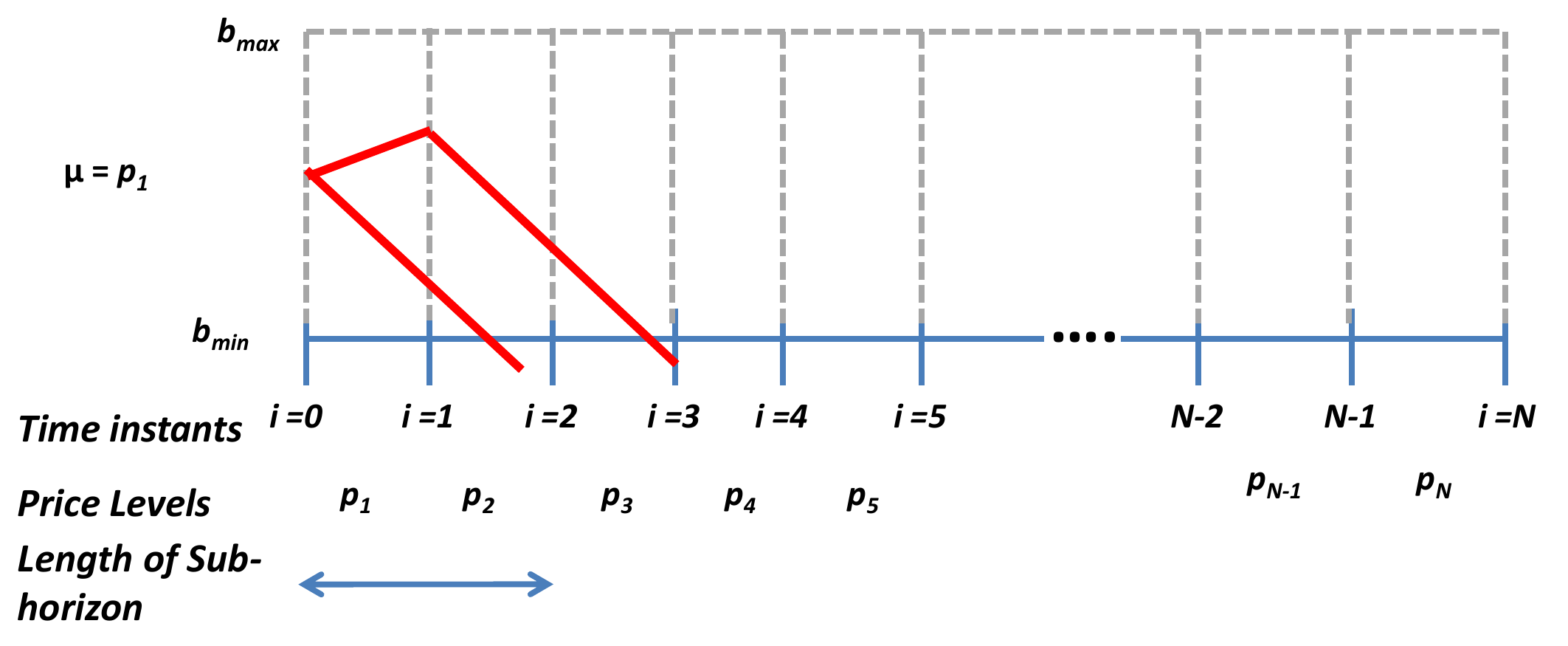}
		\caption{Battery charge level trajectory for $\mu=p_1$}\label{mu2}
	\end{figure}
	
\begin{figure}[!htbp]
		\center
		\includegraphics[width=3.0in]{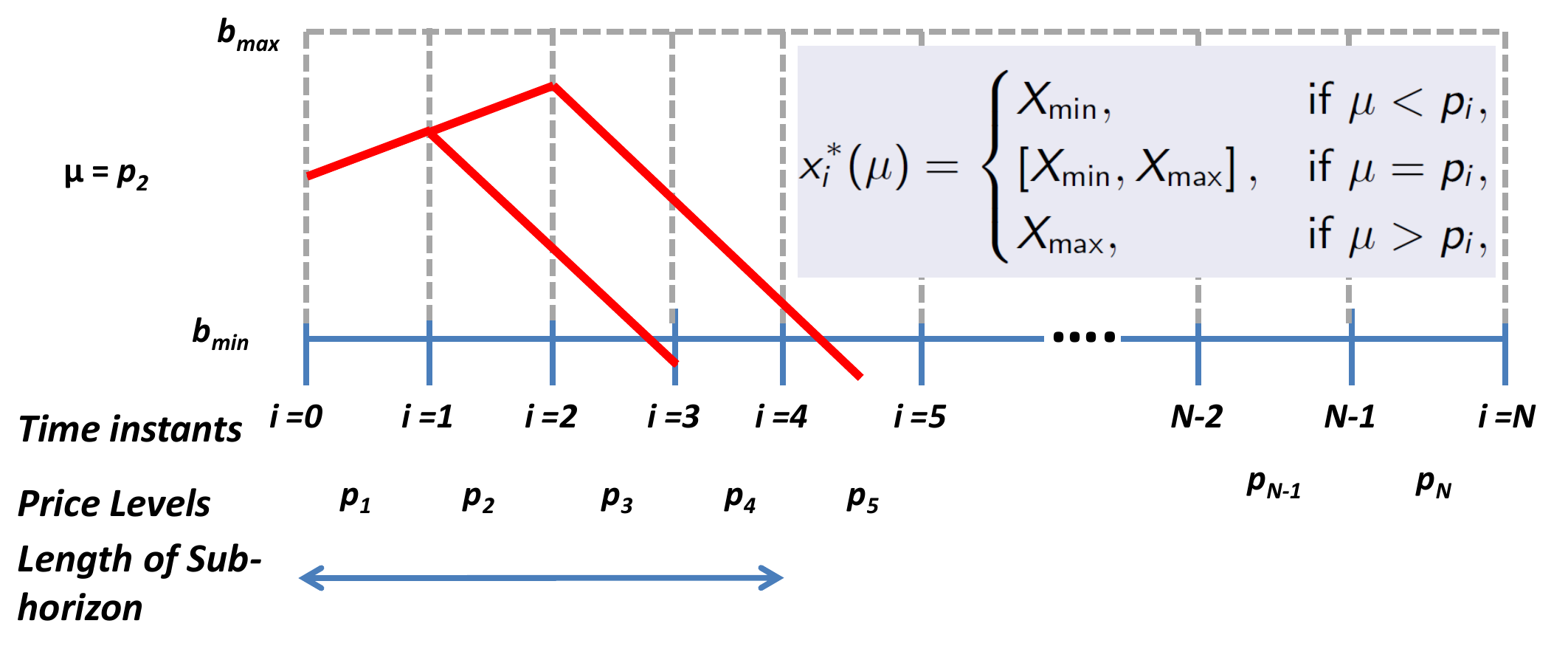}
		\caption{Battery charge level trajectory for $\mu=p_2$}\label{mu3}
	\end{figure}

	In the next iteration the value of $\mu$ is increased to $p_2$. For the first time instant the battery should charge as $\mu > p_1$. For the second time instant the battery charge level has an envelope based structure between $i=1$ to $i=2$. For the third time $\mu < p_3$ therefore, the battery should discharge. The new temporary sub-horizon has increased from $i=0$ to $i=4$, as shown in Fig.~\ref{mu3}.
	
	\begin{figure}[!htbp]
		\center
		\includegraphics[width=3.0in]{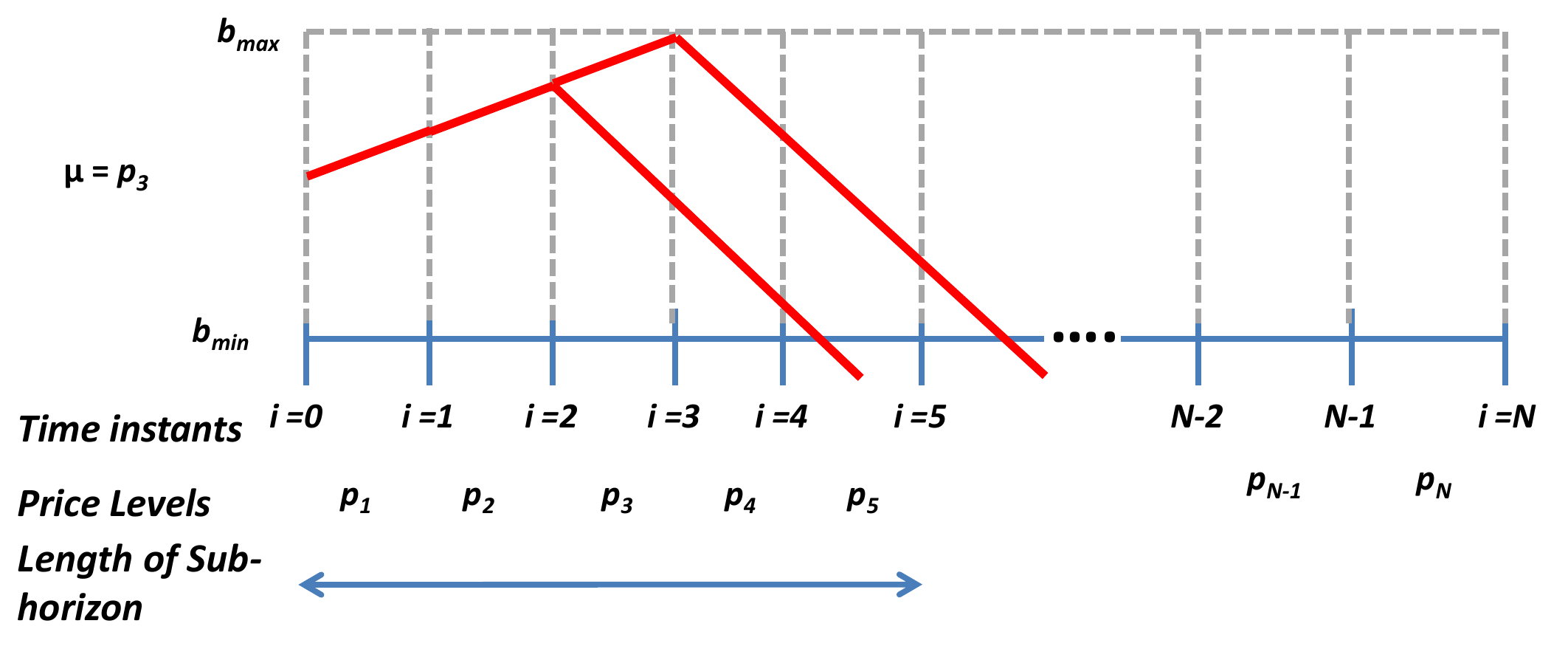}
		\caption{Battery charge level trajectory for $\mu=p_3$}\label{mu4}
	\end{figure}

	\begin{figure}[!htbp]
		\center
		\includegraphics[width=3.0in]{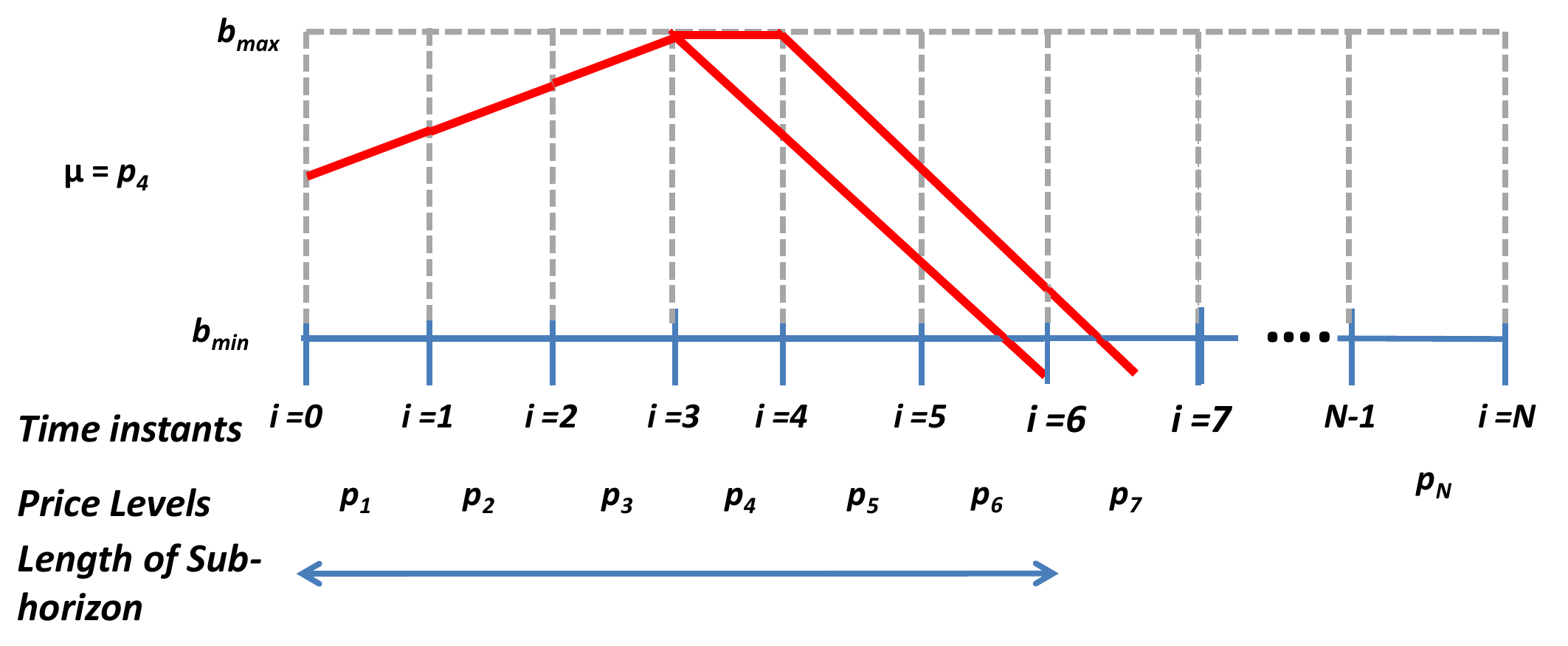}
		\caption{Battery charge level trajectory for $\mu=p_4$}\label{mu5}
	\end{figure}
	
\begin{figure}[!htbp]
		\center
		\includegraphics[width=3.0in]{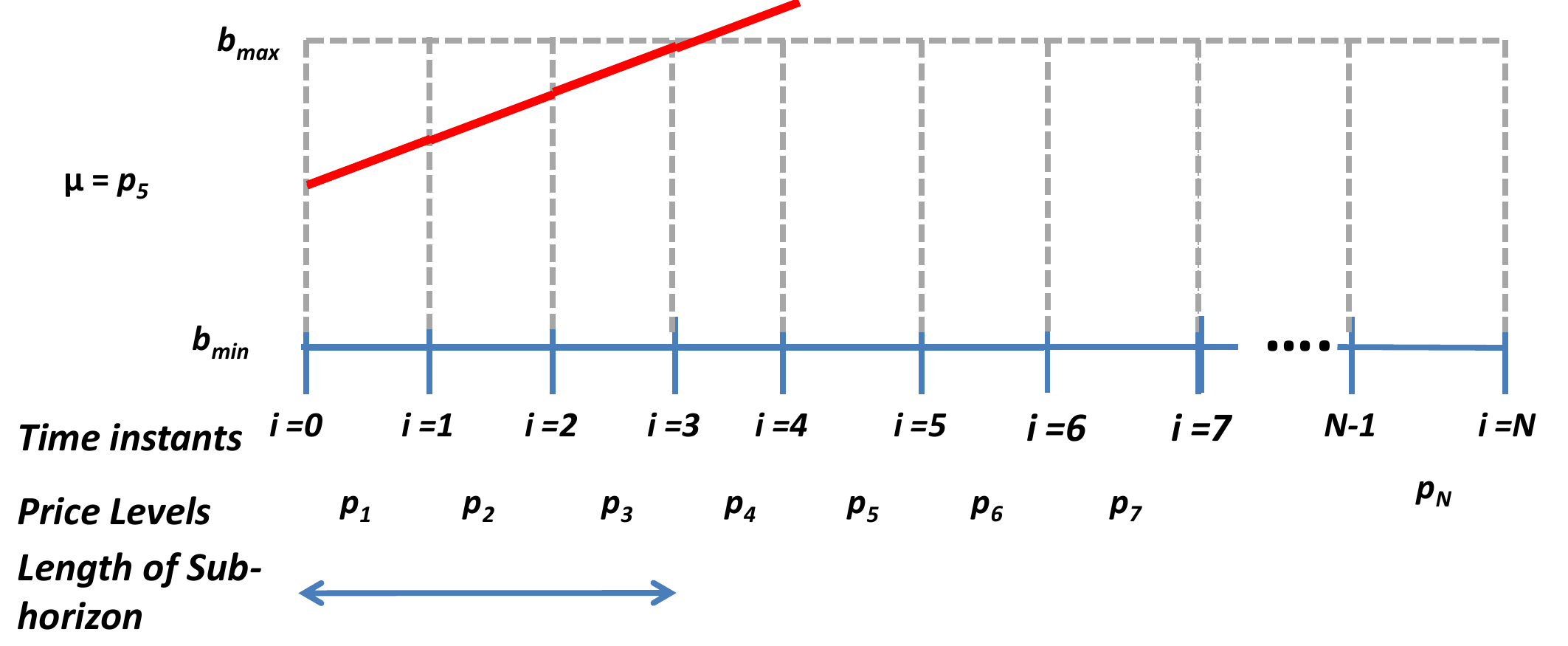}
		\caption{Battery charge level trajectory for $\mu=p_5$}\label{mu6}
	\end{figure}

	Similar adjustment of $\mu$ is performed till the sub-horizon keeps increasing, see Fig.~\ref{mu1} to ~\ref{mu5}. Any further increase in $\mu$ from the case denoted in Fig.~\ref{mu5} decreases the length of the sub-horizon, shown in Fig.~\ref{mu6}. 
	
	The value of $\mu$ associated with case shown in Fig.~\ref{mu5} is called the optimal accumulated Lagrange multiplier denoted as $\mu^*$ for the sub-horizon. For this example $\mu^*=p_4$. The value of battery charge level trajectory associated with $\mu^*$ is the input to Alg.2. Note the value of $\mu^*$ acts as the shadow price and remains constant for a sub-horizon. The value of $\mu^*$ is selected from finite number of discrete levels of electricity price in the sub-horizon. This makes the proposed algorithm computationally very efficient due to this discretization of a continuous optimization problem.

\begin{figure}[!htbp]
		\center
		\includegraphics[width=3.0in]{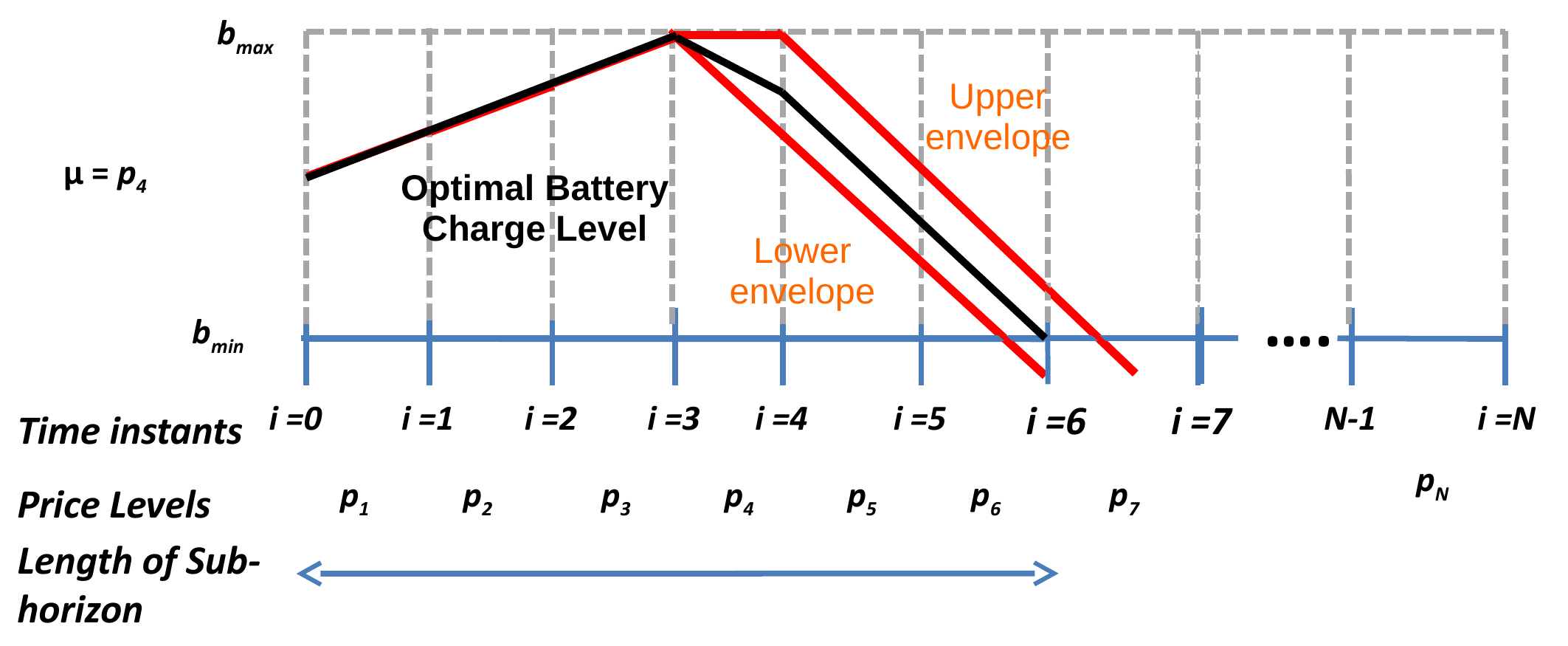}
		\caption{Optimal Battery charge level}\label{style}
	\end{figure}

	Alg. 2 takes as input the envelope of battery charge level and identifies the optimal solution. \texttt{BackwardStep} algorithm fix the last time instant in the sub-horizon at $b_{\min}$ and back-calculates the optimal battery charge level. For time instant where $\mu^*$ is equal to the price level, anything from $X_{\min}$ to $X_{max}$ is possible. Note neither charging or discharging is profitable or loss here, however, the battery could charge some more if there are possible discharging opportunities in adjacent time periods in the sub-horizon. Similarly, the battery could discharge here if there are lower discharging opportunities in adjacent time periods in the sub-horizon. This period provides a slack in adjusting battery charge level. It is essential to note that in this period intermediate ramp rate could be possible, contrary to all prior works on threshold based optimal decision making. The optimal battery capacity is shown in Fig.~\ref{style}.
	
	The next sub-horizon begins at $i=7$ as the optimal actions have been identified from $i=0$ to $i=6$.

%	
%	\section{Additional Numerical Results}
%	\label{numappendix}

\end{document}